\documentclass[fleqn,usenatbib]{mnras}
\usepackage{newtxtext,newtxmath}
\usepackage[T1]{fontenc}

\DeclareRobustCommand{\VAN}[3]{#2}
\let\VANthebibliography\thebibliography
\def\thebibliography{\DeclareRobustCommand{\VAN}[3]{##3}\VANthebibliography}

\usepackage{graphicx}	% Including figure files
\usepackage{amsmath}	% Advanced maths commands

\title[JWST WDM]{JWST Lensed quasar dark matter survey II: Strongest gravitational lensing limit on the dark matter free streaming length to date}

\author[R. E. Keeley et al.]{Ryan E. Keeley,$^{1}$ \thanks{E-mail: rkeeley@ucmerced.edu}
A.~M.~Nierenberg,$^{1}$
D.~Gilman$^{2,3,4}$,
C. ~Gannon$^1$,
S.~Birrer$^{5}$,
T.~Treu$^{6}$,
\newauthor
A.~J.~Benson$^{7}$,
X.~ Du$^{6}$,
K.~N.~Abazajian$^{8}$,
T.~Anguita$^{9,10}$,
V.~N.~Bennert$^{11}$,
S.~G.~Djorgovski$^{12}$,
\newauthor
K.~K. Gupta$^{13,14}$,
S.~F.~Hoenig$^{15}$,
A.~Kusenko$^{6,16}$,
C.~Lemon$^{17}$,
M.~Malkan$^{6}$,
V.~Motta$^{18}$,
\newauthor 
L.~A.~Moustakas$^{19}$,
Maverick~S.~H.~Oh $^{1}$,
D.~Sluse$^{13}$,
D.~Stern$^{19}$,
R.~H.~Wechsler$^{20,21,22}$
\medskip\\
$^1$ University of California, Merced, 5200 N Lake Road, Merced, CA 95341, USA \\
$^2$ Department of Astronomy $\&$ Astrophysics, University of Chicago, Chicago, IL 60637, USA \\
$^3$ Department of Astronomy and Astrophysics, University of Toronto, 50 St.\ George Street, Toronto, ON, M5S 3H4, Canada\\ 
$^4$ Brinson Prize Fellow \\
$^5$ Department of Physics and Astronomy, Stony Brook University, Stony Brook, NY 11794, USA\\
$^6$ UCLA Physics \& Astronomy, 475 Portola Plaza, Los Angeles, CA 90095-1547, USA\\
$^{7}$ Carnegie Institution for Science, Pasadena CA 91101, USA \\
$^{8}$ Department of Physics and Astronomy,  University of California, Irvine, CA 92697-4575, USA\\
$^{9}$ Instituto de Astrofisica, Departamento de Ciencias Fisicas, Universidad Andres Bello, Chile\\
$^{10}$ Millennium Institute of Astrophysics, Chile\\
$^{11}$ Physics Department, California Polytechnic State University, San Luis Obispo, CA 93407, USA \\
$^{12}$ California Institute of Technology, Pasadena CA 91125, USA \\
$^{13}$ STAR Institute, University of Li\`ege, Quartier Agora - All\'ee du six Ao\^ut, 19c B-4000 Li\`ege, Belgium \\
$^{14}$ Sterrenkundig Observatorium, Universiteit Gent, Krijgslaan 281 S9, B-9000 Gent, Belgium \\
$^{15}$ School of Physics and Astronomy, University of Southampton, Southampton SO17 1BJ, United Kingdom \\
$^{16}$ Kavli Institute for the Physics and Mathematics of the Universe (WPI), UTIAS, The University of Tokyo, Kashiwa, \\ Chiba 277-8583, Japan \\
$^{17}$Oskar Klein Centre, Department of Physics, Stockholm University, SE-106 91 Stockholm, Sweden \\
$^{18}$ Instituto de F\'{\i}sica y Astronom\'{\i}a, Universidad de Valpara\'{\i}so, Avda. Gran Breta\~na 1111, Valpara\'{\i}so, Chile \\
$^{19}$ Jet Propulsion Laboratory, California Institute of Technology, 4800 Oak Grove Dr, Pasadena, CA 91109\\
$^{20}$ Kavli Institute for Particle Astrophysics \& Cosmology, P.O. Box 2450, Stanford University, Stanford, CA 94305, USA \\
$^{21}$ Department of Physics, Stanford University, 382 Via Pueblo Mall, Stanford, CA 94305, USA \\
$^{22}$ SLAC National Accelerator Laboratory, Menlo Park, CA 94025, USA \\
}

\begin{document}
\label{firstpage}
\pagerange{\pageref{firstpage}--\pageref{lastpage}}
\maketitle

\begin{abstract}

This is the second in a series of papers in which we use JWST MIRI multiband imaging to measure the warm dust emission in a sample of 31 multiply imaged quasars, to be used as a probe of the particle nature of dark matter. We present measurements of the relative magnifications of the strongly lensed warm dust emission in a sample of 9 systems. The warm dust region is compact and sensitive to perturbations by populations of halos down to masses $\sim 10^6$ M$_{\odot}$. Using these warm dust flux-ratio measurements in combination with 5 previous narrow-line flux-ratio measurements, we constrain the halo mass function. In our model, we allow for complex deflector macromodels with flexible third and fourth-order multipole deviations from ellipticity, and we introduce an improved model of the tidal evolution of subhalos. We constrain a WDM model and find an upper limit on the half-mode mass of $10^{7.6} M_\odot$ at posterior odds of 10:1. This corresponds to a lower limit on a thermally produced dark matter particle mass of 6.1 keV. This is the strongest gravitational lensing constraint to date, and comparable to those from independent probes such as the Ly$\alpha$ forest and Milky Way satellite galaxies. 

\end{abstract}


%%%%%%%%%%%%%%%%% BODY OF PAPER %%%%%%%%%%%%%%%%%%

\section{Introduction}
Identifying the nature of dark matter (DM) is one of the most compelling endeavors of modern 
physics.  
The standard cold dark matter (CDM) paradigm accurately describes the abundance of DM, its distribution on large scales (e.g. the cosmic microwave background, ~\citealt{planck_collaboration_planck_2020}, and the cosmic web, ~\citealt{Tegmark_2004_cdm_lss}), and the profiles of DM halos on 
%cluster and 
galactic scales where DM halos host observable galaxies~\citep{white_core_1978,white_galaxy_1991,deBlok_2008_profiles,weinberg_cold_2015}. 
It is on subgalactic scales where the frontier of tests of the CDM paradigm lie \citep{bullock_small-scale_2017}.

Common tests attempt to probe DM via potential interactions with the Standard Model~\citep{2022arXiv220907426C}. Complementary to such approaches is to probe DM physics via its known gravitational interaction. The laboratories for such gravitational probes of DM are found in the Universe where gravity has collapsed DM into bound structures, referred to as halos. Characterizing the distribution and profiles of DM halos serves as a probe into the microphysics of DM, such as the DM particle mass, production mechanism, and potential self-interactions~\citep{bullock_small-scale_2017,buckley_gravitational_2018,2019BAAS...51c.207B,2019arXiv190201055D,2022JHEAp..35..112B}.

One alternative to CDM is warm dark matter (WDM). In this class of models, DM has a non-negligible velocity in the early Universe, which causes DM particles to escape the smallest peaks in the density field and prevents the formation of halos below a corresponding free-streaming length scale \citep{bode_halo_2001, schneider_non-linear_2012, bose_copernicus_2016, ludlow_mass-concentration-redshift_2016}.  
At later times, this free streaming effect causes a suppression in the abundance of DM halos below a cutoff halo mass.  Both the free-streaming length and cutoff in the halo mass function can be predicted for any DM theory for a given particle mass and production mechanism.
Another difference between CDM and WDM predictions is that WDM halos are also less concentrated than their CDM counterparts~\citep{bose_copernicus_2016,ludlow_mass-concentration-redshift_2016}.

DM halos with (infall) masses greater than a few times $10^{8}$ M$_\odot$ 
generally 
contain detectable stars and gas \citep{nadler_milky_2020}, which can provide a direct means of measuring their abundances and their internal density profiles in the Local Group \citep[albeit with some challenges, given the small baryon content of the galaxies, ][]{bullock_small-scale_2017}. 
However, below these masses, halos are decreasingly likely to host stars, and alternate tracers are required, which do not require the halos to contain stars and gas. 
In the Local Group, tidal streams probe the DM distribution on sub-galactic scale \citep{Bovy_stream,banik_stream_2018,Bonaca_stream,banik_evidence_2021,banik_evidence_2021_2}, while strong gravitational lensing can probe sub-galactic scales at cosmological distances.

Strong gravitational lensing is sensitive to the characteristics of the population of DM halos directly and thus can be used to test a range of alternative dark matter models, including WDM~\citep{treu10,vegetti_strong_2023}.
Strong lensing consists of light from a background source being multiply imaged as a result of the deflection by the gravitational potential of all matter along its trajectory, including the mass of the main lens, the DM subhalos of the main lens, and the DM halos along the line-of-sight. 
The derivative of the gravitational potential determines the positions of images, and the second derivatives determine the magnifications.
Since the overall mass distribution is mostly smooth on the scale of the galaxy, the macromodel for the main deflector primarily determines the first derivative of the gravitational potential and thus the image position.  Meanwhile, the mass distribution is clumpy on the scales of line-of-sight halos and subhalos, and thus they contribute mostly to the second derivative of the potential and thus the image magnifications.

Characterizing the population of DM halos and subhalos via the perturbations they cause in images is referred to as substructure lensing.
This substructure lensing signal does not rely on the presence of a luminous galaxy in the DM halo and thus this technique can probe halos and subhalos at masses lower below the threshold above which we expect them to form stars, and thus beyond what is feasible by counting luminous satellites. 

Previous studies have used a variety of methods to constrain DM physics from gravitational imaging of radio images~\citep{vegetti_constraining_2018,hsueh_sharp_2020,minor_unexpected_2021,laroche_quantum_2022,powell_lensed_2023}, to interpreting the abundance of gravitational lenses in cluster environments~\cite{meneghtetti_excess_2020,yang_self-interacting_2021}.

We focus on the flux ratios of quadruply lensed images of quasars. In this method, a model of the main deflector, on top of a model for the source, gives predictions for what the observed flux ratios should be.  The additional subhalos and line-of-sight halos will perturb the flux ratios away from the predictions of the main deflector. Constraints on the WDM model come in the form of a relative likelihood for a WDM model to predict observed anomalous flux ratios compared to the CDM model.  Such a likelihood is evaluated using a simulation-based inference method. 
The analysis of quasar flux ratios has led to a number of previous constraints on DM physics, including WDM
~\citep{mao_evidence_1998,dalal_direct_2002,gilman_warm_2020, zelko_constraints_2022}, 
fuzzy DM~\cite{laroche_quantum_2022},
self-interacting DM~\cite{gilman_strong_2021,
%yang_self-interacting_2021,
gilman_constraining_2023},
primordial black holes~\citep{dike_strong_2023}, sterile neutrinos \citep{zelko_constraints_2022},
as well as constraints on the primordial power spectrum~\cite{gilman_primordial_2022}. Stronger constraints can be obtained in combination with complementary methods to break degeneracies, \citep[e.g.,][]{nadler_dark_2021}.

One of the primary limitations to date of studies of DM using flux ratios has been the small sample of lenses available.  To ensure perturbations are due to dark matter halos rather than microlensing, the lens source must be $>$ milli-arcseconds in size, which is larger than $\sim 1$ pc given typical source-lens configurations. Previous studies have used radio and quasar narrow-line emission to probe the properties of dark matter. The Mid Infrared Instrument (MIRI) on JWST makes it possible to expand the sample dramatically by measuring flux ratios of the quasar warm dust region \citep{Nierenberg_0405_2023}. The warm dust region has typical sizes of $\sim1-10$ pc~\citep{burtscher_diversity_2013,leftley_parsec-scale_2019}. This size makes it insensitive to microlensing while still sensitive to perturbations from very low mass dark matter halos with masses of M$_\odot\sim 10^6$ \citep{Nierenberg_0405_2023}. 

As part of JWST-GO-2046 (PI Nierenberg), we are observing 31 lensed quasars to measure the warm dust flux ratios and infer the properties of dark matter \citep{Nierenberg_0405_2023}. In this second paper of the series, we present warm dust flux ratios for the first 9 observed systems. We then combine these flux ratios with previoulsy published radio and narrow-line measurements of other systems to constrain the free-streaming length of DM.

The outline of the paper is as follows: in Sec~\ref{sec:obs}, we describe the observations. In Sec~\ref{sec:fluxes}, we describe the procedure we use to fit the images and measure the flux of the lensed quasars and discuss these intermediate results in Sec.~\ref{sec:FluxResults}. In Sec.~\ref{sec:SED}, we describe the procedure we use to fit the spectral energy distribution of these quasars and thus measure the flux ratios of the warm dust component, the results of which are discussed in Sec.~\ref{sec:SEDResults}. In Sec~\ref{sec:DM}, we describe how we implement the WDM model and the statistical procedure we use to test it. We present the results of our WDM inference in Sec.~\ref{sec:results}, and elaborate on our uncertainty budget in Sec.~\ref{sec:uncertainty_budget}.
We compare our results to previous results in Sec.~\ref{sec:discussion}, and we summarize our findings in Sec.~\ref{sec:summary}.

\section{Observations and Initial Reduction}\label{sec:obs}

 Targets for JWST-GO-2046 were selected to be quadruply imaged quasars with detected WISE W4 fluxes (unresolved), with image separations larger than 0\farcs1 to ensure that separate image flux ratios could be well measured. A more detailed description of the observation strategy is provided by \cite{Nierenberg_0405_2023}. In this paper, we present flux-ratio measurements for the first 9 lenses that were observed in the program. Table \ref{tab:lens_list} provides a list of the targets studied in this paper as well as source and deflector redshifts, observation dates, and discovery papers.

Following ~\citet{Nierenberg_0405_2023}, the initial calibration was done with the JWST pipeline. After the completion of the paper by \cite{Nierenberg_0405_2023}, there was a significant update in the MIRI absolute flux calibration, which accounts for the time dependence observed in the detector throughput during the first year of observations as well as for a correction to the F560W absolute flux due to the cruciform artifact (see Section \ref{sec:psf_fitting}). The data presented in this paper is reduced using CRDS version 11.16.21 and supersedes those presented by \citet{Nierenberg_0405_2023}. We note that continual updates are being made to the MIRI calibration file and reduction pipeline, and therefore we anticipate that the flux values and uncertainties presented in this work may require future adjustment. However, our estimate of the residual systematics should be sufficiently large to account for future changes. Therefore we expect our precision to improve as calibrations improve, while our conclusions to remain qualitatively unchanged.
The sky subtraction was done with a customized routine based on \url{https://github.com/STScI-MIRI/Imaging_ExampleNB}. 
The pixel scale was set to 0\farcs11, the native detector scale. Reduced images are shown in Appendix~E.

\begin{table*}
    \centering
    \begin{tabular}{llllll}
    \hline
    \hline
     Lens           &Abbrev. Name. & Source z & Lens z & Obs. Date. & Discovery paper(s) \\
    \hline
     DES J040559.7$-$330851 &J0405    &1.713       &0.3$^{a,1}$   & Oct. 12 2022    & \cite{anguita_strong_2018} \\
     GraL J060710.8$-$215217 &J0607    &1.302       &0.55$^2$      & Feb. 22 2023   & \cite{Stern_2021_gral, lemon2023} \\
     GraL J060841.4+422937 &J0608     & 2.345      &              & Feb. 23 2023  &\cite{Stern_2021_gral, lemon2023} \\
     GraL J065904.1+162909 &J0659     & 3.083      & 0.766$^3$    & Feb. 27 2023  &\cite{delchambre_gaia_2019, lemon2023} \\
     W2M J104222.1+164115  &J1042     & 2.517      & 0.5985          &  Dec. 15 2022 &\cite{Glikman_1042_2023} \\
     J153725.3$-$301017  &J1537        & 1.721      &  0.592            & Mar. 7 2023  &\cite{lemon_gravitationally_2018} \cite{delchambre_gaia_2019} \\
     PS J1606$-$2333    &J1606        & 1.696      & 0.3$^{a,1}$  & Mar. 8 2023   & \cite{lemon_gravitationally_2018} \\
     WFI J2026$-$4536    &J2026        & 2.23       &              & April 15 2023  & \cite{Morgan_2026_2033_2004} \\
     DES J203802.7$-$400814 &J2038    & 0.777      & 0.230        & April 18 2023  & \cite{agnello_meets_2018} \\
    \end{tabular}
    \caption{Information about the lens systems and observation details. References are provided for deflector redshift measurements when the reference is different from the discovery paper. ($a$) photometric redshift. References: $1.$ \citep{gilman_constraints_2020}, $2.$ Mozumdar et al. (in prep.) $3.$\citep{Stern_2021_gral}}
    \label{tab:lens_list}
\end{table*}

\section{Measuring image fluxes}\label{sec:fluxes}
We follow the image-fitting procedure described by~\citet{Nierenberg_0405_2023}, aiming to measure the fluxes of the lensed images of the quasar, which appear as point sources.  Since images of the deflector galaxy and the lensed host galaxy (which appears as an arc) are also often present in the data, we simultaneously measure the fluxes of the point sources in addition to other components. When the deflector galaxy or host galaxy is not detected in the data, we do not include it in our modelling.

\subsection{Model Components}
Here we list how the individual components are modelled, when they are apparent in the data. In Section \ref{sec:image_fitting} we describe our model-fitting process, used to determine which components are necessary to fit the data.
\begin{itemize}
    \item Lensed quasar image: four point sources whose positions and fluxes are not determined by the lens model. This is to make the measurements independent of the gravitational lensing model, which is separately be fit  when inferring the presence of DM substructure. 
    \item Lensed quasar host galaxy images: modelled as a source component (an intrinsic surface brightness distribution as it would appear in the source plane) distorted by distorted by a foreground deflector. The source is modelled as a S\'ersic profile~\citep{sersic_influence_1963} with variable S\'ersic index $n$. The distortion of the deflector is modelled as arising from an elliptical power-law mass density profile \citep{tessore_elliptical_2015}, with external shear. 
    For several systems, additional source complexity was apparent, and we added shapelets implemented as a Gauss-Hermite polynomial basis with increasing order until the image likelihood was no longer improving. We found that the maximum improvement typically occurred for shapelet order $n_{\rm{max}}<5$.
    \item Deflector galaxy light: modelled as an elliptical S\'ersic profile. Because the deflector is so much fainter than the quasar point sources, we find that its light profile is poorly constrained and therefore restrict the S\'ersic index $n$ to be 4. Given that the deflector makes an extremely small contribution to the flux at the location of the quasar images, we do not expect this assumption to impact our measured flux ratios.  For J0607 a small luminous galaxy is also observed in several filters. We include this galaxy in the lens model as a Singular Isothermal Sphere and add a circular S\'ersic profile to model the light. For J0659, there is a nearby object with size and colors consistent with a star. We consider it to be a star and do not include it in our lens model. 
\end{itemize}
    
\subsection{Point Spread Function Modelling}
\label{sec:psf_fitting}
Following~\citet{Nierenberg_0405_2023}, we used \texttt{webbpsf}\footnote{Development branch 1.2.1.} \citep{perrin_simulating_2012, perrin_updated_2014-1} to model the point spread function \citep[PSF; e.g.][]{argyriou_brighter-fatter_2023}.
The PSF generated by \texttt{webbpsf} depends on both wavelength distribution of flux (i.e., the spectral energy distribution) in each band, as well as a `jitter'.  We incorporate the wavelength dependence with a blackbody at the source redshift of each lens.  The jitter is implemented in \texttt{webbpsf} by convolving the calculated PSF with a Gaussian kernel whose width is set by a \texttt{jitter\_sigma} parameter. We vary the \texttt{jitter\_sigma} parameter and the blackbody temperature for each lens for each filter separately. The jitter accounts for charge diffusion in the detector

The F560W filter of JWST contains a `cruciform' artefact~\citep{gaspar_quantum_2021, wright_mid-infrared_2023}, which adds a cross pattern to the PSF. This feature does not arise from any optical component but from the detector.
The second extension of the \texttt{webbpsf} contains a model for the cruciform artifact but frequently over-predicts this feature. To account for this, we take the weighted average between the second frame with the cruciform artifact and the $0^{\rm th}$ frame, which does not have the feature (psf = $f$*psf$_0+(1-f)$*psf$_2$, where psf$_0$ is the psf of the 0th frame and psf$_2$ is the psf of the 2nd frame). The parameter $f$, the fractional weight of the $0^{\rm th}$ frame, is varied for the F560W filter, along with the other PSF parameters.

\subsection{Image Fitting}
\label{sec:image_fitting}
We use an iterative process to fit the imaging data. The general method is the same but we have added several additional steps relative to that presented by \cite{Nierenberg_0405_2023}. 

We begin fitting the data with the simplest possible model: four unlensed point sources in the image plane. Starting with F560W, we iteratively optimize the image positions, and the PSF model parameters until both have converged. Once this is complete, we visually examine the residuals of the best-fit model to look for missing light components. 

If warranted, we add additional model components, gradually increasing the model freedom following \citet{schmidt_strides_2023}. If a lensed arc is visible, we initialize a power-law ellipse mass distribution (PEMD) lens model with a fixed power-law slope of $\gamma=2$. The quasar point-source positions are determined by the lens model in this step, and required to have the same centroid as the lensed quasar host galaxy. Light components are added with S\'ersic index fixed at 4. Still working in the single band, we iteratively optimize the lens and light model parameters, and the PSF parameters until both have converged. As expected, during this step we see significant updates to the best-fit PSF parameters.

Once the previous step has converged, we begin simultaneously fitting the data in all four filters, initializing with the best-fit model for F560W.  In each filter, we begin with only the point sources and examine the residuals to determine whether additional model components are needed in these filters. Given the very broad wavelength range we do not expect all components to be detectable or to have the same effective radii across all wavelengths, thus the model parameters for each luminous component are independent in each filter with the exception of the component centroids, which are held fixed across all filters. Naturally, the mass distribution of the lens itself is assumed to be the same across all filters.
If a galaxy is detected close in projection to the lens (as in J0607), then its mass component is included in all filters, while its light is only included in the necessary filters. We again iteratively optimize the model parameters and the PSF parameters until both have converged.

If the reduced $\chi^2$ is greater than 1 in a given filter after this step we add shapelets to the quasar host galaxy in that filter with increasing complexity until the reduced $\chi^2$ is no longer improving. We iteratively optimize model parameters and PSF parameters until both have converged.

Finally, we allow the S\'ersic indices, and then the slope of the lens mass profile $\gamma$ to vary. We continue to iteratively optimize the PSF and the model parameters. Typically, when $\gamma$ is allowed to vary, the best fit is close to $\gamma\sim2.0$, except for J1537. The S\'ersic indices are less constrained and thus vary over a larger range, yet without affecting the fit and the resulting measured flux ratios.  

In the last step, if lens modelling was used in the previous steps, we switch to having the quasars be independent point sources in the image plane.
This ensures that the measurement of the image fluxes and positions is not directly tied to a specific lens model. The point of performing our fitting procedure in the the discussed order is to assure is to ensure that the best-fit parameters lie in a physically motivated region.

\begin{figure*}
    \includegraphics[width=\textwidth]{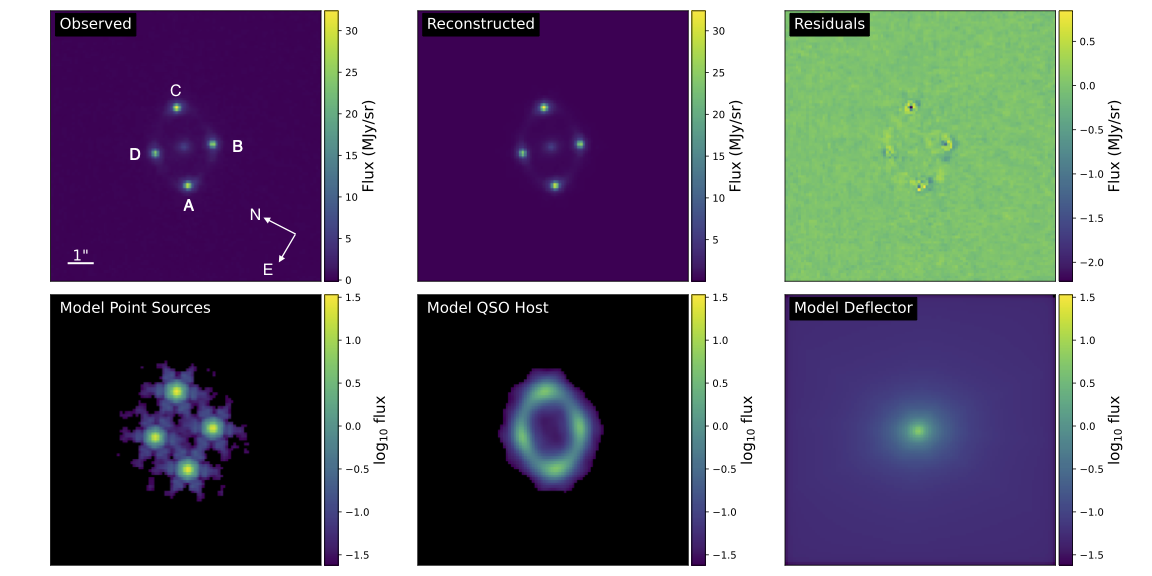}
    \caption{Example results of the image fitting procedure for J1537-3010 in the F560W band. This example shows each of the components used in the image fitting procedure, the lensed emission from the quasar that appears as point sources, the lensed extended emission from the quasar host galaxy, and light from the deflector galaxy. The top left panel shows the observed image, the top center shows the best-fit model, and the top right shows the residuals.  The bottom panels show the individual model components.}\label{fig:ex_im_fit}
\end{figure*}

Figure~\ref{fig:ex_im_fit} shows an example of the output of our modelling procedure for the F560W filter for J1537. Figures with the model output for each filter of each lens can be found in the supplementary materials section (Appendix~\ref{app:individual_lens_fits}).

\subsection{Measurement uncertainties}
\label{sec:measurement_uncertainties}

We adopt the flux ratio measurement uncertainties based on the testing of \cite{Nierenberg_0405_2023}. If the surface brightness of the lensed quasar host galaxy at the location of the point sources is greater than $\sim 50\%$ of the PSF fluxes, then we assume approximately 6\% flux-ratio uncertainties. If we have to include the lensed quasar host galaxy in the model but it is fainter than $\sim 50\%$ of the PSF fluxes, we assume 2\% flux-ratio uncertainties. If there is no detection of the lensed quasar host galaxy, then we assume 1\% flux-ratio uncertainties.  We adopt 0\farcs005 position uncertainties based on comparisons between our measured image positions and previously published observations of these systems with HST \citep{schmidt_strides_2023, shajib_is_2019, Glikman_1042_2023, nierenberg_double_2020}.

For the system J1042, we adopt different uncertainties. This system has a pair of images with a flux ratio ranging from about 10:1 in F560W to 4:1 in F2550W, and a separation of only 0\farcs5, which is smaller than the PSF FWHM (0\farcs591, 0\farcs803) in F1800W and F2550W respectively. A potential concern is that if the PSF properties vary systematically with brightness \citep[e.g. the brighter-fatter effect,][]{argyriou_brighter-fatter_2023} over this dynamic range, measuring the fluxes with a fixed PSF for all four images may yield a systematic bias. To check this, we performed a fit to the real F2550W data allowing the PSF to vary for each image, and found no trend between the inferred {\tt jitter\_sigma} and the image brightness. The inferred {\tt jitter\_sigma} varied at the 10\% level between images, although there was no trend with image flux. This variation is enough to make a significant difference in the measured image fluxes relative to holding this parameter fixed in the fitting. 
However, to be cautious, for this system we adopt a 10\% flux-ratio uncertainty for image B in all filters, and a 5\% flux-ratio uncertainty for the other images. Due to the small image separation and large flux differences, we also adopt larger astrometric uncertainties of 0\farcs01 for this system.

A new MIRI calibration pipeline was released subsequently to the analysis of \cite{Nierenberg_0405_2023}, with significant changes to the estimated zero-points, as well as estimates of the zero-point uncertainties. These calibrations also now account for the change in MIRI throughput over the observation period. Although these uncertainties are typically very small $<1\%$, given our quasar fluxes, the absolute image flux uncertainty is likely dominated by PSF modelling uncertainties. We adopt 10\% uncertainties on the absolute flux calibration based on the estimates of \cite{Nierenberg_0405_2023}. The absolute flux uncertainty is relevant for the spectral energy distribution fitting, in which we isolate light coming from specific physical regions of the quasar, as described in Section \ref{sec:SED}. In our final DM analysis paper for the full sample, we will explore how refining this uncertainty estimate would impact our DM inference.

\section{Results of image fitting}\label{sec:FluxResults}

The measured image positions and flux ratios of the point sources are presented in Table \ref{tab:data_summary}. We provide the flux ratios in the bluest (F560W) and reddest (F2100W or F2550W) filters. In Figure \ref{fig:fr_figure} we show how the flux ratios vary as a function of wavelength. We also provide the inferred point source fluxes for all wavelengths in Table ~\ref{tab:fluxes}.

The parameters of the extended emission can be found in  Table~\ref{tab:macromodel}, the parameters of the source are found in Table~\ref{tab:source}, and the parameters of the galaxy and galaxy satellite light are found in Table~\ref{tab:light}.

With the exception of J1606 and J1042, all the lenses had
  extended emission from the lensed source galaxy in the bluest filter, F560. J0607, J1537, and J1606 had extended emission from lensed source galaxy in the reddest filters (F2100W for J0607 and J1537, and F2550W for J1606). All of the lenses with source light detected in the reddest filter have sources with redshift less than 2, and the reddest filter corresponds to $\sim$ 8--10 $\mu$m  rest-frame for these systems. The corresponding physical sizes are $\sim$1--3 kpc, however we caution that these values are inferred with a lens modelling procedure optimized to accurately measure the point source image positions and flux ratios, and we leave a more robust inference of the properties of the extended source light to a separate work.

Figure \ref{fig:fr_figure} shows that many of the lenses have chromatic variations in the flux ratios. This is due to the fact that at bluer wavelengths, the quasar spectral energy distribution becomes dominated by light from the quasar accretion disk, which is small enough to be microlensed \citep{sluse_mid-infrared_2013}. In the following section, we describe how we use SED fitting to account for possible microlensing of physically smaller regions in the light source.

\section{SED Fitting}\label{sec:SED}
As described by \cite{Nierenberg_0405_2023}, our goal is to measure the light emitted from the `warm dust' region of the quasar, which is large enough to avoid contamination from microlensing while still being small enough to be sensitive to low-mass halos. Even the reddest MIRI filter contains some amount of contamination from the quasar `hot dust' component. We account for this by fitting the spectral energy distribution (SED) of the quasars based on the four image bands. This procedure is nearly identical to the one presented by \cite{Nierenberg_0405_2023}, which follows \cite{sluse_mid-infrared_2013} by modelling the quasar with a power-law continuum, hot dust blackbody, and warm dust black body.  The continuum power-law models emission from the quasar's accretion disk and the two blackbody models represent emission from the warm and hot dust regions of the quasar.
We do not include emission lines such as polycyclic aromatic hydrocarbons since their contribution to the broadband flux is expected to be below percent level for our quasar sources given the MIRI band pass widths ~\citep[e.g.][]{Tommasin_spitzer_2010,  Jensen_PAH_2017,garcia-bernete_high_2022}. The SED fitting gives us a way to propagate possible microlensing contamination into flux-ratio uncertainties, with physically motivated priors on how the different SED components might vary.

The temperature and normalization of the warm dust blackbody component for image A, as well as the flux ratios for the warm dust component in the other images (B/A, C/A, D/A) are all independent parameters that are free to vary.
To account for the fact that both the accretion disk and the hot dust region are small enough that they can be microlensed, the amplitude of each of the components in each image is independent and free to vary. Further, the slope of the continuum power-law component was free to vary between the different images, also to account for microlensing. This parameterization also accounts for the intrinsic variability of the accretion disk's luminosity, which can vary on timescales shorter than the time delay between the lensed images~\citep{schmidt_strides_2023}. 

We impose priors on several of the SED properties based on the population study by \citet{Hernan-Caballero-mid-ir-2016}. From this, the temperature of the warm dust region was allowed to vary between 100 and 800 K.  The temperature of the hot dust region was allowed to vary between 900 and 1600 K. \cite{Hernan-Caballero-mid-ir-2016} found that the flux at 3 $\mu$m coming from the accretion disk contributed a maximum of 20\% to the total. We relax this limit given the possibility of differential microlensing between the accretion disk and hot torus and allow an upper limit of 60\% on the fraction of flux coming from the accretion disk at this wavelength. 

We constrain our model with both the likelihood that the model can reproduce the absolute fluxes for image A in each filter, as well as the likelihood that the model can reproduce the flux ratios of the other images relative to A in each filter.  We transform the model SEDs into broadband fluxes following~\cite{gordon_james_2022}. We calculate the posterior probability distribution using \texttt{emcee}~\citep{foreman-mackey_emcee_2013}.

\section{Results of SED fitting}\label{sec:SEDResults}

\begin{figure*}
\includegraphics[width=\textwidth]{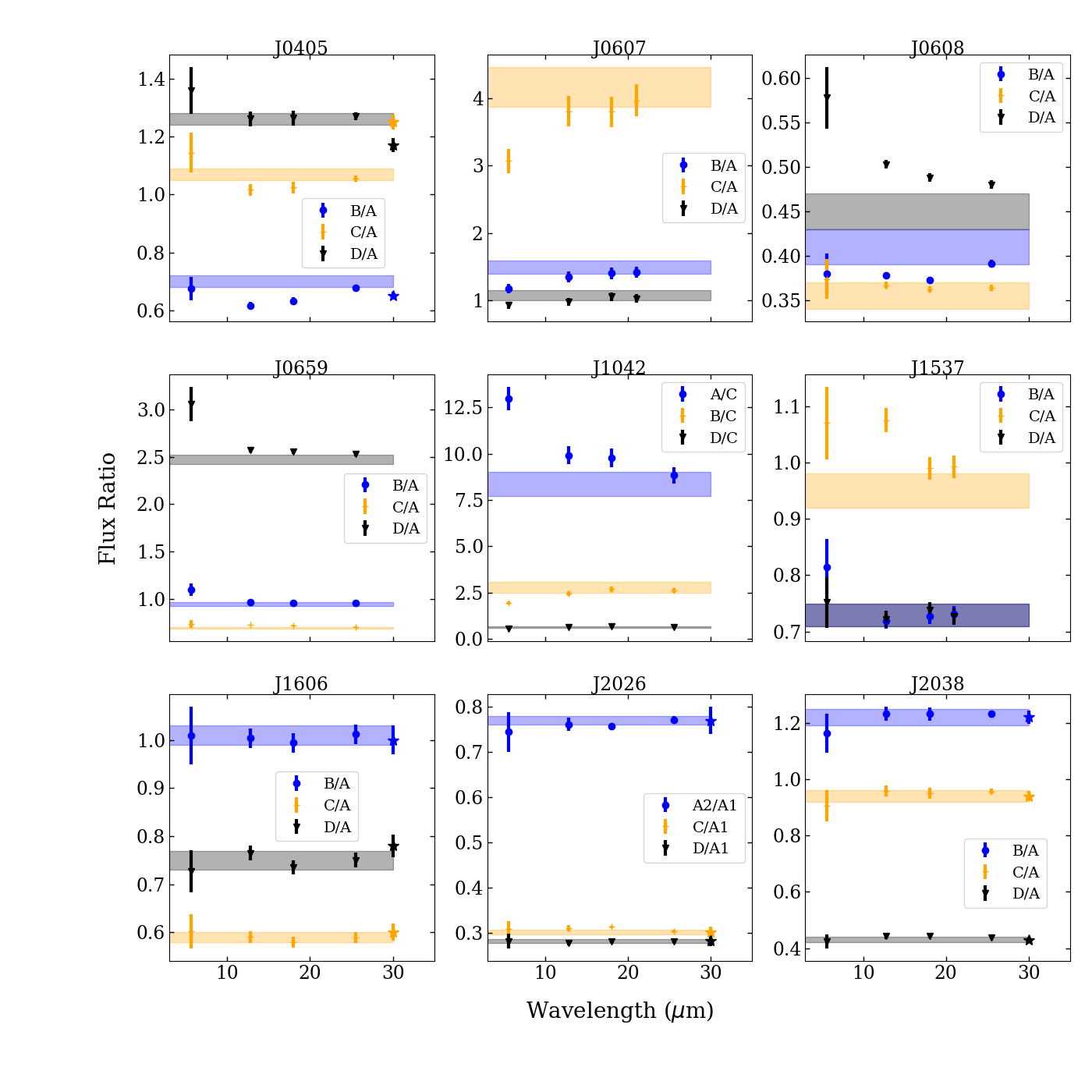}
\caption{\label{fig:fr_figure} Flux ratios as a function of wavelength for the nine lenses in this study. When available we also include [OIII] (4969$+$5007 \AA~doublet) flux ratios, plotted arbitrarily as stars at 30$\mu$m with the color corresponding to the same color as the MIRI flux ratio color scheme. Filled bands show the 68\% confidence interval for the warm-dust flux ratios. Wavelength dependence in the flux ratios (chromaticity), indicating microlensing by stars, is seen for most systems except J1606, J2026, and J2038.  The two reddest filters are always consistent within $\sim$1 $\sigma$, showing the decreased effect of microlensing at these wavelengths.
Warm dust and [OIII] are emitted from two different size scales and are not expected to be microlensed but may show differential lensing effects from small subhalos depending on the subhalo masses and locations.  }
\end{figure*}

The warm and hot dust flux ratios inferred from the SED fitting are provided in Table \ref{tab:data_summary}. Figures showing sample SEDs drawn from the posterior, can be found in the supplementary materials (Appendix E). For lower redshift sources, with $z<2$, the inferred warm dust flux ratios are virtually identical to the flux ratios measured in the reddest band. For some of the higher redshift lenses (e.g., J0659, and J1042), small and statistically insignificant differences in the flux ratios between the reddest filter and the inferred warm dust appear, at the 1 $\sigma$ level. Figure~\ref{fig:fr_figure} shows the warm dust flux ratios plotted as a band indicating the 68\% confidence interval from the SED fitting. In this figure we also show the narrow-line ([OIII], 4969$+$5007 \AA~doublet) flux ratios measured by \cite{nierenberg_double_2020} for four of the lenses. The nuclear narrow-line emission is not resolved in these lenses, but is more extended than the warm dust region \citep[e.g.][]{muller-sanchez_outflows_2011}, raising the possibility that a low mass perturber could differentially magnify the two regions, and this is a possible explanation for the difference in measured flux ratios for the case of J0405. The other three lenses show consistent flux ratios between warm dust and narrow-line emission. In future work, we will investigate the probability of differential magnifications for these systems.

Our inference allows for virtually any amount of microlensing of the hot dust, thus we obtain weak constraints on the flux ratios in the hot dust for many of our lenses particularly at lower redshifts where all four JWST filters are redward of 2 microns rest frame.  At high redshifts, this is better constrained, and we detect differential microlensing of the hot dust relative to the warm dust for J0405 (image B), J0607 (image A), J1537 (image C).  The microlensing of the hot dust could be better constrained if more realistic priors were used for, e.g., the relative amount of microlensing allowed for the continuum emission compared to the hot dust, or with microlensing simulations of the accretion disk and hot dust \citep[e.g.][]{sluse_mid-infrared_2013}.

\begin{table*}
    \label{tab:flux_results}

    \begin{tabular}{lllccllll}
    \hline
    \hline
    Lens           & image &dRA$^a$  & dDec$^a$ &F560W Flux Ratio & Hot Ratio       & Reddest Ratio $^b$ & Warm Ratio     & [OIII] Ratio \\
    \hline
    J0405    &A    & 1.065      & 0.325 & 1   & 1               & 1                  & 1               & 1                           \\
                  &B    &  0         &  0    & 0.68$\pm0.04$   & 0.58$^{+0.03}_{-0.07}$ & 0.68 $\pm0.01$    & 0.70$\pm$0.02 & $0.65\pm0.04$         \\  
                  &C    &  0.721     & 1.161 & 1.14$\pm$0.07    & 0.97$^{0.06}_{-0.1}$ & 1.06 $\pm 0.02$   & 1.07$\pm$0.02 & $1.25\pm0.03$         \\
                  &D    &  $-0.158$  & 1.022 & 1.36$\pm$0.08   & 1.19$^{0.07}_{-0.1}$ & 1.27 $\pm 0.03$   & 1.26$\pm$0.02 & $1.17\pm0.04$        \\

    \hline
    J0607         &A    & 0            & 0    &   1      & 1                         & 1                  & 1                 \\
                  &B    & 0.140       & 1.133 & 1.18$\pm$0.07    & 1.02$^{+0.2}_{-0.5}$       & 1.42 $\pm$0.08       & 1.49$^{+0.1}_{-0.09}$    \\
                  &C    & $-0.321$    & 1.531 & 3.07$\pm0.2$    & 2.7$^{+0.6}_{-1}$         & 3.97 $\pm$0.2        & 4.17$\pm$0.3      \\
                  &D    & $-1.282$    & 0.720 & 0.93$\pm0.06$&    & 0.7$^{+0.2}_{-0.4}$       & 1.03 $\pm$0.06       & 1.07$^{+0.08}_{-0.07}$     \\
                  &G2   & 0.636       & 0.965 &     &               &           &               \\

    \hline
    J0608    &A    & 0      & 0     &  1   & 1             & 1     & 1             &            \\
                  &B    & 0.613  & 0.603 &  0.38$\pm$0.02   & 0.36$\pm$0.01 & 0.391$\pm0.004$  & 0.41$\pm$0.02 &            \\
                  &C    & 1.228  &-0.273 &  0.37$\pm$0.02   & 0.36$\pm$0.01 & 0.364$\pm0.004$  & 0.36$\pm$0.01 &             \\
                  &D    & 0.156  &-0.394 &  0.58$\pm$0.03   & 0.51$\pm$0.03 & 0.479$\pm 0.005$  & 0.45$\pm$0.02 &             \\

    \hline
    J0659     &A   &  0           &  0        & 1 & 1             & 1     & 1                    \\
                   &B   &  $-4.665$    &  $-0.335$ & 1.10$\pm$0.07     & 0.99$\pm$0.04 & 0.96 $\pm 0.01$  & 0.94$\pm$0.02      \\
                   &C   &  $-0.979$    & 2.892     & 0.73$\pm$0.04    & 0.74$\pm$0.03 & 0.700 $\pm 0.007$  & 0.69$\pm$0.01      \\
                   &D   &  0.084       & 1.903     &  3.1$\pm$0.2      & 2.7$\pm$0.1   & 2.53 $\pm 0.03$ & 2.47$\pm$0.05      \\
                   &G2  & $-1.375$     & 2.442     &                 &                &                &                \\
    \hline
   
    J1042     &A   & 0      & 0      &13.0$\pm 0.6$& 10$\pm$3          & 8.8$\pm 0.4$       & 8.4$^{+0.6}_{-0.7}$          \\
                   &B   & -0.147 & -0.565 &1.9$\pm 0.2$ & 1.9$\pm$0.8      & 2.6$\pm$0.3         & 2.8$\pm$0.3     \\ 
                   &C   & -0.817 & -0.914  &1 & 1                &1                    & 1       \\ 
                   &D   & -1.584 & 0.546  &0.57$\pm$0.03 & 0.6$\pm$0.1 &    0.63 $\pm 0.03$      &  0.65$\pm0.04$    \\  
     
     \hline
    J1537           &A   & 0     & 0      &1      &    1                   & 1       & 1             &            \\
                         &B   &-1.993 &-0.329  &0.81$\pm0.05$      & 0.68$^{+0.05}_{-0.1}$     & 0.73$\pm 0.01$    & 0.73$\pm$0.02 &            \\ 
                         &C   &-2.848 & 1.644  &1.07$\pm$0.06      & 1.16$^{+0.1}_{-0.06}$     & 0.99$\pm 0.02$    & 0.95$\pm$0.03 &            \\
                         &D   &-0.750 & 1.763  &0.75$\pm$0.05     &0.70$^{+0.04}_{-0.08}$     & 0.73$\pm0.01$    & 0.73$\pm$0.02 &            \\

     \hline
    J1606                  &A   & 0      & 0      &1    & 1              & 1     & 1           &  1.00$\pm 0.03$ \\
                          &B   & -1.621 & -0.592 &1.0$\pm0.06$    & 0.98$\pm$0.06  & 1.01     & 1.01$\pm$0.02 &  1.00$\pm 0.03$ \\
                          &C   & -0.792 & -0.905 &0.60$\pm 0.04$   & 0.57$\pm$0.04  & 0.59    & 0.59$\pm$0.01 &  0.60$\pm0.02$ \\
                          &D   & -1.129 & 0.152  &0.73$\pm0.04$   & 0.76$\pm$0.05  & 0.75     & 0.75$\pm$0.02 &  0.78$\pm 0.02$ \\

  \hline
    J2026                  &A1  &  0     & 0     &1                 & 1             & 1       & 1               &  1.00$\pm$0.02    \\
                          &A2  &  0.252 & 0.219 &0.74$\pm 0.04$    & 0.75$\pm0$.03 & 0.772 $\pm0.008$   & 0.77$\pm$0.01 &  0.75$\pm$0.02 \\ 
                          &B   & -0.164 & 1.431 &0.31$\pm0.02$     &  0.31$\pm$0.01 & 0.303$\pm0.003$    & 0.302$\pm$0.005 &  0.31$\pm$0.02 \\
                          &C   & -0.733 & 0.386 &0.28$\pm 0.02$    & 0.28$\pm$0.01 & 0.280$\pm0.003$   & 0.282$\pm$0.004 &  0.28$\pm$0.02 \\

    \hline
    J2038                  &A   & 0      & 0      & 1  & 1                    & 1      & 1               &  1$\pm$0.01  \\
                          &B   & 2.307  & -1.707 & 1.16$\pm0.07$  & 1.0$^{+0.2}_{-0.5}$ & 1.23$\pm$0.01   & 1.22$\pm$0.03   &  1.16$\pm$0.02 \\
                          &C   & 0.796  & -1.678 & 0.91$\pm0.05$   & 0.8$^{+0.2}_{-0.3}$ & 0.96$\pm0.01$   & 0.94$\pm$0.02   &  0.92$\pm$0.02 \\
                          &D   & 2.178  &  0.384 & 0.42$\pm0.03$   &0.34$^{+0.08}_{-0.2}$ & 0.438$\pm0.004$   & 0.43$\pm$0.01   &  0.46$\pm$0.01 \\

    \hline
         & 
    \end{tabular}
    \caption{Results for image fitting and SED fitting. $a$) Image position uncertainties are estimated to be 0\farcs005 for all systems except for J1042, which we estimate has 0\farcs01 position uncertainties, as described in Section \ref{sec:image_fitting} $b$) Flux ratio of reddest filter, either F2550W or F2100W as listed in Table \ref{tab:fluxes}.}   
    \label{tab:data_summary}
\end{table*}

\section{Warm Dark Matter Constraint}\label{sec:DM}
As discussed in the Introduction, the flux ratios of gravitationally lensed quasars can be used to measure the properties of dark matter. In this work, we combine our measurements of the quasar warm dust with previous measurements of quasar flux ratios  to measure the half-mode mass ($M_{\rm hm}$) of the WDM halo mass function. We follow the procedure of \cite{gilman_warm_2020, gilman_probing_2019} with several updates. For convenience to the reader, we summarize some of the key parts of this analysis here. 
We begin in Subsection~\ref{sec:extralenses} by describing the full sample of lenses included in this dark matter constraint.
In Subsection~\ref{sec:macromodel} we describe the model of mass distribution of the lenses that we need to marginalize over in order to calculate the constraint on the DM parameters. 
We describe our model for the mass function of field halos in Subsection~\ref{sec:dm_model}, and for the subhalos in Subsection~\ref{sec:subhalos}. 
In Subsection \ref{sec:abc} we summarize the Approximate Bayesian Computing (ABC) method we use to infer the relative probability of different DM models.

\subsection{Full Lens Sample}
\label{sec:extralenses}

In addition to the lenses with warm dust flux ratios presented here, we add lenses that have flux ratio measurements that meet the following criteria: 1) flux ratios measured at a wavelength that is not thought to be affected by microlensing (either radio, microwave, or narrow-line emission), 2) a single, simple deflector galaxy that does not have an apparent disk. Beyond the current sample, five gravitational lenses met these criteria, four of which have narrow-line flux ratios \citep{nierenberg_probing_2017, nierenberg_double_2020}, and one lens with a CO spectral line measurement \cite{StaceyMcKean0414discovery}. A summary of information about the additional lenses is provided in Table \ref{tab:additional_lenses}.

\subsection{Macromodel Parameters}\label{sec:macromodel}
The smooth mass distribution of the lens is modelled as a PEMD and external shear. We further generalize this mass model with third and fourth order azimuthal multipoles, to account for the observed boxiness and diskiness of galaxies, specifically by using measurements from \citet{hao_isophotal_2006}. Note that during the DM inference, we do not use any of the main lens model information we derive when measuring the image fluxes as our analysis of the imaging data is focused solely on accurate measurement of the image fluxes. This is conservative and in a future work, will combine constraints from the imaging data using the method of \cite{2024GilmanTurbo}.

We sample the PEMD power law slope from a Gaussian prior distribution with $\gamma = 2.0\pm 0.1$. Multipoles are included in the convergence concentric with the PEMP centroid using:
\begin{equation}
    \kappa_m(r,\phi) = \frac{a_{m} \theta_E }{r\sqrt{q}} \cos(m(\phi-\phi_m)), 
\end{equation} 
where $\theta_E$ is the Einstein radius, $q$ is the axis ratio of the PEMD, and $r$ is the projected separation from the main deflector mass centroid. The prefactor $\theta_{\rm{E}} / \sqrt{q}$ rescales the physical amplitude of these terms such that the observed shape of the iso-density contours depends only upon $a_m$. The priors that we implement for these terms are based on the observed shapes of elliptical galaxies \citep{hao_isophotal_2006}. We adopt a Gaussian prior for $a_3/a$ with mean 0.0 and standard deviation 0.005, as well as for $a_4/a$ with mean 0.0 and standard deviation 0.01, and we use a uniform prior for $\phi_3$ that ranges from $-\pi/6$ to $\pi/6$. For $\phi_4$, we allow it to be uniform in the range $-\pi/8$ to $\pi/8$ for $a_4/a<0.02$ and keep it fixed to 0 for $a_4/a>0.02$. This prior is conservative, as it allows for more freedom for intermediate values of $0.01>a_4/a>0.02$, relative to what is actually observed in galaxy light distributions.

When additional companion galaxies are located within $\sim$5 arcseconds of the lensed images (in other filters), we include additional mass components at the location of the light centroid. We assume these objects have Singular Isothermal Sphere (SIS) mass profiles. Unless a redshift has been measured for these systems, we assume they are at the redshift of the main deflector. We estimate their masses based either on their luminosities, or in the case of J0607, by their estimated masses during the lens modelling. This perturber was massive enough to slightly deform the lensed arcs in these systems. For the DM inference, we adopt a uniform prior for the perturber masses centered at the best-fit Einstein radius from our lens modelling with a factor of two mass uncertainty.

\subsection{Source Properties}
\label{sec:source_properties}
We model the sources in our simulations as Gaussians with a width set by a source size parameter. For our warm dust flux ratios, we draw the source size from a uniform distribution in the range 1-10 pc~\citep{burtscher_diversity_2013,leftley_parsec-scale_2019}. For our lenses with narrow-line flux ratios, we draw the source size from a uniform  distribution in the range 40-80 pc~\citep{muller-sanchez_outflows_2011, nierenberg_detection_2014, nierenberg_probing_2017}. For J0414, we use the source size of 60 pc as taken from \citet{2020MNRAS.493.5290S}.

\subsection{Field halo mass function}
\label{sec:dm_model}
We use a standard cosmology of $\Omega_{\rm m} = 0.28$, $\sigma_8 = 0.82$ $h=0.7$ from \cite{Hinshaw_2013_WMAP} when calculating the distribution of halos in CDM (e.g. to calculate a CDM transfer function and CDM halo mass function), and use a halo mass definition of $M_{200}$ calculated with respect to the critical density of the Universe at the halo redshift. For subhalos (see Section \ref{sec:subhalos}), we define their mass at infall using the same definition. All models discussed in this and following section are implemented for lensing analyses with the open-source software {\tt{pyHalo}}\footnote{https://github.com/dangilman/pyHalo} \citep{gilman_warm_2020}. 

We model the line-of-sight halo mass function as
\begin{equation}
\frac{d^2N}{dM dV} = \delta_{\rm{LOS}}\left(1+\xi_{\rm 2halo}\left(M_{\rm{host},z}\right)\right)\frac{d^2N}{dM dV} \bigg |_{\rm{ST}}
\end{equation}
where $M_{\rm{host}}$ is the host halo mass, and $\frac{dN}{dM dV}|_{\rm ST}$ is the mass function model presented by \citet{Sheth_Torman_2001}. The term $\xi_{2\rm{halo}}$ accounts for correlated structure around the main deflector, as described by \citet{gilman_probing_2019}. In our calculation of $\xi_{\rm{2halo}}$, we also include the modification calibrated against N-body simulations by \citet{Lazar++21}, which increases the number of halos introduced through this term by a factor of $\sim 2$.  $\delta_{\rm LOS}$ is introduced to account for theoretical uncertainties in calculating the normalization of the halo mass function~\citep{2016MNRAS.456.2486D}.

The WDM mass function is calculated as a suppression applied to the Sheth-Torman mass function, with the following fitting formula
\begin{equation}
    \frac{d^2N_{\rm WDM}}{dM dV} = \frac{d^2N_{\rm CDM}}{dM dV} f_{{\rm{WDM}}}\left(M, M_{\rm{hm}}\right)
\end{equation}
where 
\begin{equation}
\label{eqn:mfuncsuppression}
f_{{\rm{WDM}}}\left(M, M_{\rm{hm}}\right) = \left(1 + \left(\alpha \frac{M_{\rm hm}}{M}\right)^{\beta}\right)^{\gamma}
\end{equation}
with $\alpha = 2.3$ $\beta = 0.8$ $\gamma = -1.0$ \citep{Lovell+2020}. We model halos as truncated Navarro-Frenk-White (NFW) profiles \citep{navarro_universal_1997}
\begin{equation}
\label{eqn:densityprof}
\rho\left(r\right) = \frac{\rho_s}{x \left(1+x\right)^2}\frac{\tau^2}{x^2 + \tau^2}
\end{equation}
where $x \equiv r / r_s$, $\tau \equiv r_t / r_s$, $r_s$ is the halo scale radius, and $r_t$ is a truncation radius. For field halos, we set $r_t = r_{\rm{50}}$, where $r_{50}$ is the radius that encloses 50 times the critical density, in order to keep the mass rendered along the line of sight finite. 

In CDM, we calculate $r_s$ for a halo of mass $m$ using the concentration--mass relation presented by \cite{Diemer_2019_CDM_concentration}. To model the concentrations of WDM halos, we use the fitting function given by \cite{bose_copernicus_2016}
\begin{equation}
\label{eqn:concentrationwdm}
    \frac{c_{\rm WDM}(M,z)}{c_{\rm CDM}(M,z)} = (1+z)^{\beta(z)}\left(1 + 60 \frac{M_{\rm hm}}{M} \right)^{-0.17}
\end{equation}
where $\beta(z) = 0.026 z - 0.04$. The suppression of the concentration of WDM halos results from the delayed formation time of structure in these models~\cite{2002ApJ...568...52W,ludlow_mass-concentration-redshift_2014}.  

\subsection{Subhalos}
\label{sec:subhalos}
When a halo in the field accretes onto the host halo of the main deflector, it decouples from the background density of the Universe and evolves in the gravitational tidal field of the host. This section details how we model the population of main deflector subhalos, and implements several improvements relative to the work by \citet{gilman_warm_2020}. 
\subsubsection{The projected number density of main deflector subhalos}
We model the infall subhalo mass function with a mass function of the form \citep{gilman_warm_2020}
\begin{equation}
\label{eqn:shmf}
    \frac{d^2N_{\rm sub}}{dM dA} = \frac{\Sigma_{\rm sub}}{10^8} \left(\frac{M}{10^8}\right)^\alpha \mathcal{F}(M_{\rm{host}},z) f_{{\rm{WDM}}}\left(M, M_{\rm{hm}}\right),
\end{equation}
where $\Sigma_{\rm sub}$ and $\alpha$ are the normalization and logarithmic slope of the subhalo mass function at infall. The slope $\alpha$ is taken to be in the range -1.95 to -1.85, as found in simulations~\citep{Springel++08,fiacconi_cold_2016}. The function $\mathcal{F}$, given by 
\begin{equation}\label{eq:mathcalF}
\log\mathcal{F}(M_{\rm{host}},z) = k_1 \log \left(\frac{M_{\rm{host}}}{10^{13}M_{\odot}}\right) + k_2 \log \left(z + 0.5\right)
\end{equation}
which factors out the evolution of $\Sigma_{\rm{sub}}$ with host halo mass and redshift such that we can combine inferences of $\Sigma_{\rm{sub}}$ from lenses with different host halo masses and different redshifts. Based upon an updated calibration of the semi-analytic model {\tt{galacticus}} \citep{benson_g_2012}, we use $k_1 = 0.5$ and $k_2 = 0.3$ (Gannon et al, in prep). For this analysis, we assume all of the deflectors have the strong lens population average halo mass $M_{\rm{host}} = 10^{13.3}$ M$_\odot$ measured by \cite{Lagattuta_2010}. We place lenses that do not have spectroscopic or photometric redshift measurements at $z=0.5$. This uncertainty primarily affects the inference through the dependence of the subhalo population on the host mass and, to a much smaller extent, the host redshift. Using Equation~\ref{eq:mathcalF}, the number of subhalos varies by a factor of $2$ when varying the halo mass in the range $M_{\rm halo}\in 10^{13}, 10^{13.6}$ and by a factor of 1.2 when varying the lens redshift over the range $z\in 0.3, 1.0$. Both of these factors are much smaller than the 1.5 dex range used for the $\Sigma_{\rm sub}$ prior.   

\subsubsection{The evolved subhalo mass function}
By definition, the amplitude of the projected infall subhalo mass function (Equation \ref{eqn:shmf}) does not depend on tidal stripping and heating. However, this is not what we measure with strong lensing. Instead, lensing measures the {\textit{evolved}} subhalo mass function in addition to the mass function of halos along the line of sight. The evolved subhalo mass function depends on tidal stripping and heating by the host halo and the main deflector galaxy. A common approach to modeling the evolved subhalo mass function is to fit a mass function to the output of numerical simulations, and draw subhalo masses from this model. The drawback of this approach is that it changes the mass definition of subhalos, and formally requires the re-calibration of the concentration--mass relation for these objects. It also obscures the connection between the physical properties of subhalos at infall, which depend on DM physics, and the properties of the evolved subhalo mass function. Put differently, this approach conflates the modification to halo density profiles from the cosmological effects of DM free-streaming with tidal stripping and heating by a central potential. 

To model the evolved subhalo mass function, we derive a transfer function that establishes a probabilistic mapping between properties of halos at infall and the bound mass of a subhalo at the time of lensing. We can phrase the task at hand as assigning bound masses to subhalos, or equivalently, assigning truncation radii $r_t$ (Equation \ref{eqn:densityprof}), to individual subhalos in such a way that the statistical properties of the evolved subhalo population match the statistical properties of evolved subhalo populations output by numerical simulations. This approach assumes that the lensing signal depends primarily on the tidal truncation $r_t$, and not on the less substantial evolution of other structural parameters, such as $\rho_s$ and $r_s$, that can in principle also change due to tidal heating. 

The structural parameters of NFW halos evolve along `tidal tracks' \citep{Errani++21,Du++24}, a self-similar trajectory along which halo density profiles evolve in a tidal field. The rate at which halos traverse tidal tracks depends on the concentration of the halo and the concentration of the host, and the position of a halo along the tidal track depends on the elapsed time since infall \citep{Stucker++23}. Because strong lensing probes a region projected down the center of the host halo that is very small compared to the virial radius of the host, the bound mass function in this regime does not have a strong dependence on the projected two-dimensional position of a subhalo (\citealt{Xu++15}; Gannon et al. in prep). Thus, we will seek a mapping to the bound mass of subhalos only in terms the subhalo concentration at infall, $c$, the host halo concentration, $c_{\rm{host}}$, and the time since infall, $t_{\rm{infall}}$. As the number of subhalos is typically $\mathcal{O}\left(100\right)$, from the central limit theorem we model the probability distribution of $m_{\rm{bound}} / m_{\rm{infall}}$ as a Gaussian, and introduce dependence on $c$, $c_{\rm{host}}$, and $t_{\rm{infall}}$ through the mean and standard deviation, i.e. $m_{\rm{bound}} / m_{\rm{infall}} \sim p\left(\mu\left(c, c_{\rm{host}},t_{\rm{infall}}\right), \sigma\left(c, c_{\rm{host}},t_{\rm{infall}}\right) \right)$. From $m_{\rm{bound}} / m_{\rm{infall}}$, we can then calculate $r_t$ given the infall mass of the subhalo using Equation \ref{eqn:densityprof}. 
\begin{figure}
\includegraphics[width=0.49\textwidth]{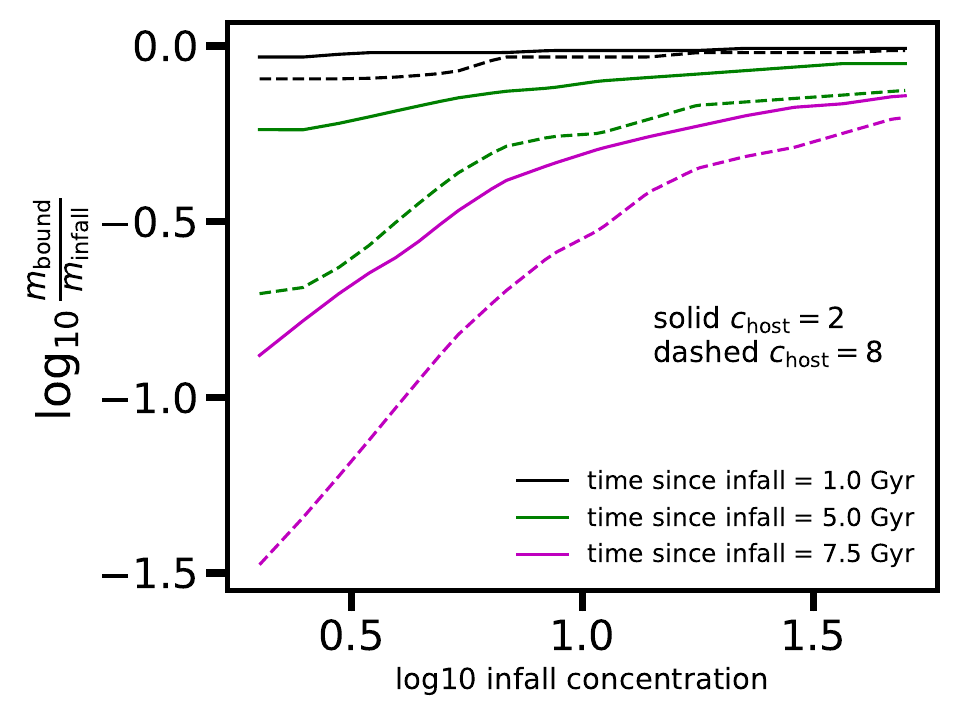}
\includegraphics[width=0.49\textwidth]{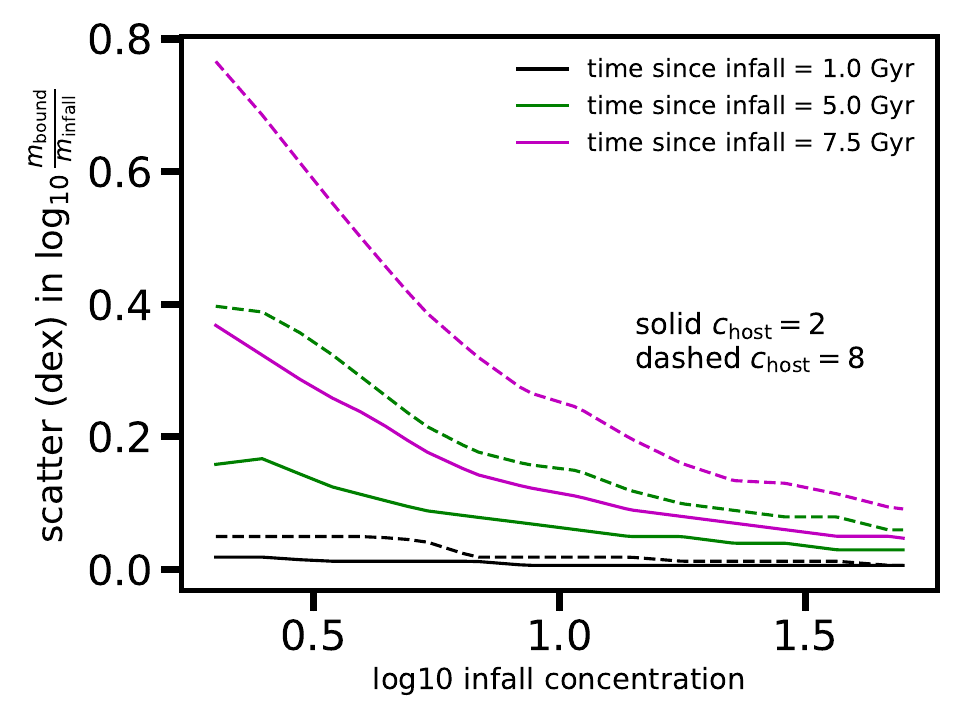}
\caption{\label{fig:musigma}The mean (top) and standard deviation (bottom) of the bound mass of an individual subhalo as a function of the infall concentration, the time since infall, and the host halo concentration. }
\end{figure}

To compute the functions $\mu\left(c, c_{\rm{host}},t_{\rm{infall}}\right)$ and $\sigma\left(c, c_{\rm{host}},t_{\rm{infall}}\right)$, we perform a series of calculations with {\tt{galacticus}} in which we inject individual subhalos into a static host potential, and evolve them for $t_{\rm{infall}}$ Gyr. Subhalos are initially placed at the virial radius of the host, and are assigned initial orbital parameters by drawing from the distribution of subhalo infall orbits measured in cosmological N-body simulations by \cite{2015MNRAS.448.1674J}. Each subhalo is then evolved under the physical models (including dynamical friction, tidal heating, and tidal stripping) described by \cite{2020MNRAS.498.3902Y}, \cite{Benson++2022}, and \cite{Du++24}. We then record the bound mass of the subhalo at the end of the simulation, and repeat the calculation thousands of times for each cell on a grid of $c_{\rm{host}}$, $c$, and $t_{\rm{infall}}$. We determine $\mu$ and $\sigma$ for the distribution of bound subhalo masses in each cell, and perform a spline interpolation of the $\mu$ and $\sigma$ values to derive continuous functions for these quantities. The $\mu$ and $\sigma$ functions have a negligible dependence on infall subhalo mass. 
\begin{figure}
\includegraphics[width=0.49\textwidth]{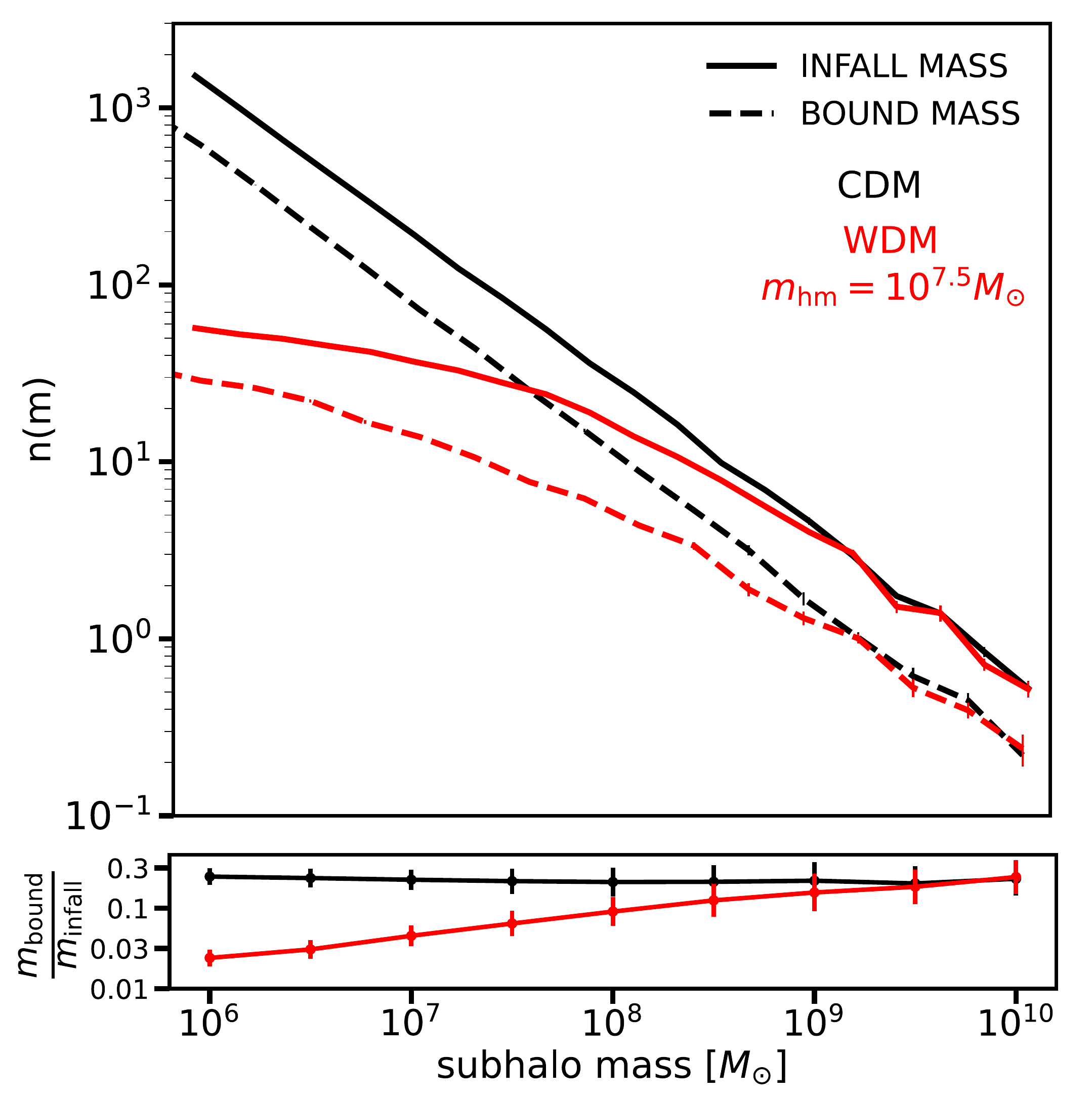}
\caption{\label{fig:mfunc} Infall and bound subhalo mass functions in CDM (black) and for a WDM model with $m_{\rm{hm}} = 10^{7.5} M_{\odot}$. The top panel depicts the median and scatter in the number of halos averaged over 100 realizations according to the infall mass definition (solid lines) in comparison with the number of halos binned according to their bound mass calculated according to the procedure discussed in Section \ref{sec:subhalos}. The bottom panel shows the average bound mass of surviving subhalos as a function of their infall mass, illustrating the differential suppression of the WDM bound subhalo mass function due to the suppression of halo concentrations in WDM and the dependence on the tidal stripping model on subhalo concentration at infall (see Equation \ref{eqn:concentrationwdm} and Figure \ref{fig:tidalstripping})}
\end{figure}
\begin{figure}
\includegraphics[width=0.49\textwidth]{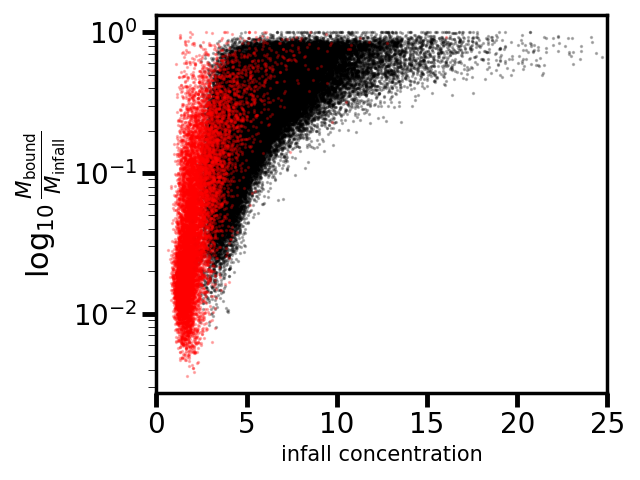}
\caption{\label{fig:tidalstripping} The distribution of infall concentration versus bound mass for a CDM subhalo population (black) and WDM subhalo population with $m_{\rm{hm}} = 10^{7.5} M_{\odot}$ (red) using the tidal truncation model discussed in Section \ref{sec:subhalos}.}
\end{figure}

Figure \ref{fig:musigma} shows the functions for $\mu$ and $\sigma$ that result from these calculations. The mean captures the leading-order effects of tidal stripping, reflecting the fact that the amount of mass loss depends, to leading order, on $c$, $c_{\rm{host}}$, and $t_{\rm{infall}}$. The standard deviation accounts for all other effects that we do not explicitly include in the model, such as the orbital pericenter of subhalo orbits. The average bound mass relative to infall mass decreases for lower infall concentration, reflecting the fact that less concentrated subhalos lose mass more rapidly than more concentrated subhalos. The scatter in bound masses includes the dependence on orbital pericenters. The scatter increases at lower infall concentration because subhalos with small pericenters get rapidly disrupted if they have low concentrations, while subhalos that appear in projection near the Einstein radius but have large orbital pericenters retain most of their mass. In contrast, more concentrated subhalos are more resilient to tidal effects, and therefore they have a weaker dependence on the orbital pericenter, which is a latent variable in these calculations. The model predicts that halos accreted at earlier times lose more mass than recently-accreted subhalos. We assign values of $t_{\rm{infall}}$ to individual subhalos in {{\tt{pyHalo}}} using the distribution of infall times predicted by {\tt{galacticus}} for subhalos that appear within a 20 kpc aperture of the host center. 

Figure \ref{fig:mfunc} shows the result of applying this tidal stripping model to subhalo populations in CDM and WDM generated with {\tt{pyHalo}}. In CDM, the model predicts halos lose $\sim 70\%$ of their mass on average, with little to no dependence on the infall mass. On the other hand, in WDM the amplitude of the bound mass function becomes increasingly suppressed at lower masses due to the suppressed halo concentrations on scales $m \lesssim m_{\rm{hm}}$. Figure \ref{fig:tidalstripping} more clearly shows the effect of infall concentration on the distribution of subhalo bound masses. The explicit dependence on infall concentration introduces an additional lever with which to distinguish between CDM and WDM models, as WDM subhalo populations will systematically have a lower bound mass function amplitude due to the explicit dependence on infall concentration, which becomes suppressed in WDM models (Equation \ref{eqn:concentrationwdm}).

\begin{figure}
    \centering
    \includegraphics[width=0.49\textwidth]{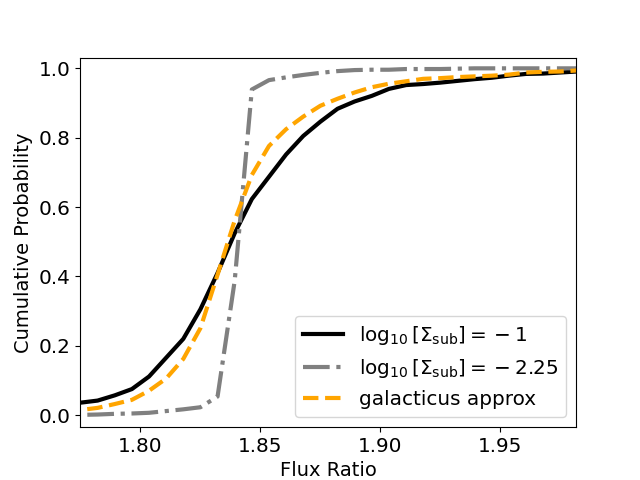}
    \caption{Cumulative probability distribution of flux ratios produced by dark matter subhalos for a lens with halo mass $10^{13}$ M$_\odot$ at redshift 0.5 for different choices of the normalization of the subhalo mass function, $\Sigma_{\rm sub}$. Comparison of the orange, grey and black lines shows that the prior on $\Sigma_{\rm{sub}}$ used in our measurement encompasses predictions from {\tt galacticus}, while also accommodating a broad range of theoretical uncertainties. In reality, subhalos make up only $10-20\%$ of the halos in projection near lensed images, so this figure exaggerates the impact of uncertainties in the properties of subhalos.}
    \label{fig:flux_ratio_compare}
\end{figure}

In Figure \ref{fig:flux_ratio_compare}, we compare the cumulative distribution for the flux ratio of a lensed image for varying values of $\Sigma_{\rm{sub}}$, with no line of sight halos included. These simulations are for a lens with a halo mass of $10^{13}$ M$_\odot$ at a redshift of 0.5, for subhalos with bound masses between $10^7$ and $10^8$ M$_\odot$. One should interpret the width of the distribution of flux ratios (or the slope of the CDF) as a proxy for the lensing signal that statistically differentiates various dark matter models. As expected, as $\Sigma_{\rm{sub}}$ increases, the width of the distribution also increases. The yellow dashed line shows the cumulative distribution of flux ratios for subhalos with bound masses and concentrations consistent with direct output from the {\tt galacticus simulations}. The prior on $\Sigma_{\rm{sub}}$ is large enough to accommodate a large range of theoretical uncertainty on the properties of dark matter subhalos. We emphasize that these distributions are for flux ratios produced only by subhalos, which make up only about 10-20\% of the halos near the lensed images in projection, thus differences in the flux ratio distributions for full realizations of subhalos and line-of-sight halos will depend significantly less on $\Sigma_{\rm{sub}}$ than the results shown here.

\subsection{The calculation of the likelihood function}
\label{sec:abc}
Here we describe the statistical procedure we use to constrain WDM using the flux-ratio measurements of the warm dust region. This procedure follows the methodology described in\footnote{\citet{2024GilmanTurbo} present a method to incorporate constraints from lensed arcs simultaneously with flux ratios, which is not implemented in this analysis. However, this work follows the the most update to date statistical framework presented in \citet{2024GilmanTurbo} for analyzing image positions and flux ratios,  which we follow here.}~\citet{2024GilmanTurbo} (see also \citet{gilman_warm_2020}), and yields accurate and unbiased results in both WDM and CDM mock data sets. 

Our methods require simulation-based likelihoods because the fluxes of each image depends on the specific realization of DM structure; the model is stochastic.
Different parameter values for the DM model predict different distributions of flux ratios. The colder the DM model, the more halos the model predicts, and thus the wider the distribution of predicted flux ratios. These distributions of flux ratios are very correlated and non-Gaussian and so to estimate the likelihood of observing a set of measured flux ratios, given a set of model parameters, we use Approximate Bayesian Computing (ABC)~\citep{Rubin1984,Marin++11,Lintusaari++17}. The ABC method works first by sampling parameters from the prior, computing the observables for each sample, and then selecting the samples that best fit the data, as defined by a summary statistic.
For the parameters of our DM model, 
$\log_{10} M_{\rm hm}$, $\log_{10}\Sigma_{\rm sub}$, and $\delta_{\rm LOS}$ are sampled from uniform priors with ranges $\in {4.0,10.0}$,  $\in {-2.5,-1.0}$, $\in 0.8,1.2$, respectively and $\alpha$ has a Gaussian prior with $-1.90\pm0.05$.
The prior on $\Sigma_{\rm sub}$ was chosen to span a wide range centered on the predictions of {\tt galacticus}, while the prior of $M_{\rm hm}$ was chosen such that the upper end was the maximum value of the halos we explicitly render with {\tt pyHalo} while the lower end extends below where we were forecasted to constrain by a few dex. 
The prior on $\delta_{\rm LOS}$ was chosen to account for theoretical uncertainties associated with calculating the halo mass function~\cite{2016MNRAS.456.2486D}. The prior on $\alpha$ was taken from results of N-body simulations~\citep{Springel++08,fiacconi_cold_2016}.

With these parameters, we use \texttt{pyHalo} to populate our lens with DM subhalos and line-of-sight halos, drawn from randomly sampled 
DM parameters. Halos with masses above $10^{10} M_\odot$ are relatively rare, and expected to contain galaxies with absolute magnitudes of approximately $M_v\sim -16 $ \citep[e.g.][]{Tollerud_mw_lf_2008}. Such galaxies are detectable in single orbit Hubble Space Telescope imaging to a redshift of $z\sim2$ \citep{Koekemoer_cosmos_2007,Nierenberg2016}. The majority of our systems have such imaging with the exception of J0607 and J0608. For J0607, a luminous companion is detected in the JWST MIRI imaging. When detected, we explicitly include the companions in the lens model as singular isothermal spheres (SIS) assumed to be at the redshift of the main deflector. We include these massive subhalos in J0607,  J1042, and J1606.  

Given a realization of DM halos, as well as a subset of randomly-sampled macromodel parameters (the slope of the lens mass distribution $\gamma$, the multipole amplitudes $a_3$, $a_4$, and angles, $\phi_3$, $\phi_4$) we apply a non-linear solver to the remaining portion of the lens macromodel(Einstein radius $\theta_E$, ellipticity, orientation, external shear, and source position) such that the multi-plane lens equation is satisfied for each realization of dark matter subhalos and line-of-sight halos. In the next step, we calculate the flux ratios for a sampled realization of DM halos and macromodel parameters, with source properties sampled as described in~\ref{sec:source_properties}.

Now we can compare the predicted flux ratios, for a given DM halo realization, from a given set of model parameters, to the observed flux ratios. This is the step where the ABC method is relevant.  The ABC method approximates a posterior by selecting the parameters from the prior which predict observables close to the data, as measured by a summary statistic.
We choose the following summary statistic, $S =  \sqrt{\sum_i(f_{i,\rm{obs}} - f_{i,\rm{pred}})^2}$ where the $f$ variable are the observed and predicted flux ratios and the sum is over the three flux ratios. 
We choose the 1000 prior samples corresponding to the smallest summary statistic for each lens. The acceptance rates vary between lenses, but this choice typically results in flux ratios that match the data to well-within the measurement uncertainties of the flux ratios.   We have verified this way of estimating the posterior probability distribution returns accurate results with mock data sets \citep{2024GilmanTurbo}.

We use \texttt{samana}\footnote{https://github.com/dangilman/samana} \citep{2024GilmanTurbo} to sample the prior, vary the macromodel parameters, and wrap the functionality of \texttt{pyHalo}\footnote{https://github.com/dangilman/pyHalo} and \texttt{lenstronomy}~\citep{birrer_lenstronomy_2018,birrer_lenstronomy_2021}  for the lensing calculations.

\section{Results}\label{sec:results}
 
\begin{figure*}
    \centering
    \includegraphics[width=\linewidth]{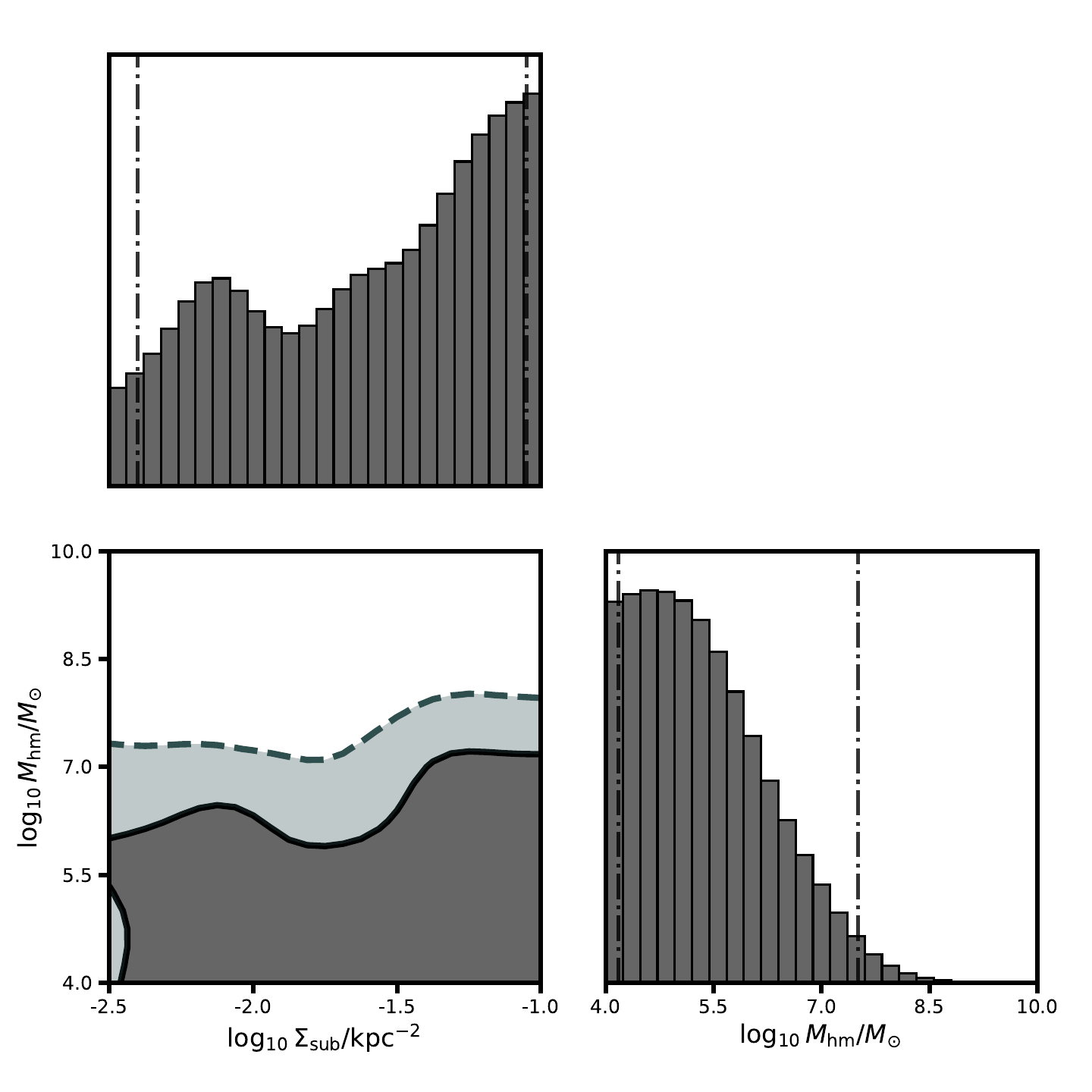}
    \caption{Posterior probability distribution for the DM model parameters $M_{\rm{hm}}$ and $\Sigma_{\rm{sub}}$. The data favors models which yield high amounts of low-mass structure over those that do not. 
    We rule out a half-mode mass greater than $10^{7.6} $M$_{\odot}$ (posterior odds 10:1) corresponding to a WDM particle mass of 6.1 keV. }
    \label{fig:WDMconstraint}
\end{figure*}

The results of the ABC inference for these lenses is shown in Fig.~\ref{fig:WDMconstraint}.  There, the posterior indicates that our results constrain the half-mode mass to be below $\log_{10} M_{\rm hm}<7.6$ (posterior odds 10:1, relative to the peak of the posterior). In Table~\ref{tab:odds}, we show the posterior odds for a few additional example values of the half-mode mass.

We use the following equation to interpret our constraints on the half-mode mass as a constraint on the mass of the DM particle,
\begin{equation}
    M_{\rm hm} = 3 \times 10^8 \left( \frac{m_{\rm WDM}}{3.3\,\rm{keV}}\right)^{-3.33} M_\odot,
\end{equation}
as shown by \cite{schneider_non-linear_2012,schneider_wdm_2012,2023PhRvD.108d3520V}. Thus, our constraint on the half-mode mass corresponds to a constraint on the DM particle mass of $m_{\rm WDM}>6.1$ keV (posterior odds 10:1). 

\begin{table}
    \centering
    \begin{tabular}{llll}
    \hline
    \hline
     $\log_{10} M_{\rm hm}$ & $m_{\rm th}$ (keV) & Odds & Gilman 2020a\\
    \hline
    $7.0$ & 9.2 & 3.6 & 1.2\\
    $7.5$ & 6.5 & 8.7 & 1.6\\
    $8.0$ & 4.6 & 24. & 3.2\\
    $8.5$ & 3.2 & 44. & 9.6
    \end{tabular}
    \caption{Posterior odds evaluated for example values of $\log_{10} M_{\rm hm}$. We include a comparison with the odds calculated for the same half-mode masses from ~\citet{gilman_warm_2020}.}
    \label{tab:odds}
\end{table}

\section{Uncertainty Budget}\label{sec:uncertainty_budget}
Here we summarize some of the key contributors to the uncertainty budget of our measurement. 

\subsection{Warm Dust Measurements} The uncertainty associated with the flux ratio measurements was discussed extensively by \cite{Nierenberg_0405_2023}. Lenses with a bright arc in the reddest filter will have the largest measurement uncertainty ($\sim$6\%), while the absolute flux uncertainty of $\sim10\%$ also contributes through the SED fitting to the measurement uncertainty. With improvements to the PSF model as well as better characterization of the MIRI detector it is possible that the absolute flux uncertainty may be reduced in the future. Typically, however, the inferred warm dust flux ratio uncertainties are dominated by the flux ratio uncertainties in the reddest filter so we do not anticipate this will have a significant affect on the DM measurement based on this data.

\subsection{Lens Macromodel}
We adopt a uniform prior on the ellipticity and orientation of the macromodel. The multipole moments are drawn from optical measurements of field ellipticals \citep{hao_isophotal_2006}. \cite{2024GilmanTurbo} demonstrated that the inclusion of mass distribution information from the lensed quasar host galaxy from Hubble Space Telescope imaging can directly improve the sensitivity to WDM models by directly constraining the macromodel for each lens. For the sample of mock lenses studied by \citet{2024GilmanTurbo}, this improved the measurement by a factor of $\sim 3$ in the half-mode mass for a CDM truth for a sample of 25 lenses with HST imaging of the lensed quasar host galaxy relative to a measurement without this information.

\subsection{Dark Matter Model}
The most significant uncertainties in our DM model are related to the model for tidal stripping of the subhalos. 
For a typical lens, with source redshift of 1.5, deflector redshift of 0.5 and a halo mass of $10^{13}$ M$_\odot$, the ratio of subhalos to field halos near the lensed images is approximately 1:5 based on estimates with pyHalo \footnote{This ratio can vary significantly depending on how the volume in which field halos are chosen is determined.}.
Although the subhalos are subdominant, they comprise a large enough fraction of the population that uncertainties in modelling must be carefully accounted for. 
We enable a broad range of $\Sigma_{\rm{sub}}$ values relative to what is seen in simulations of these populations; this reflects our uncertainty in the pre-infall normalization of the subhalo mass function as well as our much larger uncertainty in the amount of tidal stripping that subhalos undergo. 
Our prior on $\Sigma_{\rm{sub}}$ is chosen to match the bound mass fraction measured from {\tt galacticus} when the {\tt pyHalo} tidal stripping model is applied to a population of subhalos.
This yields a bound mass fraction which is consistent with not only {\tt galacticus} but also a range of N-Body simulations \citep{fiacconi_cold_2016,griffen2016caterpillar,10.1111/j.1365-2966.2012.21564.x} for halos at this mass.
\cite{nadler_constraints_2021} showed that combining luminous satellite galaxies of the Milky Way could provide a constraint on $\Sigma_{\rm{sub}}$, and that a combination of gravitational lensing and luminous satellite counts can provide a stronger constraint on a turnover in the halo mass function than either method on its own. We will incorporate such constraints in a future paper, when we infer the properties of dark matter for the whole sample of lenses.

The model for the suppression of the warm dark matter halo mass function implements the most recent calibration by \citet{Lovell+2020}. The suppression term (Equation \ref{eqn:mfuncsuppression}) in this model has a logarithmic slope at $m < m_{\rm{hm}}$ of -0.8, relative to the -1.3 logarithmic slope of mass function suppression presented by \citet{Lovell_2014_WDM} that was used in previous studies \citep{gilman_warm_2020,hsueh_sharp_2020}. For a given $m_{\rm{hm}}$, the updated model predicts more low-mass halos than the previous WDM fit, which leads to weaker constraints on the free-streaming length for a given dataset.

\section{Discussion}\label{sec:discussion}

By combining our new measurements of the warm dust with the narrow-line and CO measurements, we placed the tightest gravitational lensing constraint to date on a possible turnover in the halo mass function, and thus on free streaming length of warm dark matter.

Our limit is consistent and more stringent than that by \citet{gilman_warm_2020} which found $M_{\rm{hm}}<10^{7.8}$ $M_{\odot}$ for a sample of 8 lenses with narrow-line flux ratios. 
The improved limit is a result of several differences with respect to \citet{gilman_warm_2020}. First, the sample is larger, including five lenses that were not part of the previous sample. Thus, we expect that differences in the strength of the constraint will arise from sample variance.  Second, we have updated the treatment of tidal stripping of subhalos. Third, the intrinsic source sizes differ in the two works, with the nuclear narrow-line region having a characteristic scale of $\sim 50-80$ pc, and the warm dust region having sizes of order $1-10$ pc; this analysis combines measurements from both source types while the prior measurement used only narrow-line sources. Finally, we have now included flexible higher order multipole perturbations to the macromodel, following \cite{2024GilmanTurbo}.

Several recent works have highlighted the existence of multipoles in lens galaxy isophotes \citep[e.g.][]{Stacey_complex_2024, He_lens_light_complexity_2024,2024arXiv240308895C, 2024GilmanTurbo, Oh_improving_2024}, and considered the impact of these features on the measurement of the properties of low mass halos.  \citet{2024GilmanTurbo} tested the impact of these multipoles on mock simulations of quasar lenses similar to the sample considered in this paper, and demonstrated that it was possible to derive correct constraints on the half-mode-mass for both warm and cold ground truths even in the presence of random, unknown $a_3$ and $a_4$ multipoles in the deflector macromodels with an appropriate statistical treatment of the problem. Our results in the current work further demonstrate that strong constraints on the nature of dark matter are possible, while simultaneously accounting for multipole deviations in the deflector mass distribution based on measurements of the properties of populations of elliptical galaxies. 
In Figure~\ref{fig:DM-pole}, we show the joint posterior probability distribution for the multipole amplitudes and the DM half-mode mass for two of the lenses that are typical of our sample, in that there is no significant correlation between a cutoff in the half mode mass and the amplitude of the multipole moments.

Our constraints are comparable to those based on Lyman-$\alpha$ \citep{Viel_2013_Lya,Irsic_2017_Lya,Villasenor_2023_Lya}, which correspond to $m_{\rm WDM}>3.3,5.3,3.1$ keV, respectively.  
Similarly, counting satellite galaxies of the Milky Way provide constraints in the range $m_{\rm WDM}>2.02$ keV~\citep{Newtson_wdm_2021}, $m_{\rm WDM}>3.6$--$5.1$ keV (depending on the mass of the Milky Way)~\citep{dekker_warm_2022}, and 6.5 keV~\citep{nadler_constraints_2021}, and 7.4 keV when combined with strong lenses~\citep{nadler_dark_2021}. 
Substructure lensing, Milky Way satellite counts, and the Lya forest probe independent scales of the matter power spectrum at different redshifts, so most of the systematic uncertainties are independent. It is reassuring that these probes all disfavor WDM models in similar portions of parameter space and each probe makes the others' conclusions more robust.    

\section{Summary}\label{sec:summary}

Here we provide a summary of the key results of our paper. 

 \begin{itemize}
     \item We present rest-frame mid-IR fluxes measured with JWST MIRI for a sample of 9 lenses. Using SED fitting, we isolate the light coming from the warm dust region. The SED fitting gives inferred warm-dust flux ratios  consistent with those measured in the reddest filters, where the effects of microlensing are minimized. All systems show some degree of microlensing in at least one image in the bluest filter relative to the reddest filter, highlighting the importance of having multiband MIRI band imaging.

     \item We present an updated treatment of subhalo tidal stripping and evolution within the host potential.

     \item We use the flux ratios of the warm dust region, in combination with previously published flux ratios at other wavelengths, to calculate the posterior probability distribution of the WDM model.  We find that the half-mode mass is constrained to be less than $10^{7.6}$ $M_\odot$ 
     at posterior odds of 10:1.  This corresponds to constraint on the WDM particle mass to be above $m_{\rm WDM} > 6.1$ keV. 
    \item Our new limit on the free streaming length of dark matter is the strongest calculated from gravitational lensing to date and improves upon and is consistent with previous results from flux-ratio anomalies \citep{gilman_warm_2020}. Importantly, by virtue of the larger sample and new measurements, it is more stringent than the previous measurement even if it allows for departures from ellipticity in the mass distribution of the lenses.
    \item Our new limit is consistent with independent limits based on the Ly-$\alpha$ forest and Milky Way Satellites.

 \end{itemize}

\section*{Acknowledgements}
We thank Crystal Mannfolk, Greg Sloan, Blair Porterfield, and Henrik R. Larsson for help with observation planning. We thank Karl Gordon, Mattia Libralato, Jane Morrison, and Sarah Kendrew for their help in answering questions about the data reduction. We thank Marshall Perrin for helpful conversations about {\tt webbPSF}.

This work is based on observations made with the NASA/ESA/CSA James Webb Space Telescope. The data were obtained from the Mikulski Archive for Space Telescopes at the Space Telescope Science Institute, which is operated by the Association of Universities for Research in Astronomy, Inc., under NASA contract NAS 5-03127 for JWST. These observations are associated with program \#2046. 
Support for program \#2046 was provided by NASA through a grant from the Space Telescope Science Institute, which is operated by the Association of Universities for Research in Astronomy, Inc., under NASA contract NAS 5-03127. 

AN and TT acknowledge support from the NSF through AST-2205100 "Collaborative Research: Measuring the physical properties of DM with strong gravitational lensing". 
The work of LAM and DS was carried out at Jet Propulsion Laboratory, California Institute of Technology, under a contract with NASA. TA acknowledges support from the Millennium Science Initiative ICN12\_009, the ANID BASAL project FB210003 and ANID FONDECYT project number 1240105. DS acknowledges the support of the Fonds de la Recherche Scientifique-FNRS, Belgium, under grant
No. 4.4503.1. K.K.G. thanks the Belgian Federal Science Policy Office (BELSPO) for the provision of financial support in the framework of the PRODEX Programme of the European Space Agency (ESA). VM acknowledges support from ANID FONDECYT Regular grant number 1231418 and Centro de Astrof\'{\i}sica de Valpara\'{\i}so. 
VNB gratefully acknowledges assistance from a National Science Foundation (NSF) Research at Undergraduate Institutions (RUI) grant AST-1909297. Note that findings and conclusions do not necessarily represent views of the NSF.
KNA is partially supported by the U.S. National Science Foundation (NSF) Theoretical Physics Program, Grants PHY-1915005 and PHY-2210283. AK was supported by the U.S. Department of Energy (DOE) Grant No. DE-SC0009937, by the UC Southern California Hub, with funding from the UC National Laboratories division of the University of California Office
of the President,  by  the World Premier International Research Center Initiative (WPI),  MEXT,  Japan, and by Japan Society for the Promotion of Science (JSPS) KAKENHI Grant No. JP20H05853.
SB acknowledges support from Stony Brook University. DG acknowledges support for this work provided by the Brinson Foundation through a Brinson Prize Fellowship grant, and from the Schmidt Futures organization through a Schmidt AI in Science Fellowship.

%%%%%%%%%%%%%%%%%%%%%%%%%%%%%%%%%%%%%%%%%%%%%%%%%%
\section*{Data Availability}
This work was based on JWST MIRI imaging which becomes publicly available after a one-year proprietary period. All software used in the DM inference is publicly available, and intermediate data products may be made available upon reasonable request.

%%%%%%%%%%%%%%%%%%%% REFERENCES %%%%%%%%%%%%%%%%%%

\bibliographystyle{mnras}
\bibliography{example}

%%%%%%%%%%%%%%%%%%%%%%%%%%%%%%%%%%%%%%%%%%%%%%%%%%

%%%%%%%%%%%%%%%%% APPENDICES %%%%%%%%%%%%%%%%%%%%%

\clearpage

\appendix

\section{Multipoles}

\begin{figure*}
    \centering
    \includegraphics[width=0.49\textwidth]{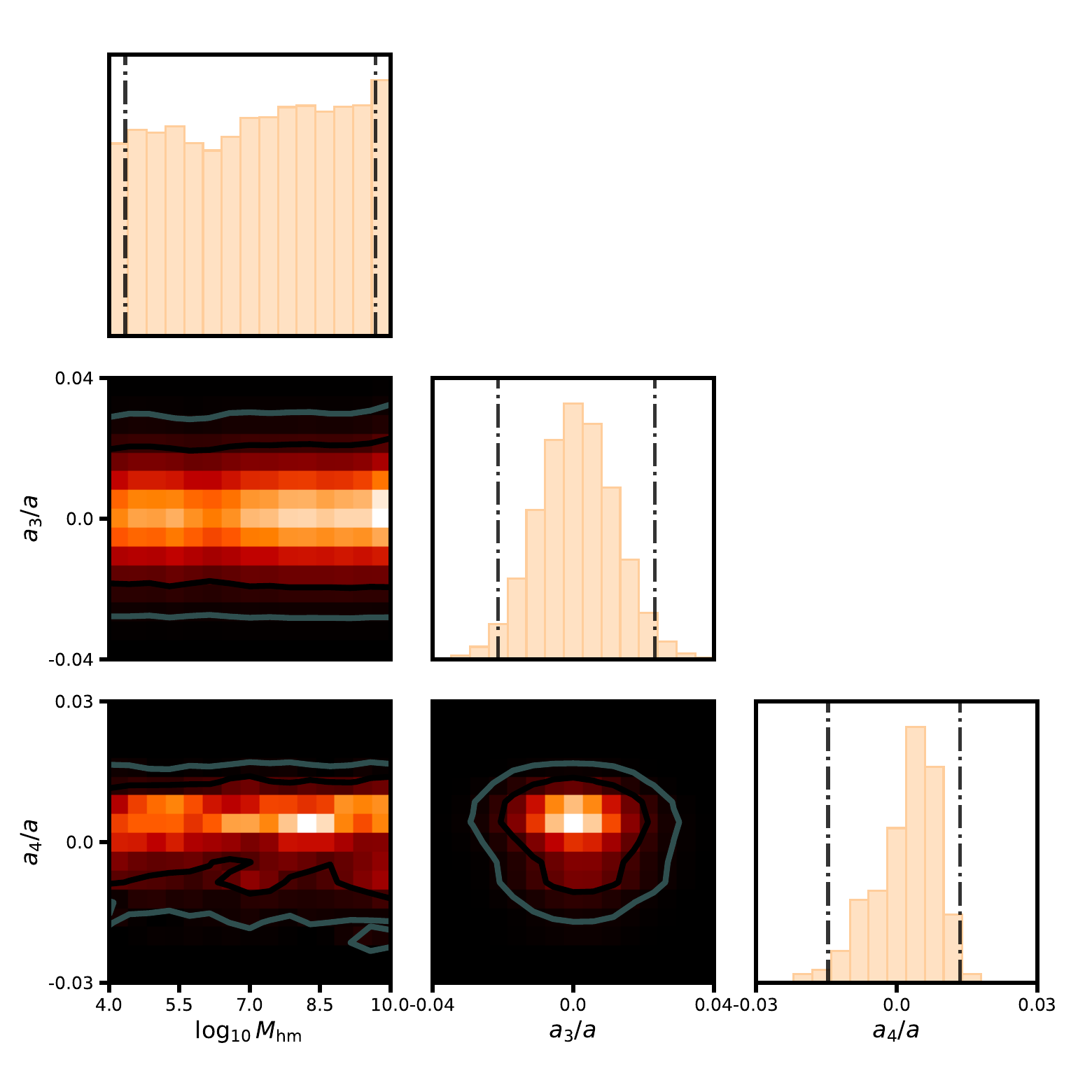}
    \includegraphics[width=0.49\textwidth]{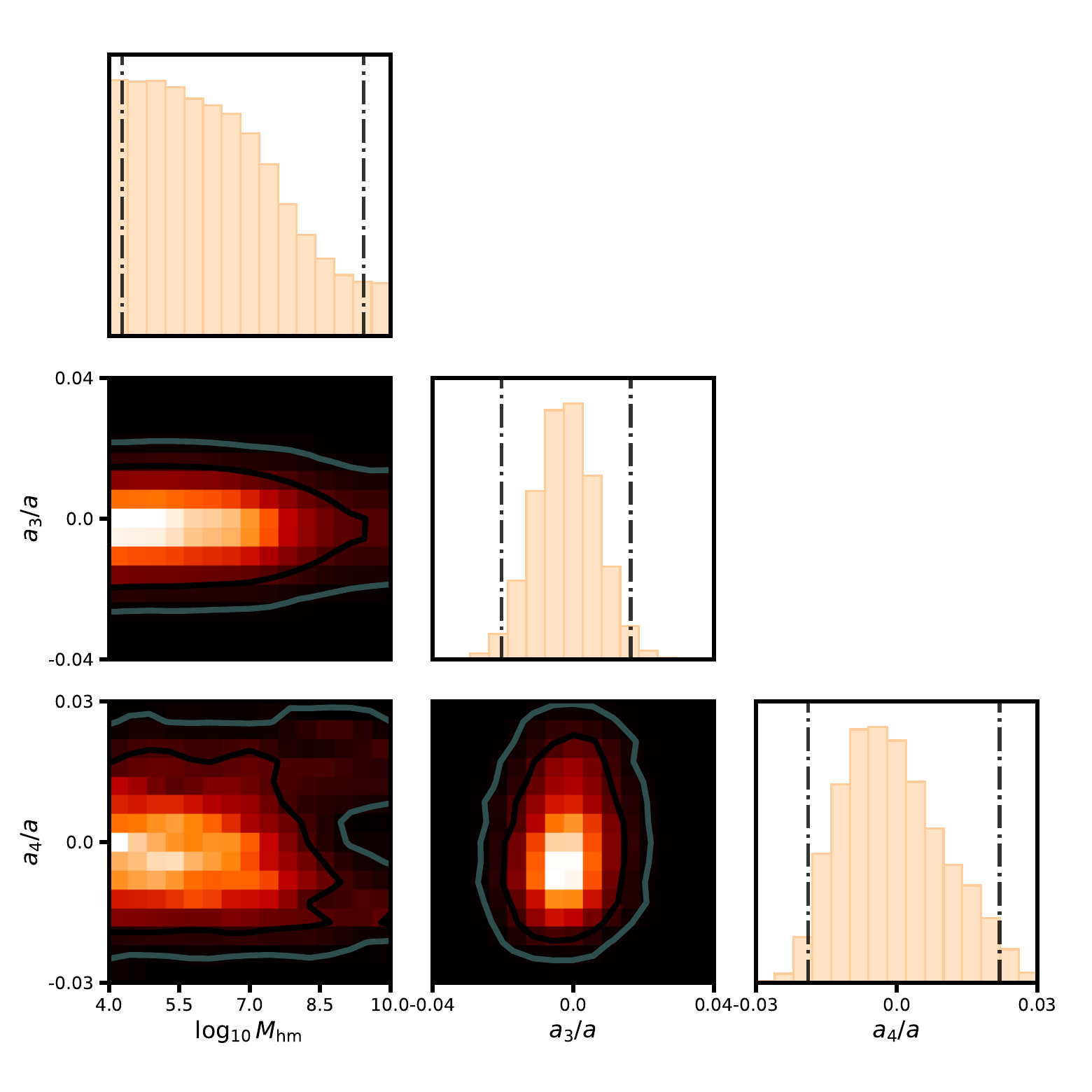}
    \caption{Example posteriors in the space of $M_{\rm hm}$, and the multipole amplitudes, $a_3/a$ and $a_4/a$, for the individual lenses 0405 (left), and 0659 (right).}
    \label{fig:DM-pole}
\end{figure*}

To illustrate the effects of multipoles, we show the posteriors for two example lenses in Fig~\ref{fig:DM-pole}.  For 0405, we see that the half-mode mass and multipole parameters are uncorrelated and weakly constrained, relative to the prior. For 0659, based on the structure of the 95 $\%$ confidence interval in the joint $a_4 / a$ and $M_{\rm{hm}}$ distribution, non-zero $a_4$ slightly weakens the constraint on $M_{\rm{hm}}$ relative to $a_4 = 0$. This figure clearly illustrates how additional degrees of model freedom affect ones posterior beliefs regarding the allowed values of certain model parameters, in this case the free streaming length. 

Distinguishing between various models that could explain the same dataset involves Bayesian model comparison. To illustrate this concept, we have analyzed the fourteen lenses in this work with lens models that only includes $m=3$ and $m=4$ multipole terms. We compute the posterior odds ratio, or Bayes factor, between lens models that include only these multipole terms and with lens models that include both multipoles and substructure (for additional details about these tests, see Appendix B in \citet{2024GilmanTurbo}). The posterior odds enables a Bayesian model comparison between two competing models that could both give rise to an observed dataset. These odds can be calculated from simulation-based inference methods by quantifying the frequency with which a the observed dataset emerges from a given model. The joint Bayes factor for the 14 lenses considered in this work exceeds 1,000\footnote{The exact number fluctuates based on the stochastic effect of the statistical measurement uncertainties of the flux ratios, but the inferred Bayes factor always exceeds 1,000.}. This indicates that a lens model that fits the data with substructure is overwhelmingly preferred relative to model that explains the data with only perturbations to the shape of the main deflector. Analyzing the data within the context of the model that includes both sources of perturbation leads to the statements regarding the free-streaming length of dark matter presented in this work. 

\section{Image fluxes}

Here we provide the measured image fluxes for our systems.

\begin{table*}
    \centering
    \begin{tabular}{lllllll}
    \hline
    \hline
    Lens           &image &  F560W &  F1280W & F1800W& F2100W & F2550W   \\
    \hline
     J0405         &A    &  0.365    &  1.05    & 1.38     &        & 2.66   \\
                  &B    &  0.246    &  0.646   & 0.871    &        & 1.80  \\
                  &C    &  0.418    &  1.07    & 1.41     &        & 2.80  \\
                  &D    &  0.497    &  1.32    & 1.74     &        & 3.37  \\
    \hline
    J0607          &A    & 0.0785    & 0.161    &0.223    & 0.261   &       \\
                  &B    & 0.0923    & 0.217    &0.313    & 0.370   &        \\
                  &C    & 0.241     & 0.613    &0.849    & 1.038    &        \\
                  &D    & 0.03    & 0.157    &0.236    & 0.269   &        \\

    \hline
    J0608          &A     & 1.2154   & 5.5023   & 6.9931  & & 8.0359   \\
                  &B     & 0.46174  & 2.0791   & 2.6063  & & 3.1453   \\
                  &C     & 0.454099 & 2.0185   & 2.5336  & & 2.9256   \\
                  &D     & 0.702176 & 2.7666   & 3.4098  & & 3.8568   \\
    \hline
    J0659          &A    &0.131      & 0.675    &0.999    &          &1.237   \\
                  &B    &0.145      & 0.648    &0.953    &          &1.181   \\
                  &C    &0.0963     & 0.486    &0.713    &          &0.867   \\
                  &D    &0.402      & 1.737    &2.556    &          &3.313   \\
                  \hline
    J1042          &A    &5.33       & 19.98    &31.66    &          &42.78  \\
                  &B    &0.797      & 4.94     &8.69     &          &12.74  \\
                  &C    &0.411      & 2.015    &3.24     &          &4.84   \\
                  &D    &0.235      & 1.26     &2.17     &          &3.08   \\
                  \hline
    J1537          &A    &0.0632  & 0.242   &0.386    &0.483  &            \\
                  &B    &0.0515  & 0.174   &0.281    &0.353  &            \\
                  &C    &0.0677  & 0.260   &0.382    &0.479  &            \\
                  &D    &0.0476  & 0.174   &0.285    &0.351  &            \\    
                  \hline
    J1606          &A    &0.516     &1.817     &2.640     &        &4.950  \\
                  &B    &0.521     &1.825     &2.625     &        &5.010  \\
                  &C    &0.310     &1.072     &1.530     &        &2.914  \\
                  &D    &0.375     &1.389     &1.942     &        &3.714  \\                
                  \hline
    J2026          &A1   &1.999     &8.581     &12.59     &        &20.23  \\
                  &A2   &1.488     &6.534     &9.529     &        &15.62  \\
                  &B    &0.615     &2.661     &3.936     &        &6.136  \\
                  &C    &0.562     &2.382     &3.525     &        &5.666  \\ 
                  \hline
    J2038          &A    &3.799     &8.798     &15.22     &        &21.36  \\
                  &B    &4.419     &10.85     &18.74     &        &26.31  \\
                  &C    &3.440     &8.434     &14.46     &        &20.43  \\
                  &D    &1.608     &3.891     &6.756     &        &9.362  \\
    
    \end{tabular}
    \caption{Measured image fluxes in units of mJy}
    \label{tab:fluxes}
\end{table*}

\clearpage
%%%%%%%%%%%%%%%%%%%%%

\section{Lens Model Parameters}

Here we provide the best-fit lens model parameters (Table \ref{tab:macromodel}), light parameters, and PSF parameters found in the final step of our fitting, where the point sources are treated as free foreground objects. These macromodels are provided to aid reproduction of our flux-ratio measurements. Given that our goal was to optimize the measurement of the image positions and fluxes, the fact that we do not use the image positions to constrain the lens model in the final fitting step, the ongoing room for improvement in the MIRI PSF models, particularly in F560W which typically has the brightest arc, and therefore the strongest constraint on the lens model, we do not recommend these values should be interpreted as robust physical constraints of the properties of the lenses.

\begin{table*}
    \centering
    \begin{tabular}{lllllllll|lll}
    \hline
    \hline
    Lens          & $\theta_{\rm{E}}$ & $\gamma_p$ & $q$ &  $\phi$ (degree) & $\gamma_{\rm{ext}}$ & $\phi_\gamma$ (degree) & dRA& dDec&    $\theta_{\rm{G2}}$ & dRA$_{\rm{G2}}$ &dDec$_{\rm{G2}}$  \\
    \hline
     J0405         &  0.72  & 2.02     &0.55      & 260     & 0.077       & 250     & 0.362        & 0.542       &      &       &  \\  
     J0607       & 0.77 & 2.00 & 0.71 & 142 & 0.077 & 143 & -0.532 & 0.717 & 0.118 & 0.687 & 0.984 \\
     
     J0608         & 0.68 & 2.00 & 0.34 & 184 & 0.147 & 182 & 0.558 & 0.062 \\
     J0659         & 2.40 & 2.01 & 0.57 & 210 & 0.010 & 150 & -1.699 & 0.983 & &  &  \\
     J1537         &1.39 & 2.02 & 0.89 & 357 & 0.133 & 466 & -1.444 & 0.766 \\
     J1606         &0.67 & 1.93 & 0.47 & 389 & 0.154 & 351 & -0.826 & -0.451  \\
     J2038         &1.38 & 2.33 & 0.54 & 305 & 0.047 & 365 & -0.721 & -0.914 \\

    \end{tabular}
    \caption{Best fitting lens model parameters, with the image positions not used in the lens model. Position angles are given in degrees East of North, all other angles are in units of arcseconds. Lens centers given relative to coordinate system in Table \ref{tab:flux_results}.
    Note that lenses 1042, 2026 did not have detected extended arcs in any filter and thus no lens model was applied.}
    
    \label{tab:macromodel}
\end{table*}

\begin{table*}
    \centering
    \begin{tabular}{lllll|llll}
    \hline
    \hline
    Lens    &Filter & r$_{\rm{s}}$ (arcsec)& n$_{\rm{s}}$ & n$_{\rm{max}}$ &  $\beta$ (arcsec) & dRa & dDec\\
    \hline

     J0405   &F560W  &  0.25  & 7     & 5      & 0.11  &  0.245 & 1.282   \\
            &F1280W &  0.30 &0.5      & --     & --    &         &   \\
            &F1800W & 0.07 & 1     &  --    &  --    \\
    \hline
     J0607  & F560W    & 1.081 & 5.95 & 3 & 0.11   & -0.180 & 0.017 \\
            & F1280W   & 0.318 & 1.17 & --  &  --  &        &       \\
            & F1800W   & 0.314 & 1.15 & --  &  --  &       &       \\
            & F2100W   & 0.306 & 1.05 & --  &  --  &       &        \\
    \hline
    J0608   &F560W &  0.55 & 6.00 & & & 0.720  & -0.468 \\

    \hline
    J0659  & F560W  &  0.21 & 5.95 &  -- & --  & $-2.284$ & 0.726 \\
    \hline
    J1537  & F560W   &0.12 & 1.00 & & & $-1.203$  & $-0.254$ \\
          & F1280W  &0.02 & 1.00 & & &   & \\
          & F1800W  &0.13 & 1.00 & & &   & \\
          & F2100W  &0.16 & 1.00 & & &   & \\
    \hline
    J1606  & F560W  &0.34  & 1.0 & & & $-0.631$  & 0.085 \\
          & F1280W  &0.34  & 1.0 & & &   &  \\
          & F1800W  &0.35  & 0.9 & & &   &  \\
          & F2550W  &0.36  & 0.9 & & &   & \\
    \hline       
    J2038  & F560W & 0.25 & 1.00 & & & $-$1.060  & $-$0.987 \\
          & F1280W & 0.25 & 1.00 & & &   &  \\

    \end{tabular}
    \caption{Best fitting source parameters with positions given in the coordinate system of Table \ref{tab:flux_results}. The source centroids are restricted to be the same in all filters. Lenses 1042, 2026 and filters which are not listed did not contain a detected lensed quasar host galaxy.}
    
    \label{tab:source}
\end{table*}

\begin{table*}
    \centering
    \begin{tabular}{llllllll}
    \hline
    \hline
    Lens    &Filter & R$_{\rm{s}}$ (arcsec) & $q$ & $\theta$ (degrees) &dRA & dDec & R$_{\rm{s, sat}}$ \\
    \hline
    J0405   & F560W  & 0.16   & 0.87    & 0 & 0.375  & 0.542 &         \\
           & F1280W  & 0.04   & 0.59    & -40 &       &     &     \\
    \hline
     J0607   & F560W  & 0.39 & 0.75 & 86 & -0.549 & 0.752 & 0.075 \\
     \hline
     J0608 & F560W & 0.30 & 0.44 & 173 & 0.453 & -0.010 \\

    \hline
    J0659    & F560W  &  0.59 & 0.95 & 162 & -1.795 & 0.860 & \\
            & F1280W &0.27 & 0.60 & 172 &  &  & \\
            & F1800W & 0.88 & 0.93 & 158 &  &  &  \\
            & F2550W & --   & -- & --    &  &  & \\
    \hline 
    J1537    &F560W   &0.66 & 0.79 & 372 & -1.434 & 0.754 \\
            &F1280W  &1.45 & 0.52 & 391 &  &  & \\
    J1606    & F560W  &0.11 & 0.41 & 353 & -0.845 & -0.395 \\
    \hline
    J2038    & F560W &6.95 & 0.53 & 308 & -0.721 & -0.85\\

    \end{tabular}
    \caption{Best fitting deflector and satellite light parameters with positions given relative to the coordinate system in Table \ref{tab:flux_results}. Deflector light centroids are constrained to be the same in all filters. S\'ersic indices are held fixed to 4 for the deflector and satellite light. }
    
    \label{tab:light}
\end{table*}

\begin{table*}
    \centering
    \begin{tabular}{llllll}
    \hline
    \hline
    Lens    &Filter & {\tt{jitter\_sigma}} (arcsec) & T (K) &  f$^a$ \\
    \hline
 
     J0405   &F560W  & 0.063  & 1690    & 0.41    \\
            &F1280W & 0.062  & 756    &      &     \\
            &F1800W & 0.073 &  744    &      &       \\
            &F2550W & 0.081  & 280    &       &              \\
    \hline

     J0607   &F560W  & 0.068  & 704    & 0.18    \\
            &F1280W & 0.069  & 820    &      &     \\
            &F1800W & 0.070 &  207    &      &       \\
            &F2100W & 0.080  & 230    &       &              \\
    \hline
    J0608 &F560W  & 0.058 & 2092  &  0.71  \\
         &F1280W & 0.062 & 661   &      &     \\
         &F1800W & 0.066 & 414   &      &       \\
         &F2550W & 0.072 & 433   &       &              \\        
    J1042  & F560W  & 0.053 & 2550 & 0   \\
           & F1280W & 0.059 & 498 &  \\
           & F1800W & 0.069 & 288 & \\
           & F2550W & 0.079 & 340 & \\
    \hline
    J1537    &F560W  & 0.063  & 509 & 1.0    \\
            &F1280W & 0.063  & 837 &      &     \\
            &F1800W & 0.062  & 338 &      &       \\
            &F2100W & 0.073  & 299 &       &              \\           
    \hline
    J1606    &F560W  & 0.064  & 1163    & 0    \\
            &F1280W & 0.063  & 787    &      &     \\
            &F1800W & 0.077 &  912    &      &       \\
            &F2550W & 0.058  & 367    &       &              \\        
    \hline
    J2026    &F560W  & 0.060  & 812   & 0.09    \\
            &F1280W & 0.061  & 685    &      &     \\
            &F1800W & 0.065 &  473    &      &       \\
            &F2550W & 0.077  & 715    &       &              \\    
    \hline
    J2038    &F560W &0.047 &1448 & 0.22 \\
            &F1280W &0.049 &281 & \\
            &F1800W &0.069 &215 & \\
            &F2550W &0.092 &186 & \\
    \end{tabular}
    \caption{Best fitting PSF parameters. a) Fractional weighting of the zero and second fits extensions in the PSF model. 
    Set to 1 for all but F560W. }
    
    \label{tab:psf}
\end{table*}

\section{Additional Lenses}
Information about additional  lenses included in the DM constraint.
\begin{table*}
    \centering
    \begin{tabular}{lllll}
    \hline
    \hline
     Lens           & Source z & Lens z &  Discovery paper(s)               & Quasar Region \\
    \hline
     MG J0414+0534  & 2.64     &  0.96  & \citet{StaceyMcKean0414discovery} & CO spectral line (quasar nucleus)     \\
     HE J0435-1223  & 1.69     &  0.45  & \citet{nierenberg_double_2020}    & narrow-line region   \\
     RX J0911+0551  & 2.76     &  0.77  & \citet{nierenberg_double_2020}    & narrow-line region   \\
 %    PG J1115       & 1.72     &  0.31  &                                   & radio jet  \\
     B J1422+231    & 3.67     &  0.36  & \citet{nierenberg_double_2020}    & narrow-line region   \\
     WFI J2033-4723 & 1.66     &  0.66  & \citet{nierenberg_double_2020}    & narrow-line region   \\
    \end{tabular}
    \caption{Information for additional lenses used in the DM constraint.}
    \label{tab:additional_lenses}
\end{table*}

%%%%%%%%%%%%%%%%%%%%%

\section{Individual Lens Fits}\label{app:individual_lens_fits}
Here we present the lens models with separate components.
\begin{figure*}
    \includegraphics[width=\textwidth]{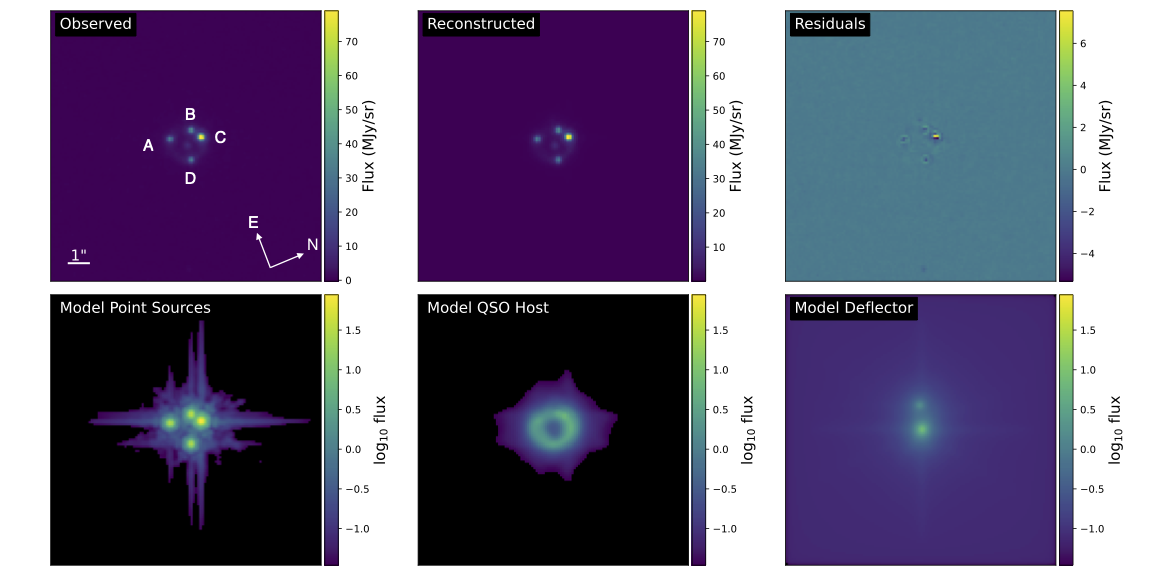}
     \caption{J0405 F560W.}
\end{figure*}
\begin{figure*}
    \includegraphics[width=\textwidth]{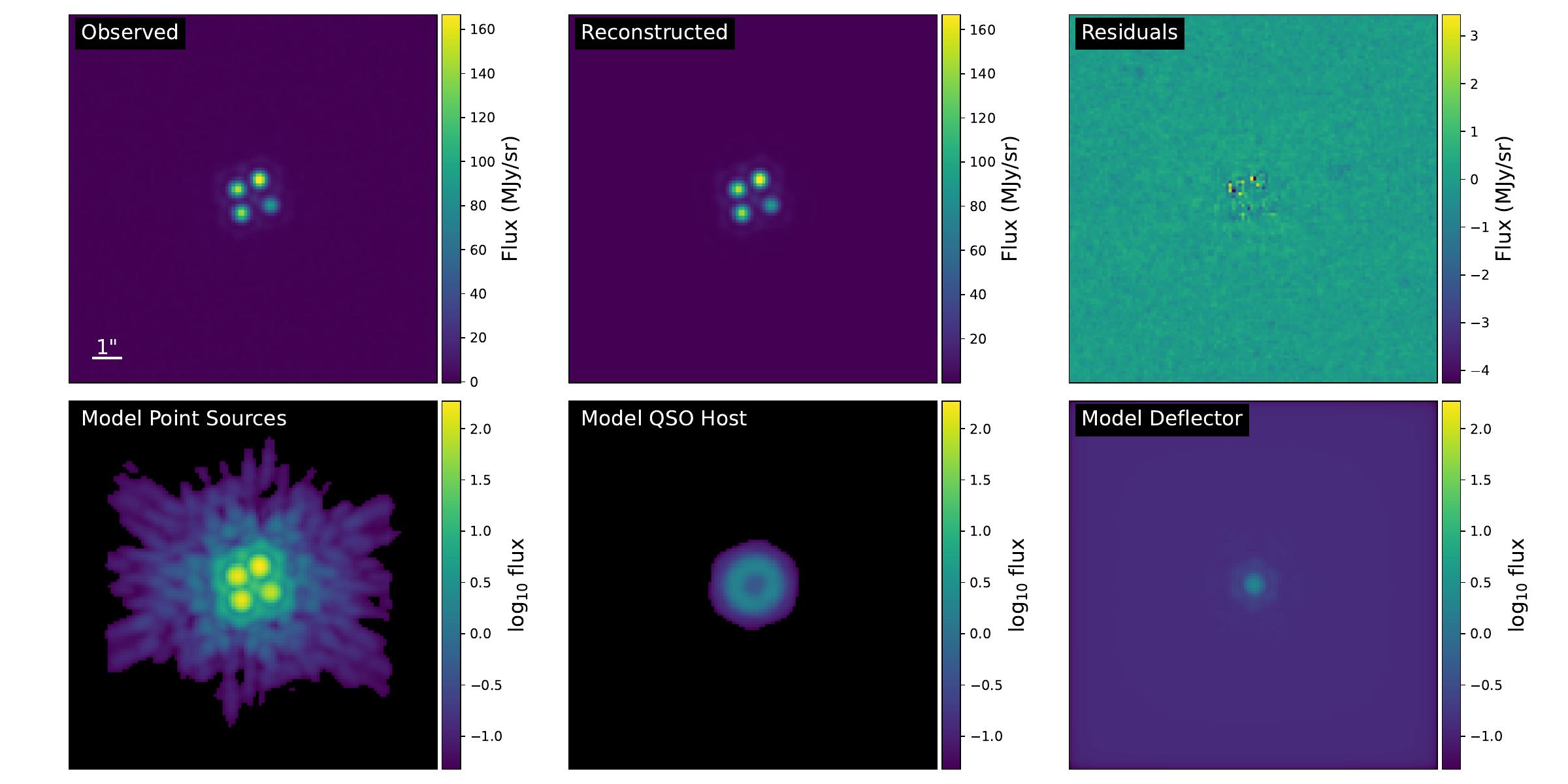}
    \caption{J0405 F1280W.}
\end{figure*}
\begin{figure*}
    \includegraphics[width=\textwidth]{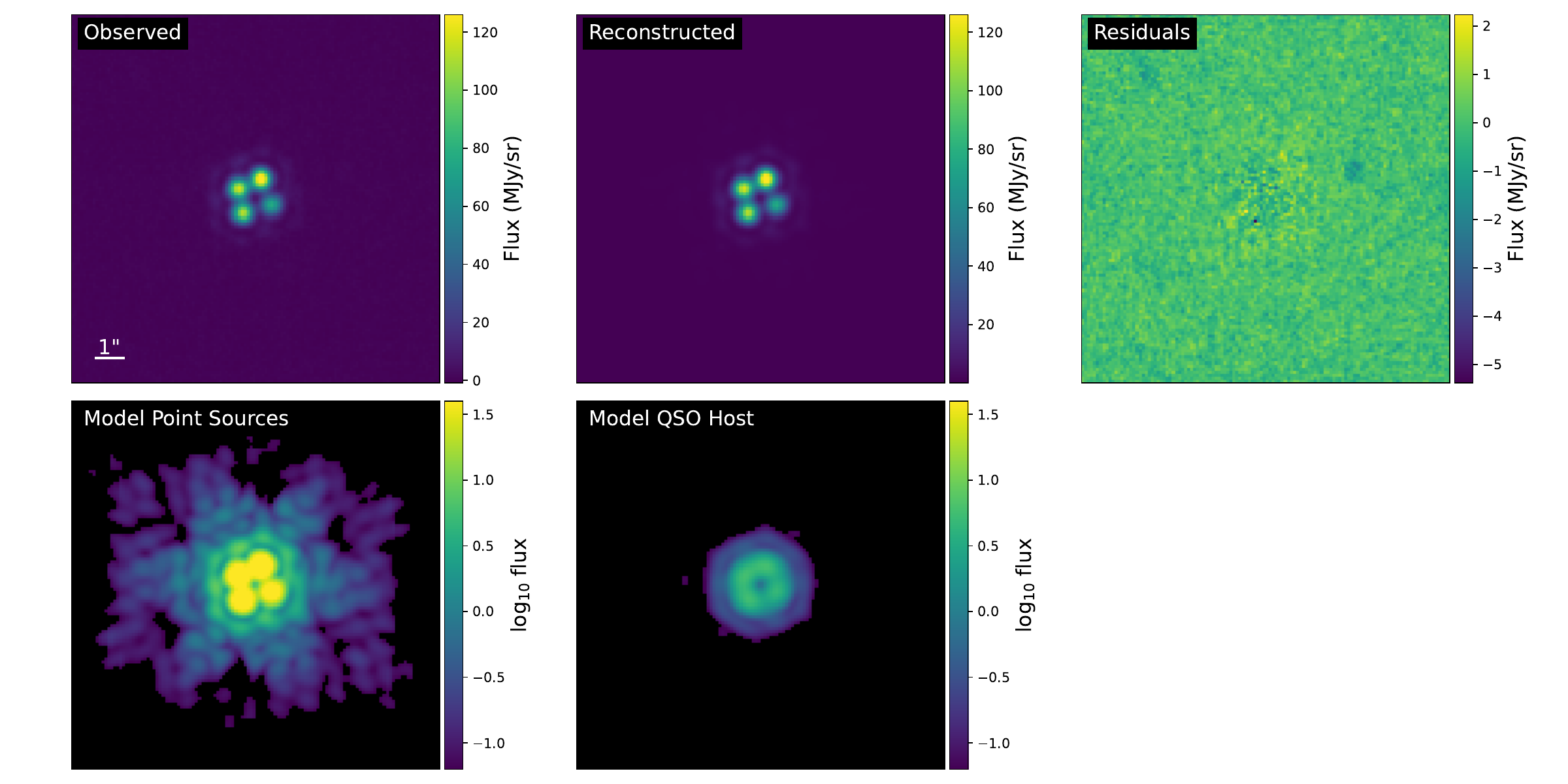}
    \caption{J0405 F1800W.}
\end{figure*}
\begin{figure*}
    \includegraphics[width=\textwidth]{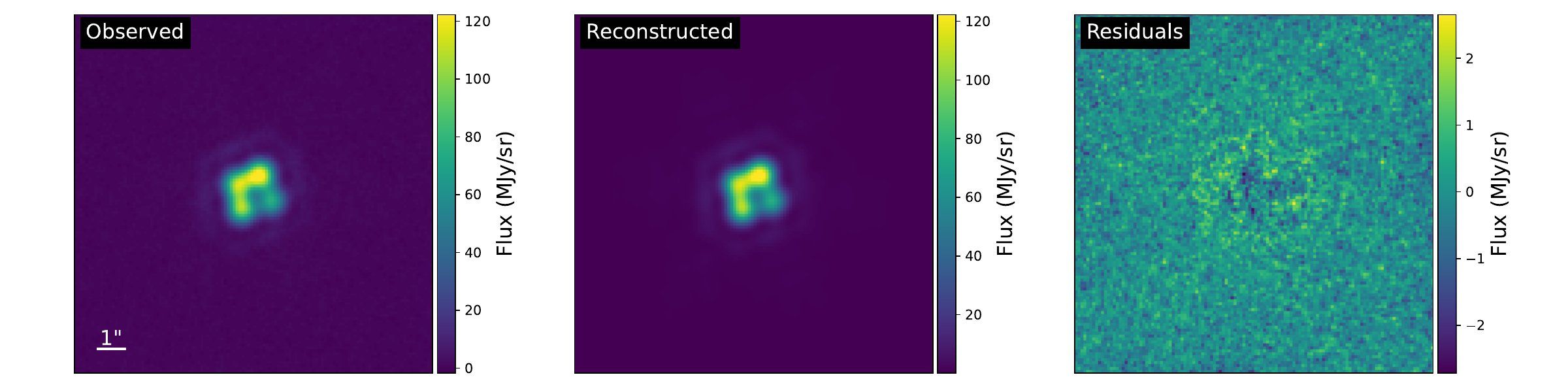}
    \caption{J0405 F2550W.}
\end{figure*}

\begin{figure*}
    \includegraphics[width=0.49\textwidth]{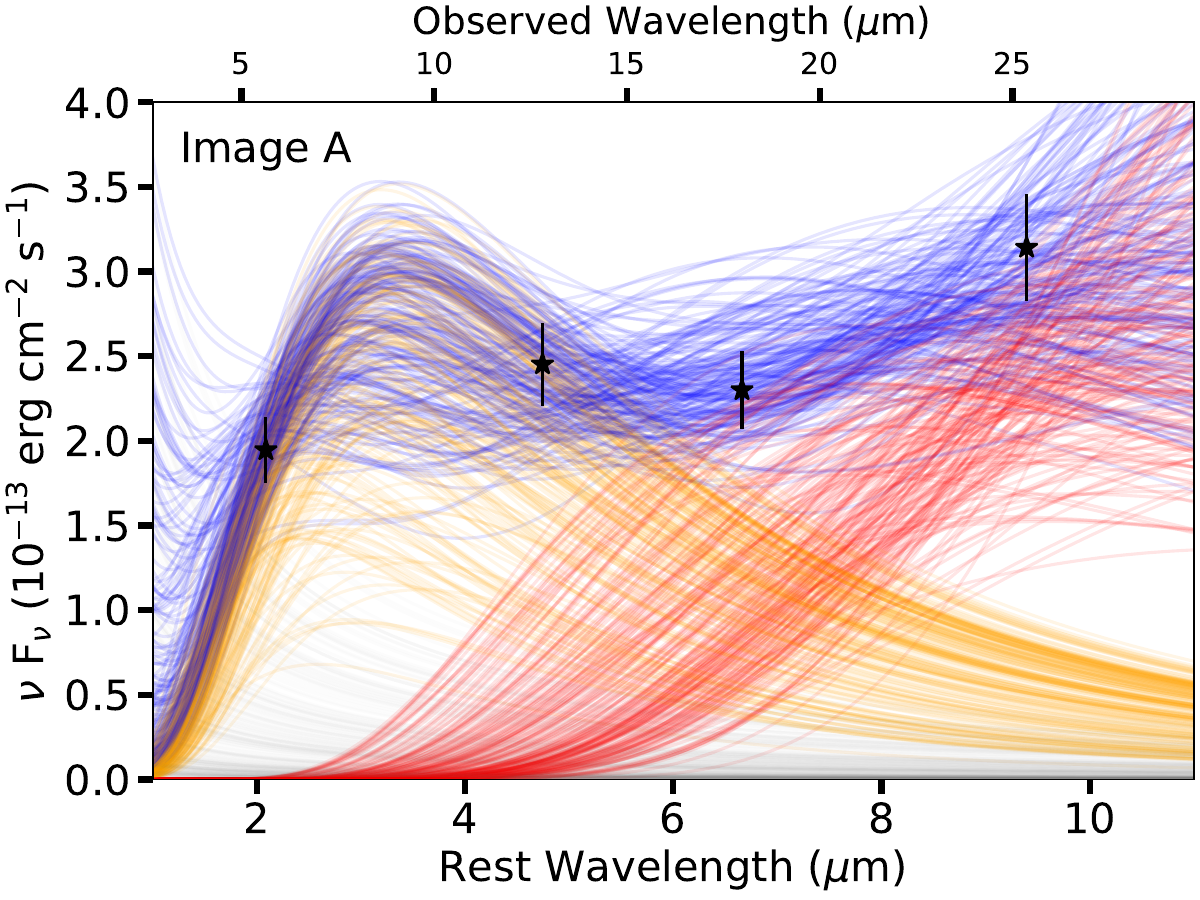}
    \includegraphics[width=0.49\textwidth]{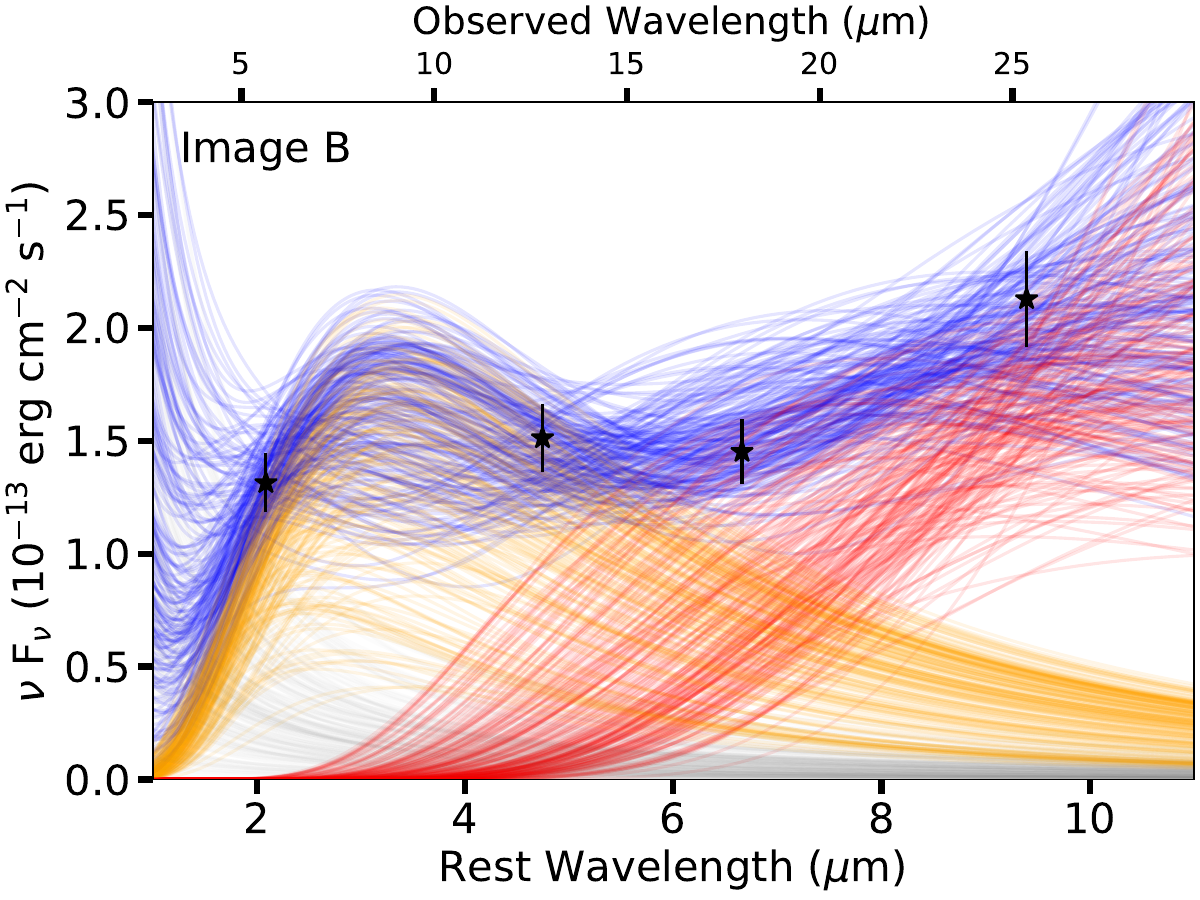}
    \includegraphics[width=0.49\textwidth]{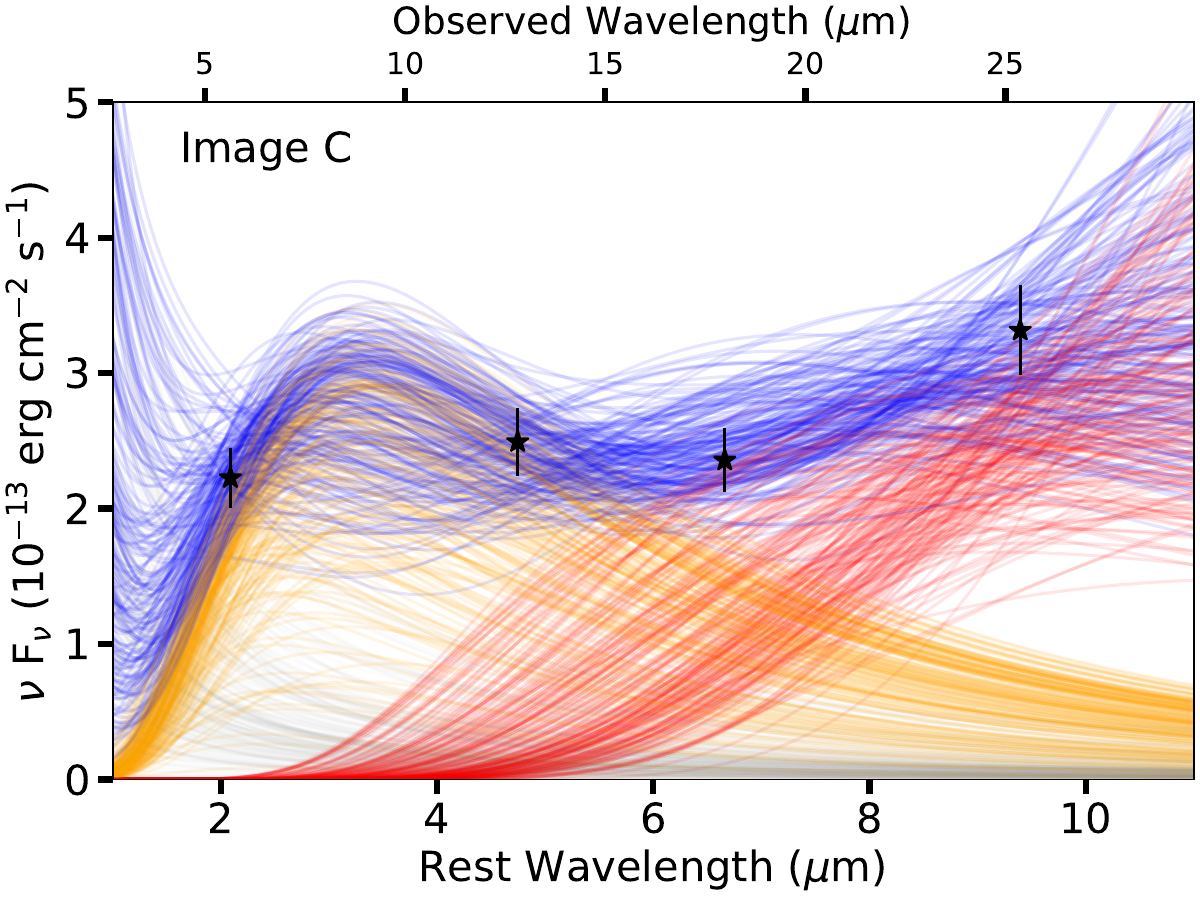}
    \includegraphics[width=0.49\textwidth]{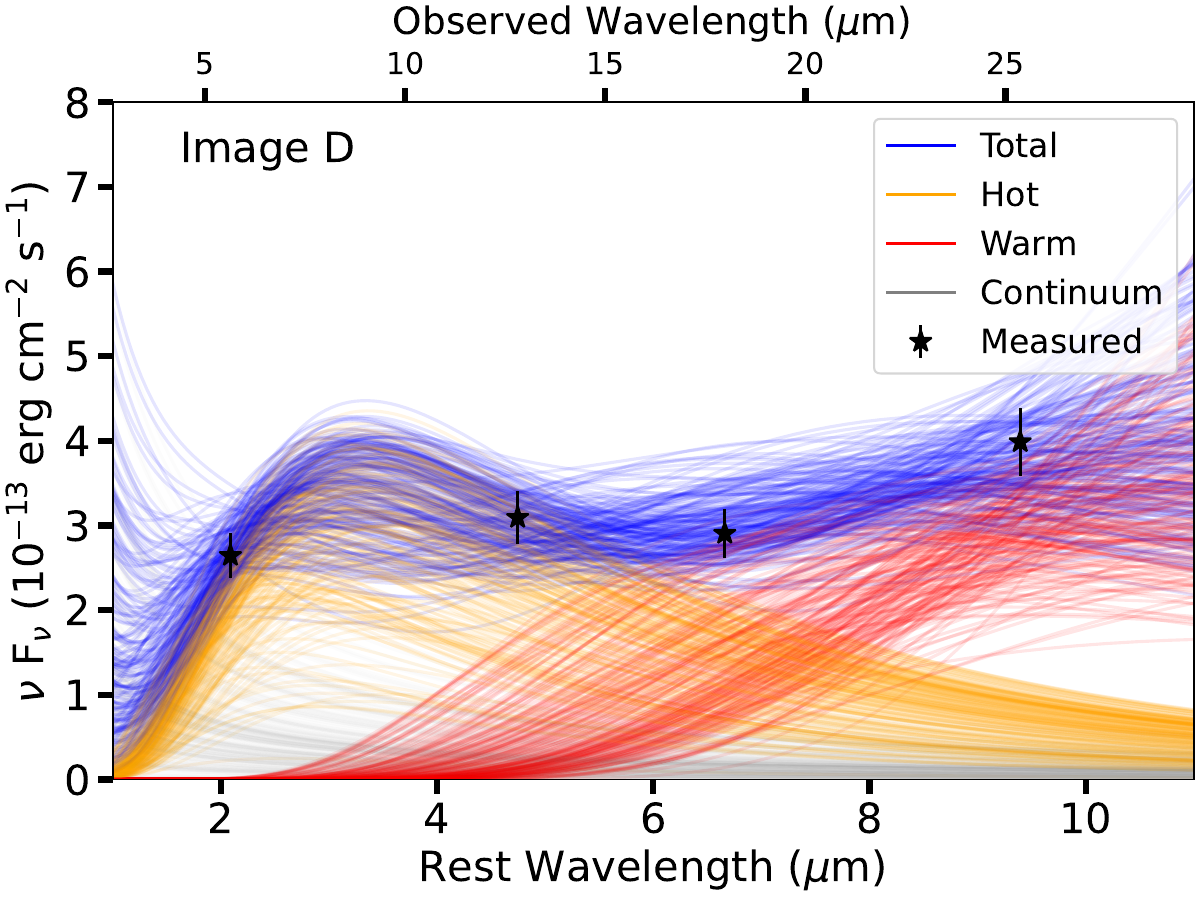}
    \caption{Posterior predictive distributions for the SED fitting for J0405.}
\end{figure*}

\begin{figure*}
    \includegraphics[width=\textwidth]{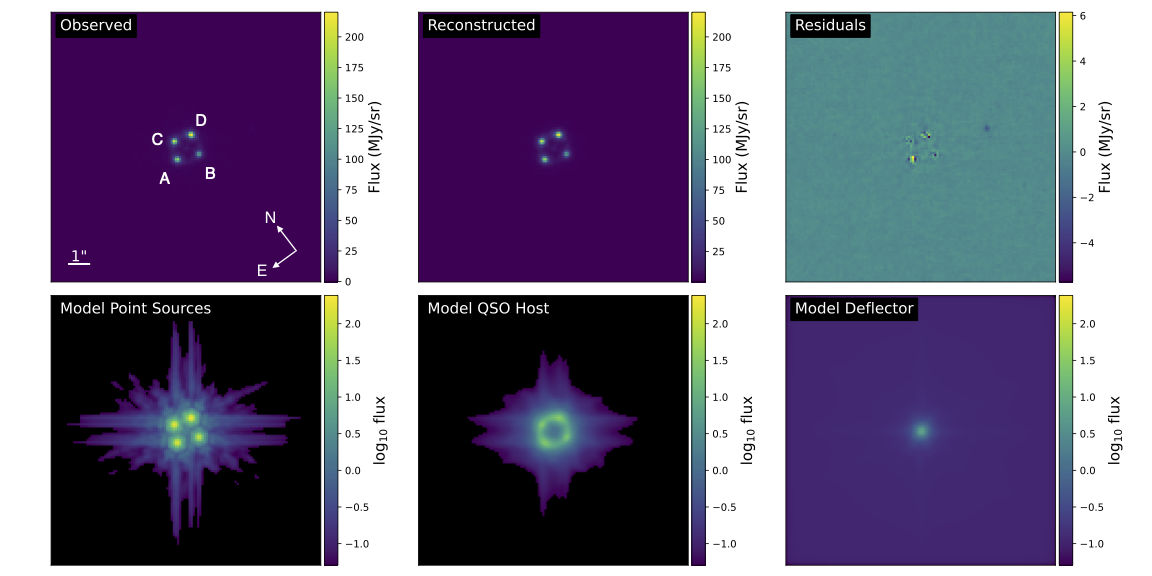}
     \caption{J0607 F560W.}
\end{figure*}
\begin{figure*}
    \includegraphics[width=\textwidth]{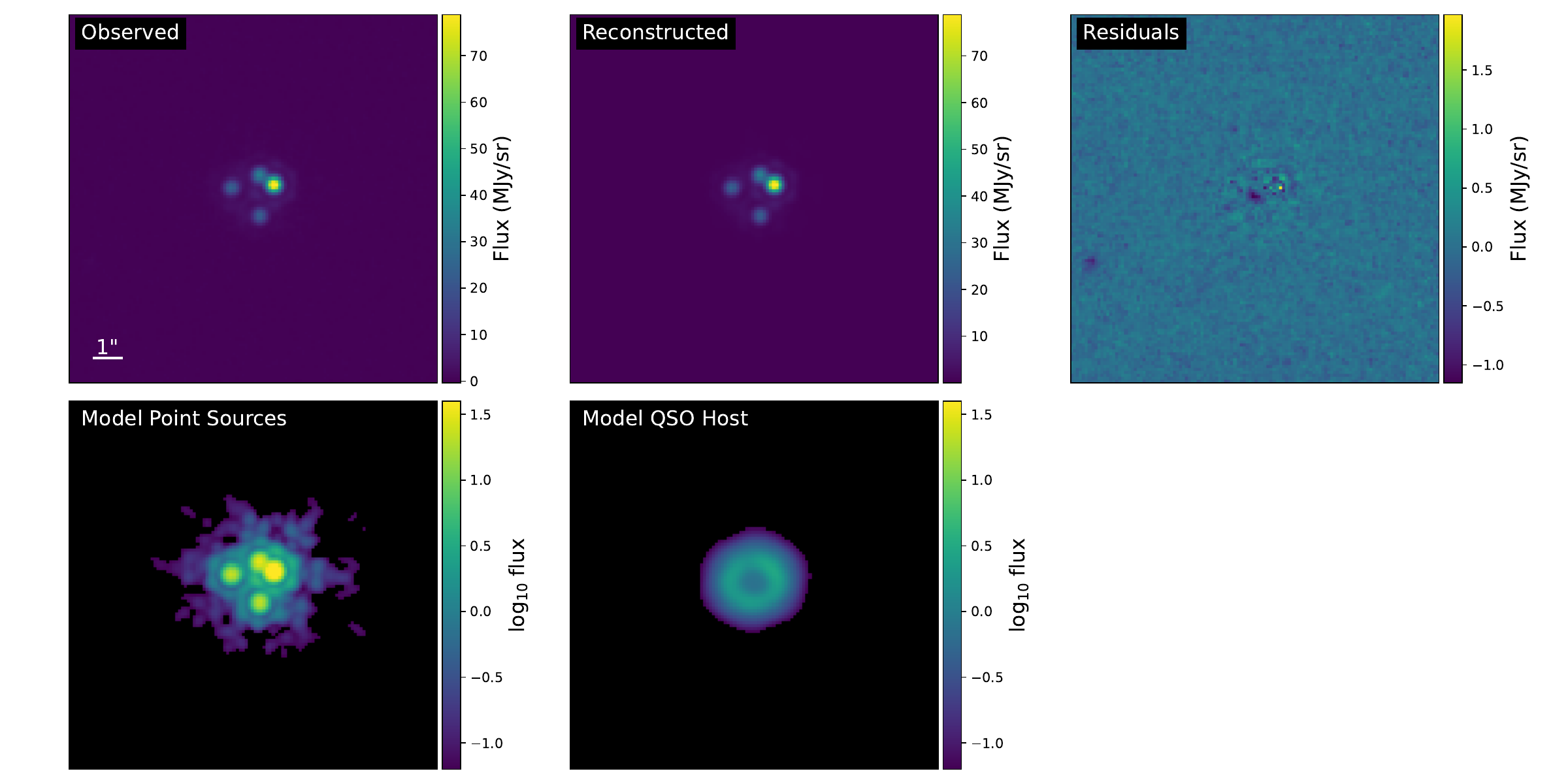}
    \caption{J0607 F1280W.}
\end{figure*}
\begin{figure*}
    \includegraphics[width=\textwidth]{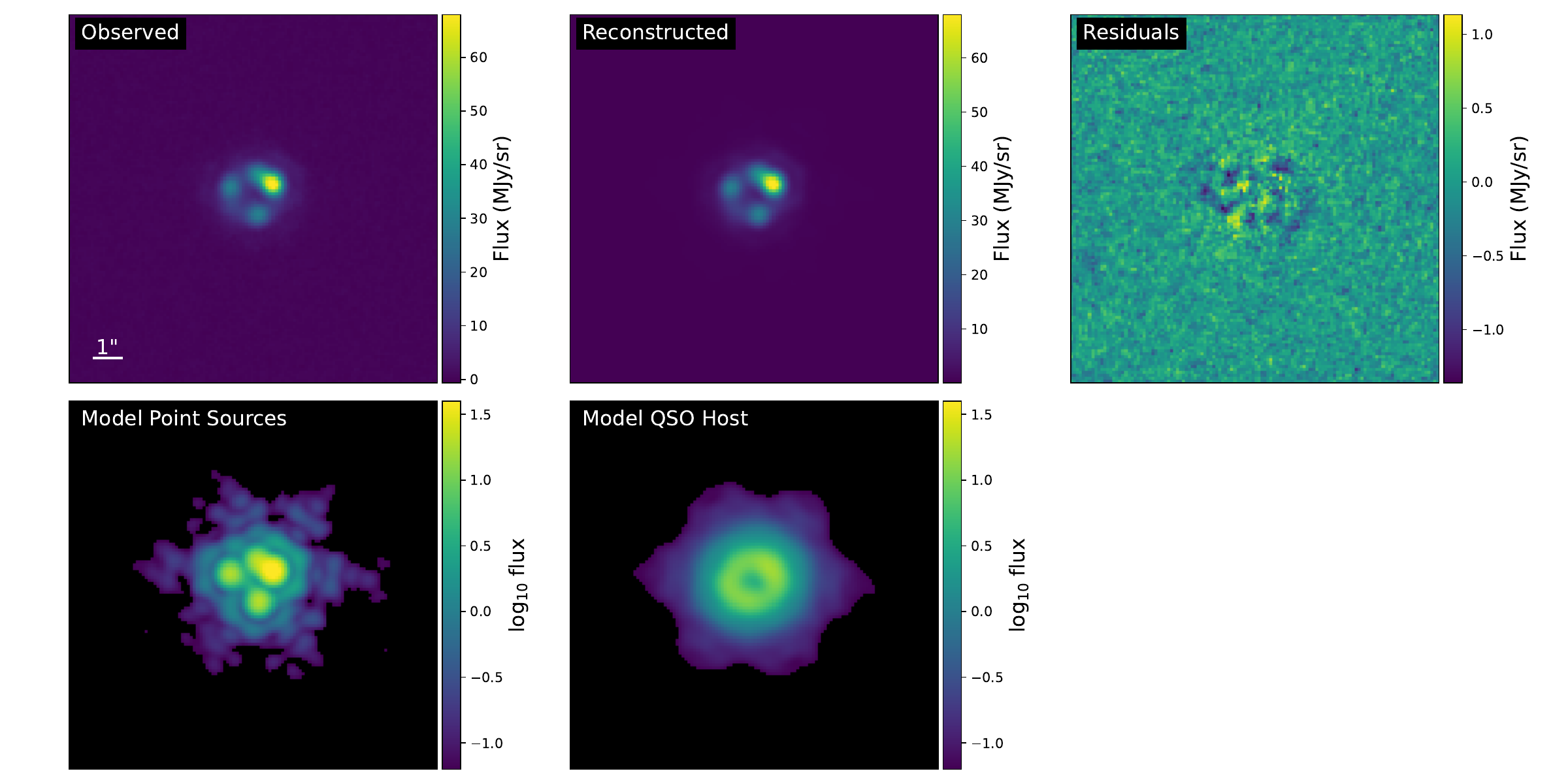}
    \caption{J0607 F1800W.}
\end{figure*}
\begin{figure*}
    \includegraphics[width=\textwidth]{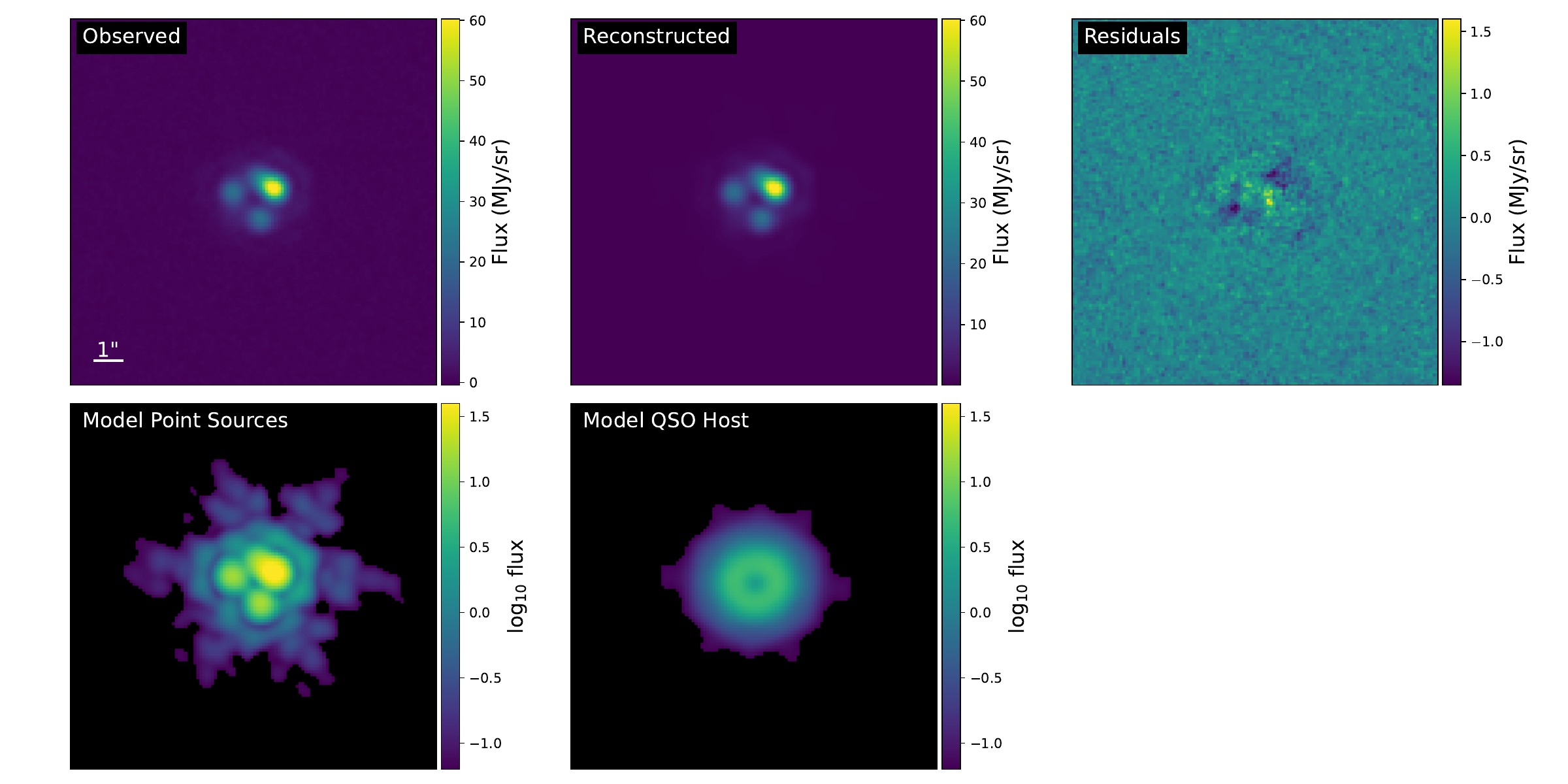}
    \caption{J0607 F2100W.}
\end{figure*}

\begin{figure*}
    \includegraphics[width=0.49\textwidth]{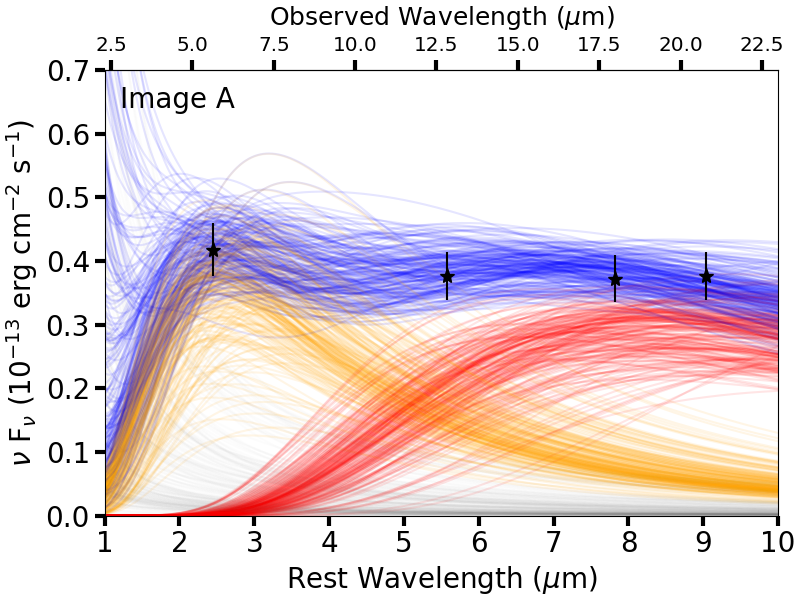}
    \includegraphics[width=0.49\textwidth]{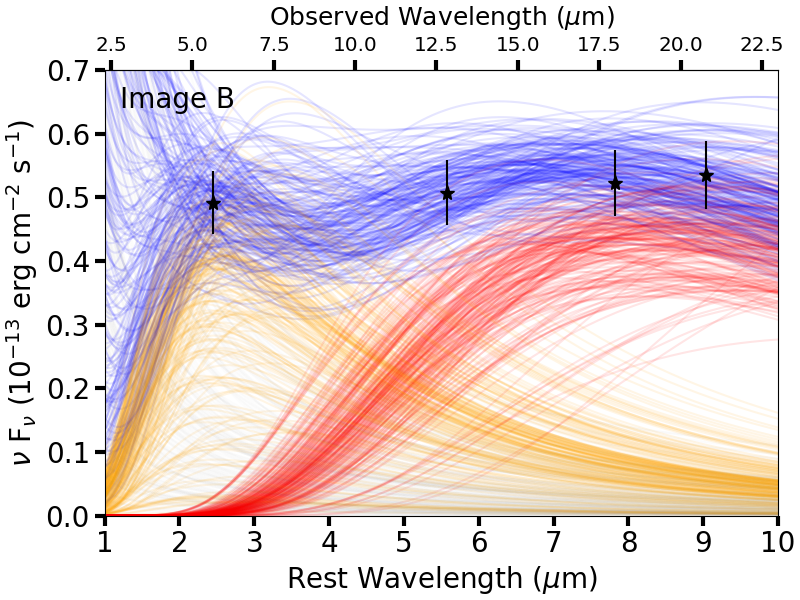}
    \includegraphics[width=0.49\textwidth]{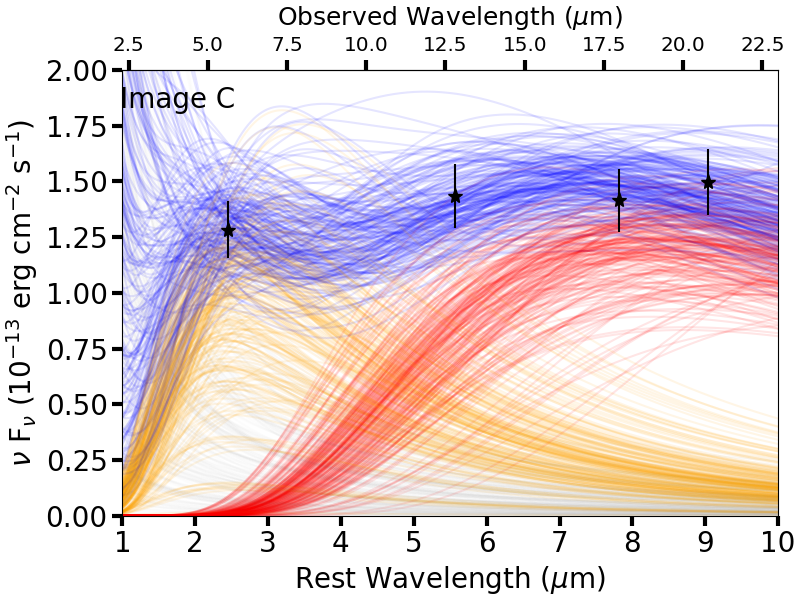}
    \includegraphics[width=0.49\textwidth]{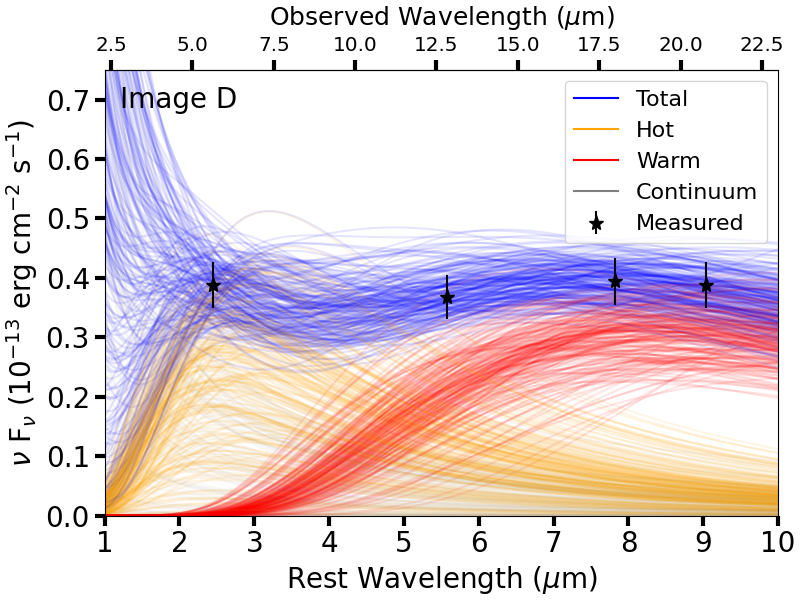}
    \caption{Posterior predictive distributions for the SED fitting for J0607.}
\end{figure*}

\begin{figure*}
    \includegraphics[width=\textwidth]{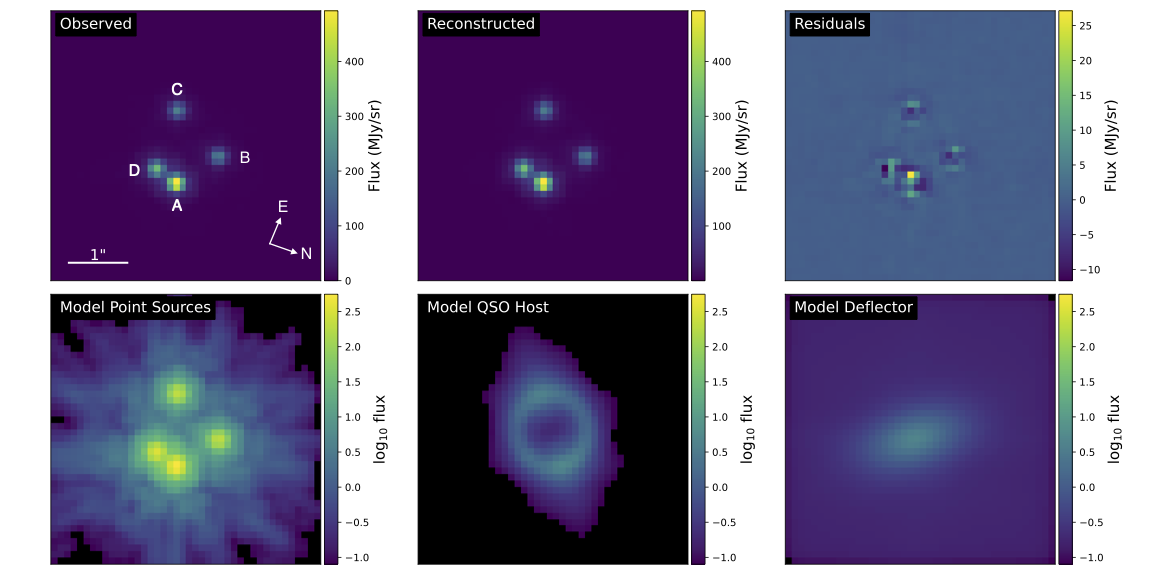}
     \caption{J0608 F560W.}
\end{figure*}
\begin{figure*}
    \includegraphics[width=\textwidth]{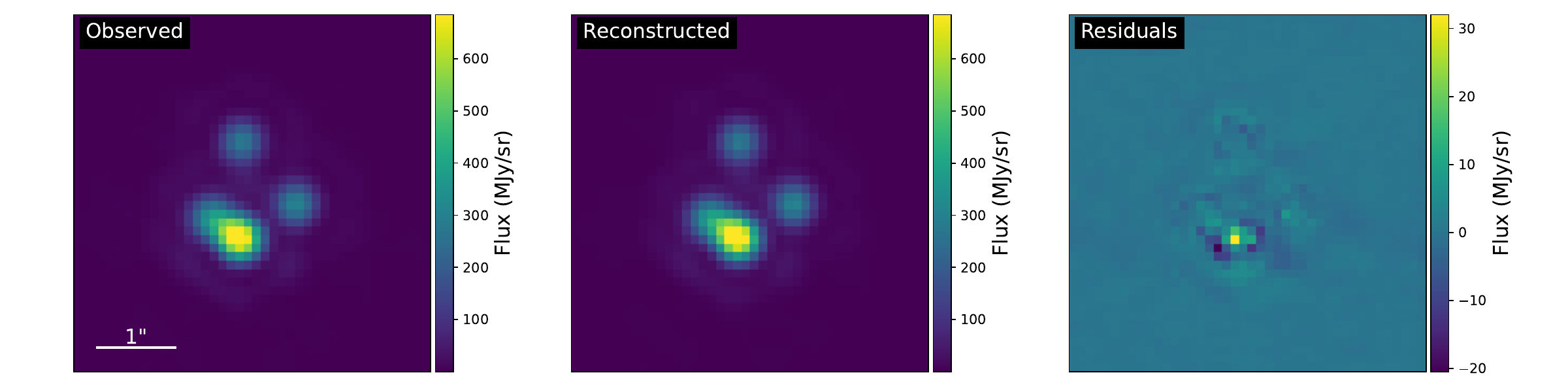}
    \caption{J0608 F1280W.}
\end{figure*}
\begin{figure*}
    \includegraphics[width=\textwidth]{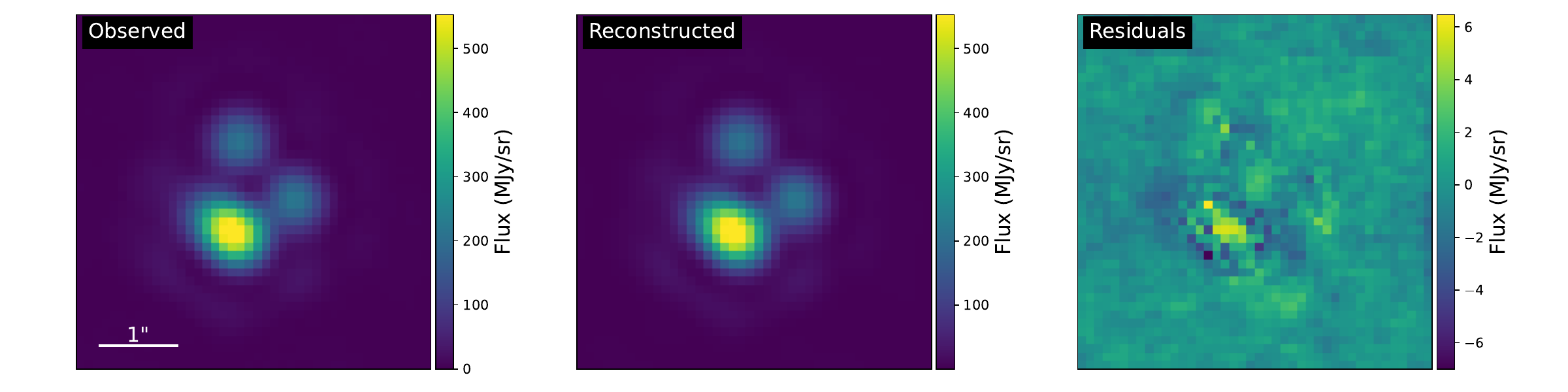}
    \caption{J0608 F1800W.}
\end{figure*}
\begin{figure*}
    \includegraphics[width=\textwidth]{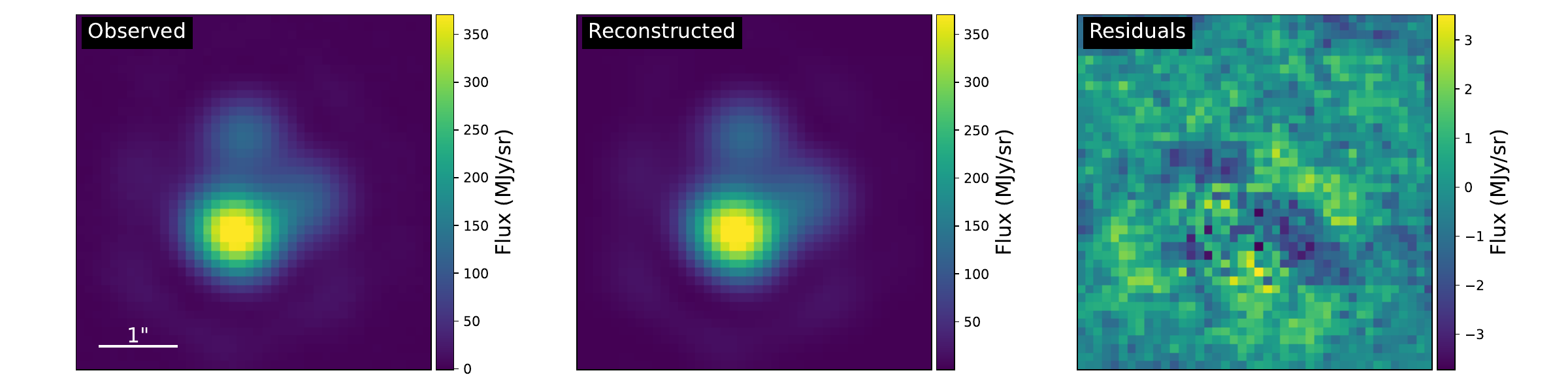}
    \caption{J0608 F2550W.}
\end{figure*}

\begin{figure*}
    \includegraphics[width=0.49\textwidth]{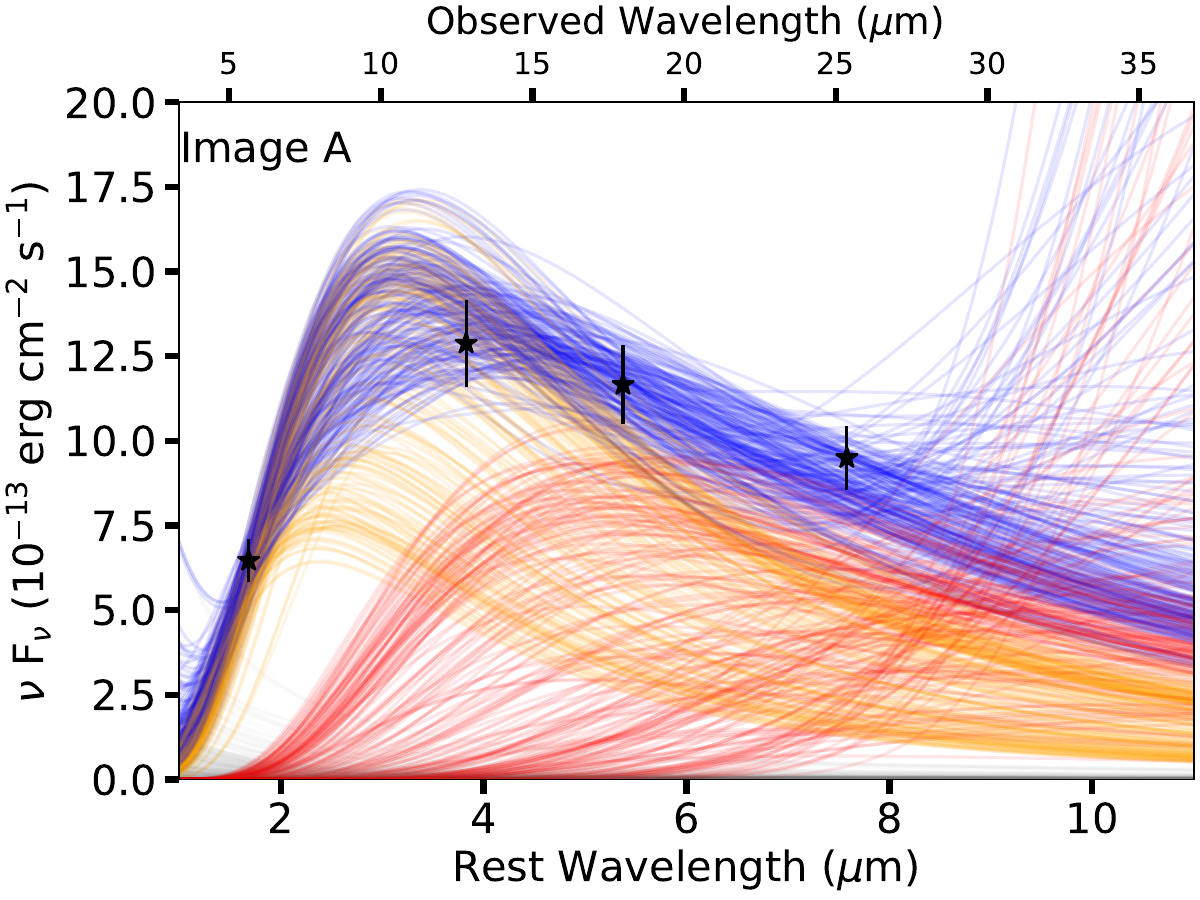}
    \includegraphics[width=0.49\textwidth]{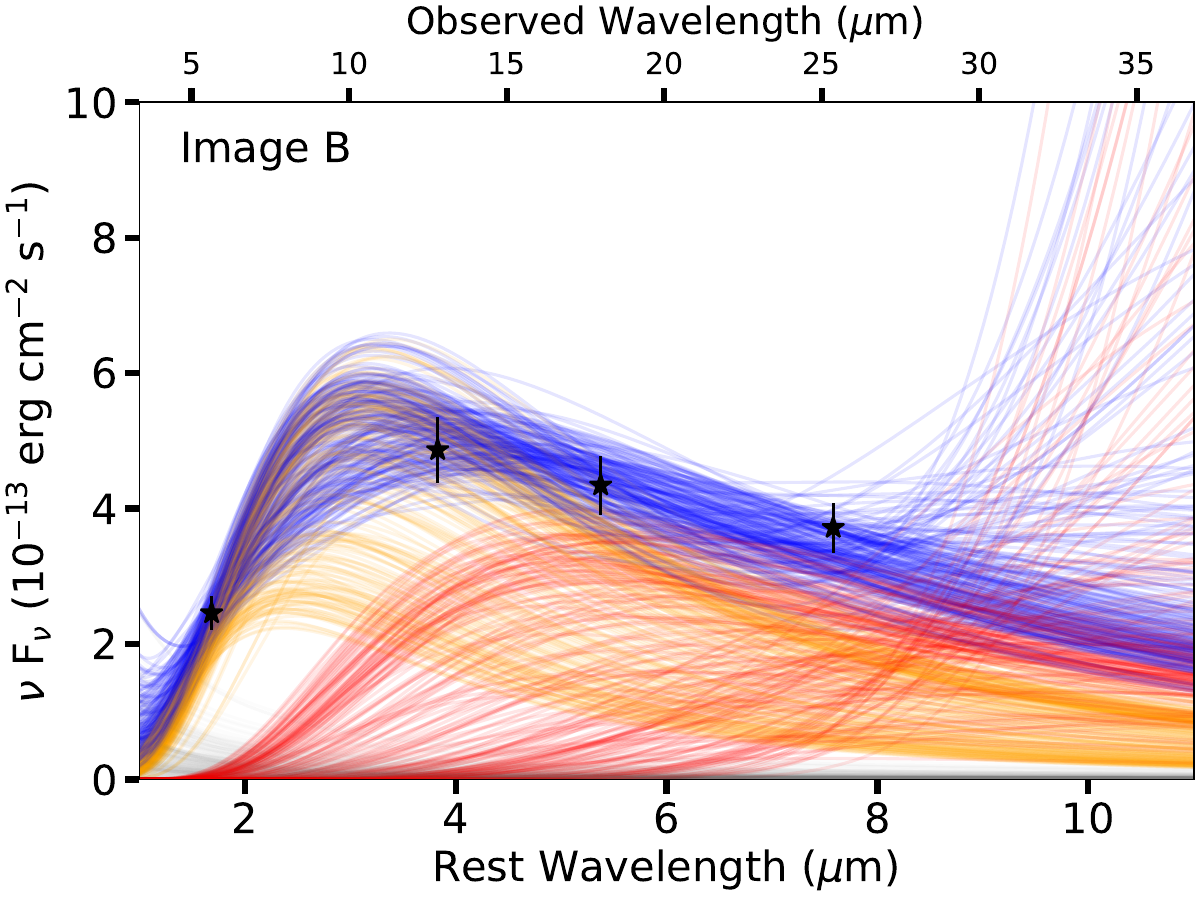}
    \includegraphics[width=0.49\textwidth]{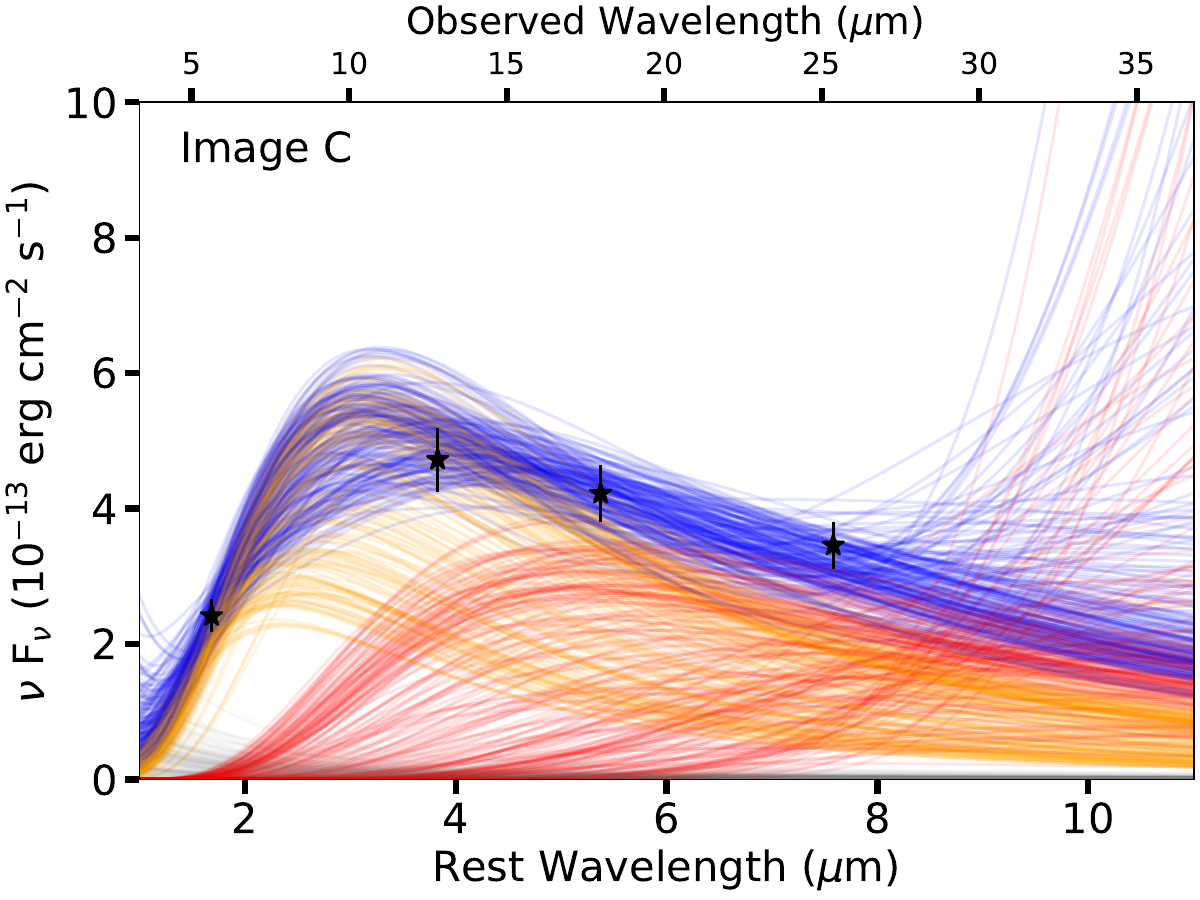}
    \includegraphics[width=0.49\textwidth]{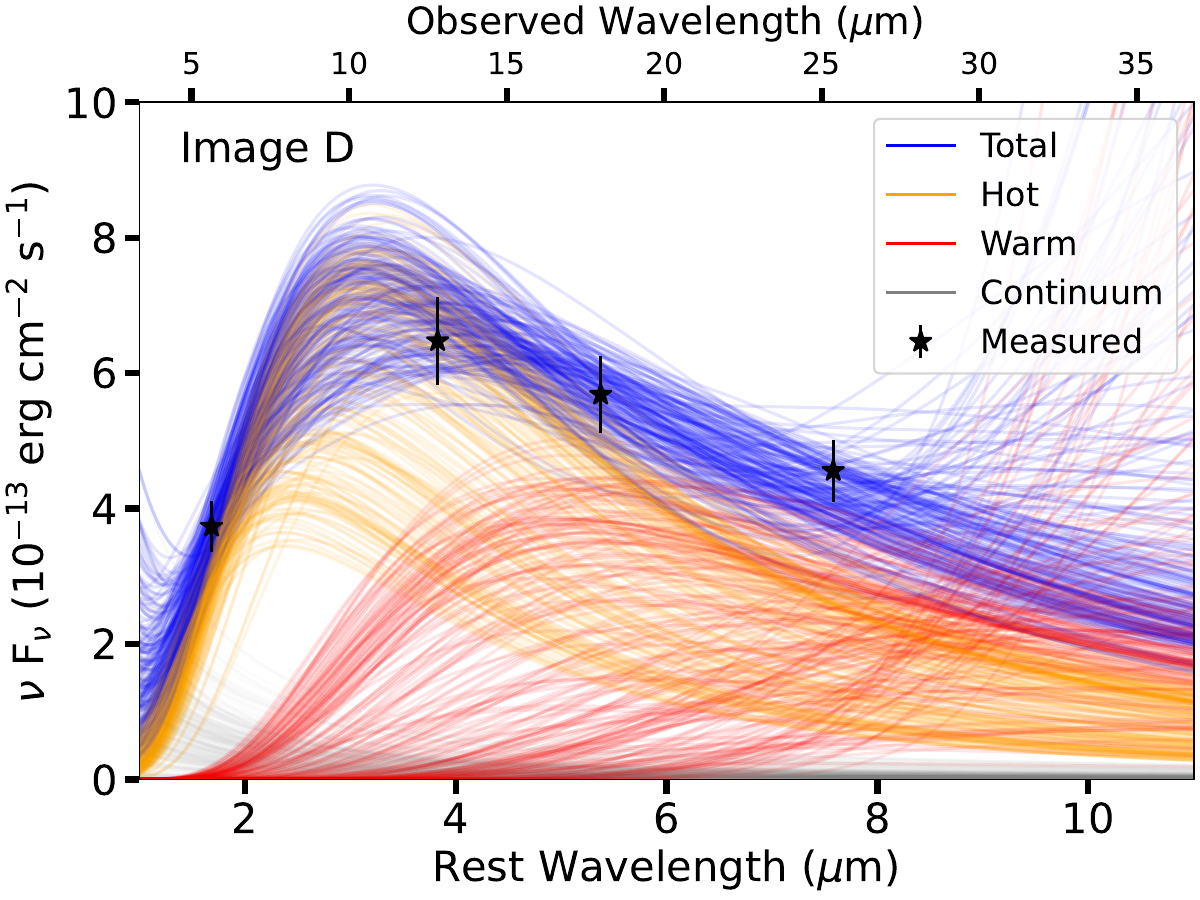}
    \caption{Posterior predictive distributions for the SED fitting for J0608.}
\end{figure*}

\begin{figure*}
    \includegraphics[width=\textwidth]{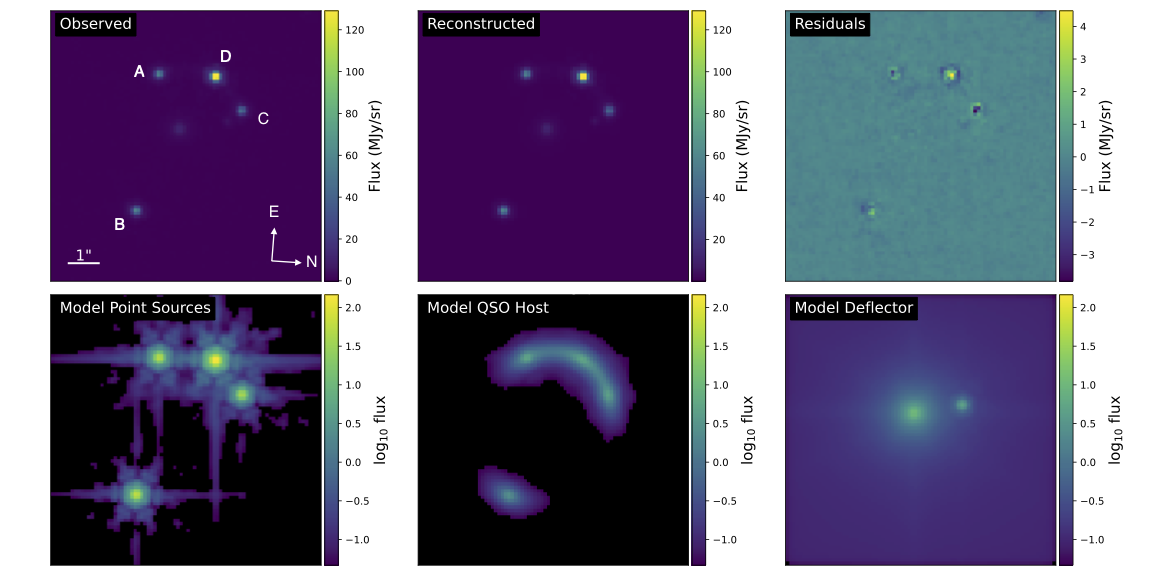}
     \caption{J0659 F560W.}
\end{figure*}
\begin{figure*}
    \includegraphics[width=\textwidth]{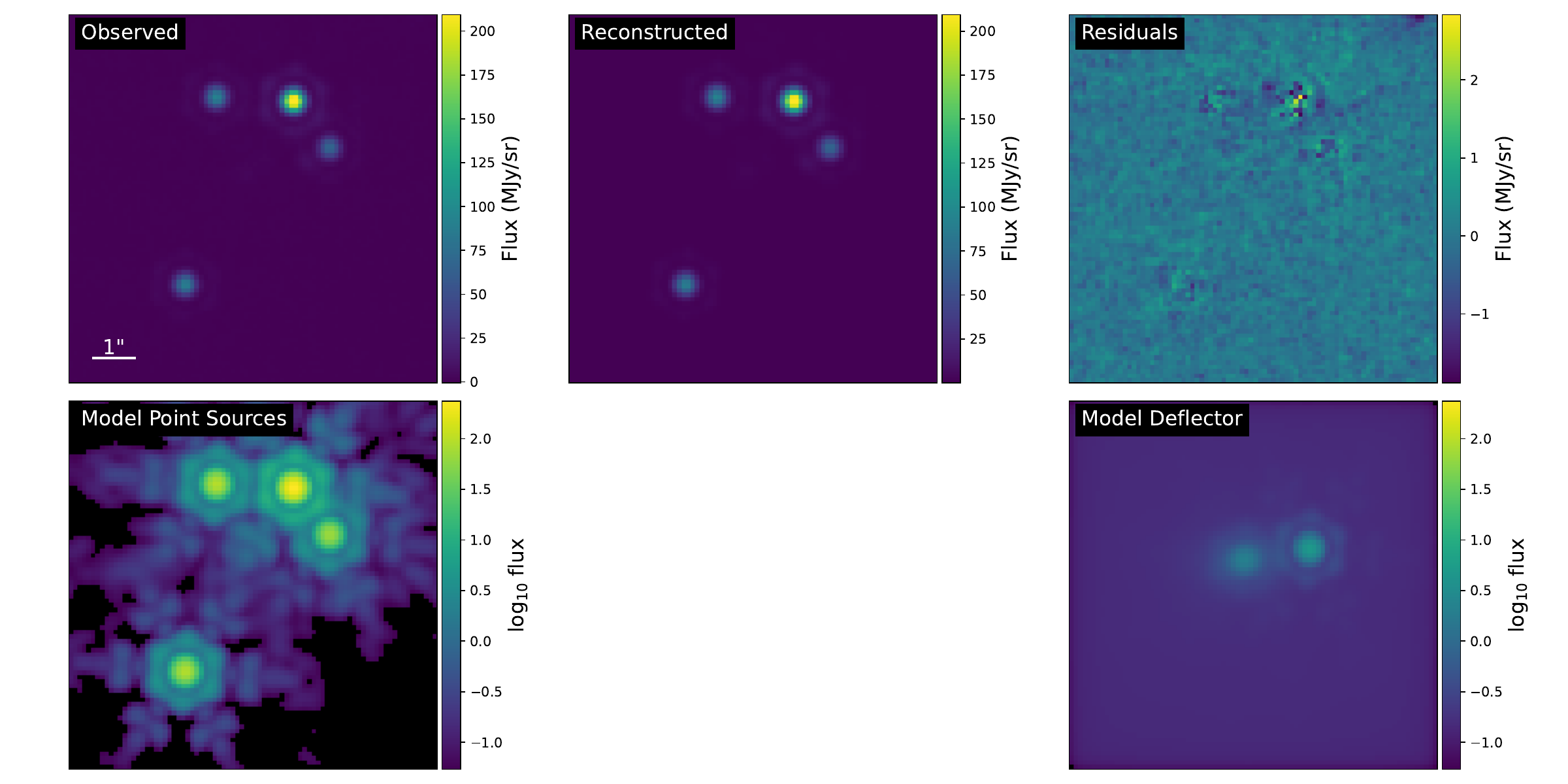}
    \caption{J0659 F1280W.}
\end{figure*}
\begin{figure*}
    \includegraphics[width=\textwidth]{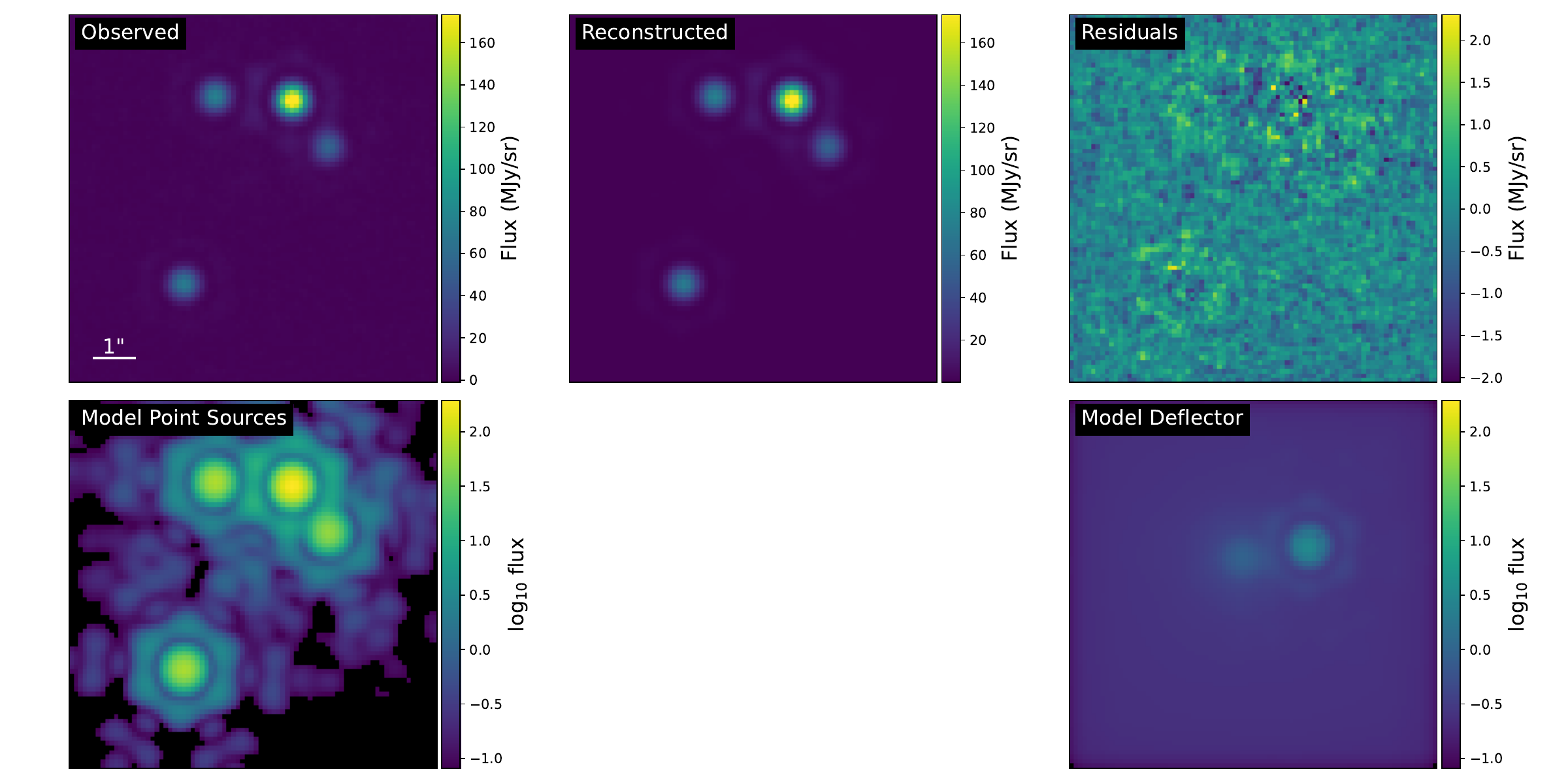}
    \caption{J0659 F1800W.}
\end{figure*}
\begin{figure*}
    \includegraphics[width=\textwidth]{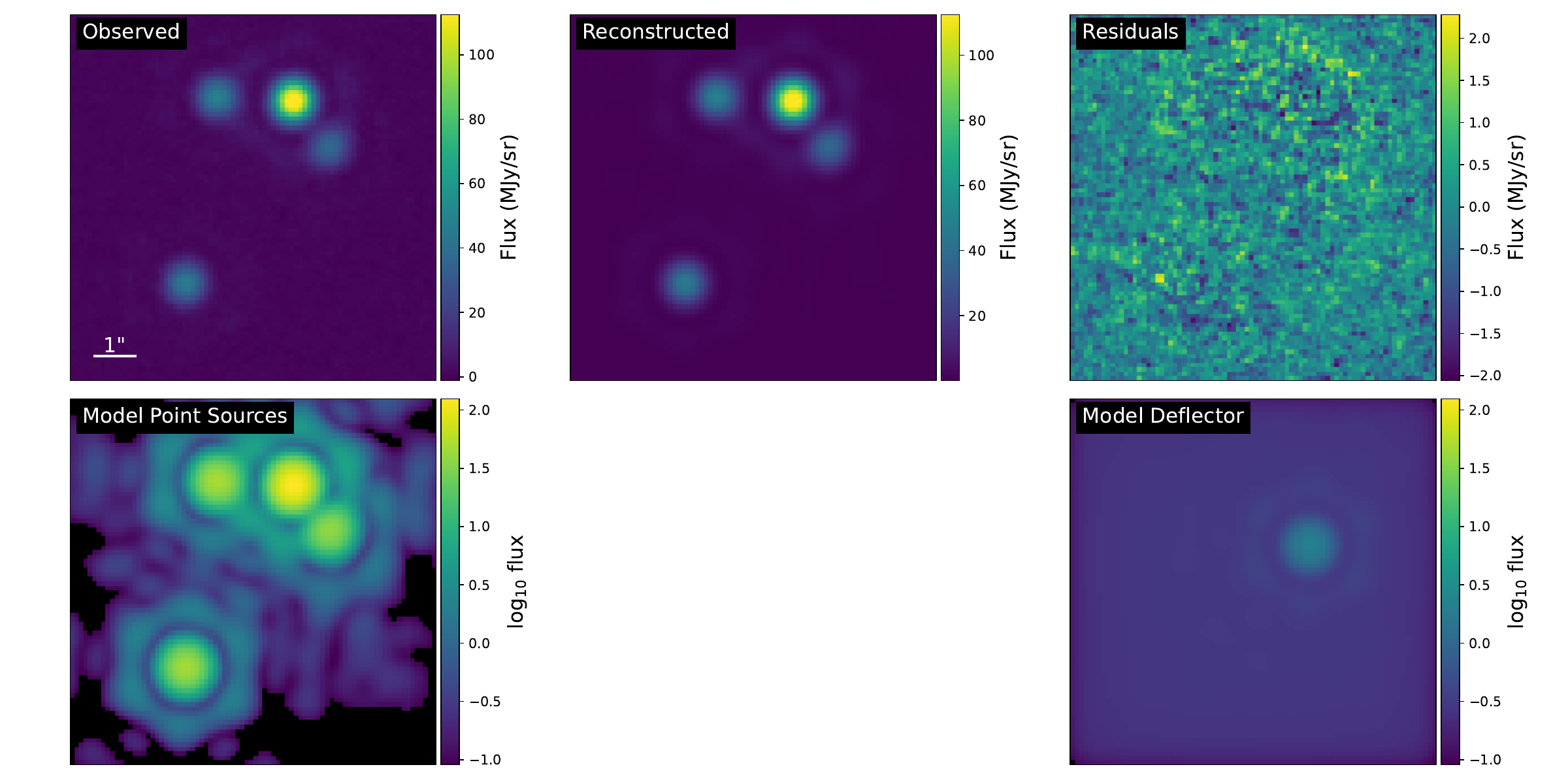}
    \caption{J0659 F2550W.}
\end{figure*}

\begin{figure*}
    \includegraphics[width=0.49\textwidth]{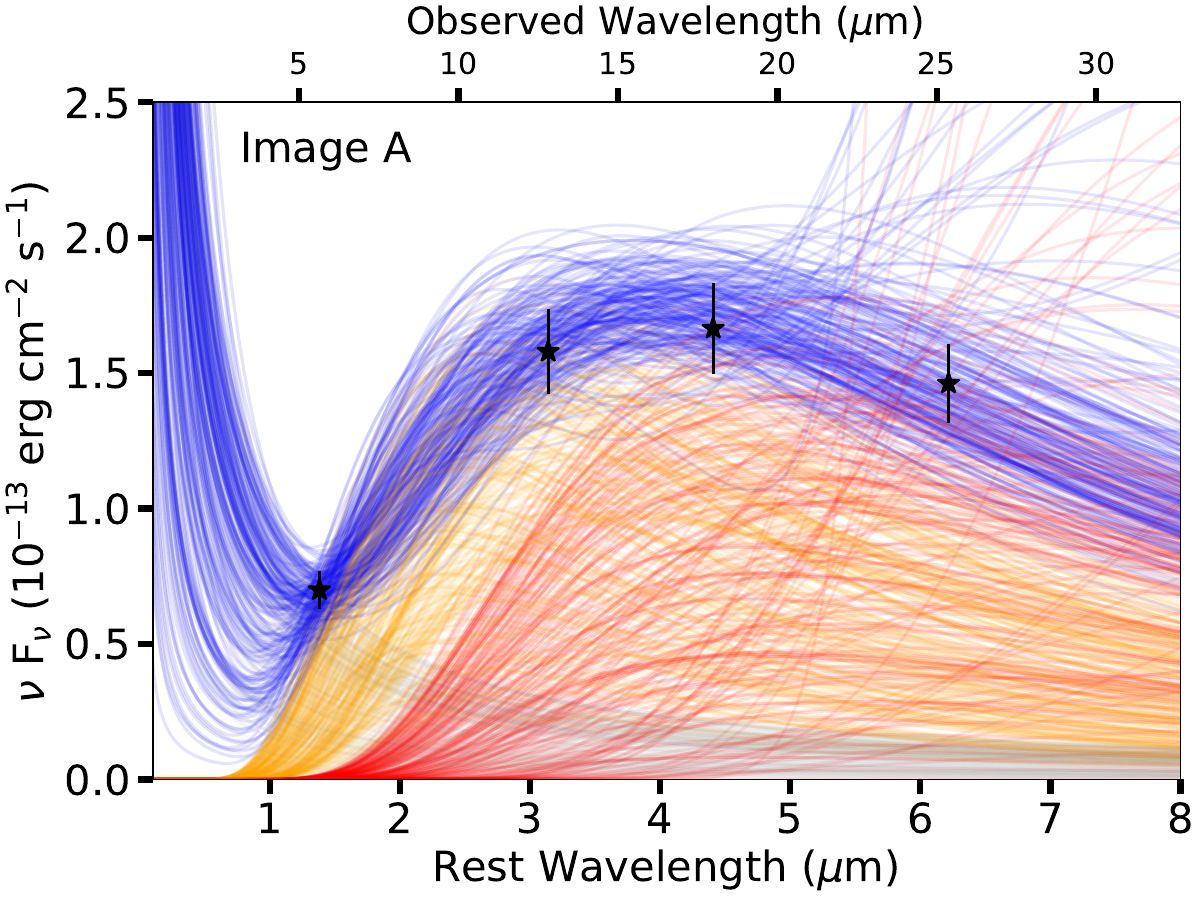}
    \includegraphics[width=0.49\textwidth]{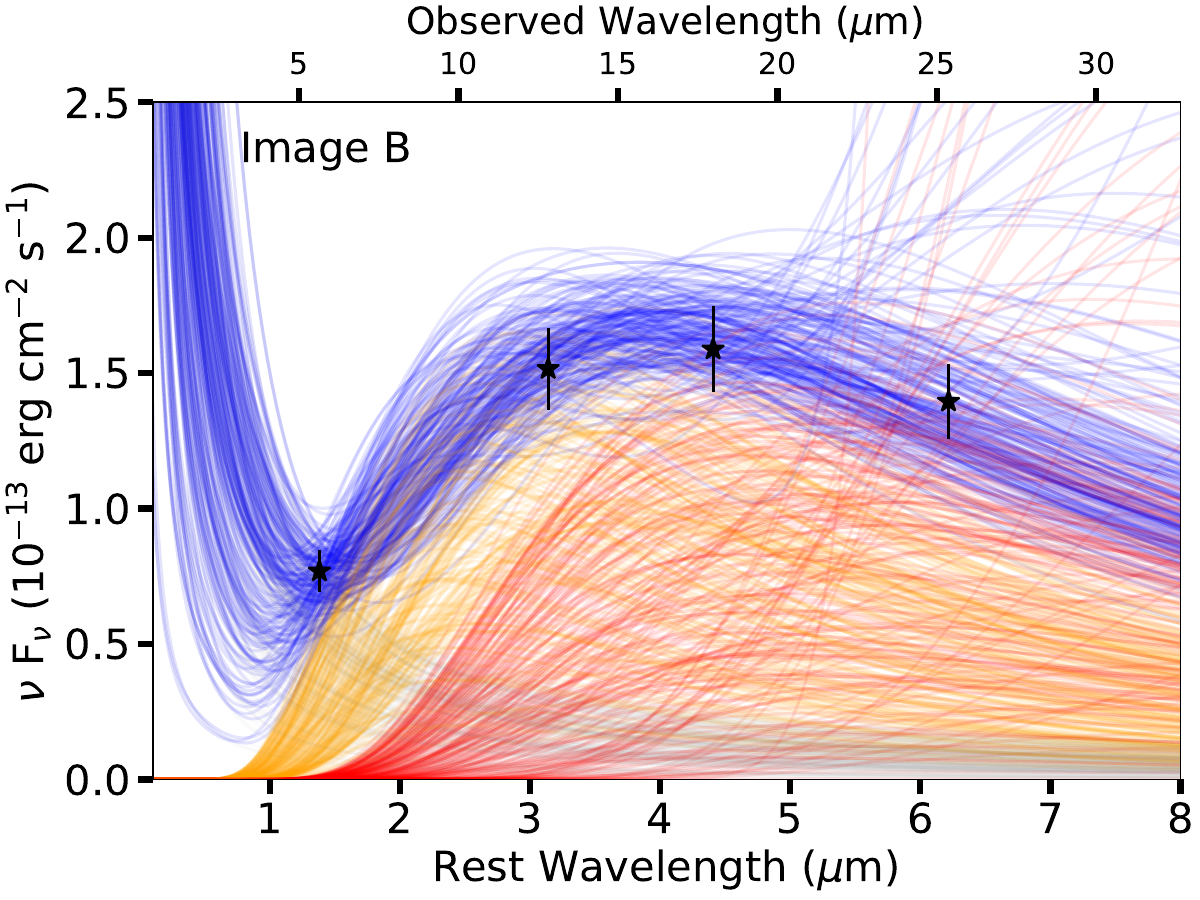}
    \includegraphics[width=0.49\textwidth]{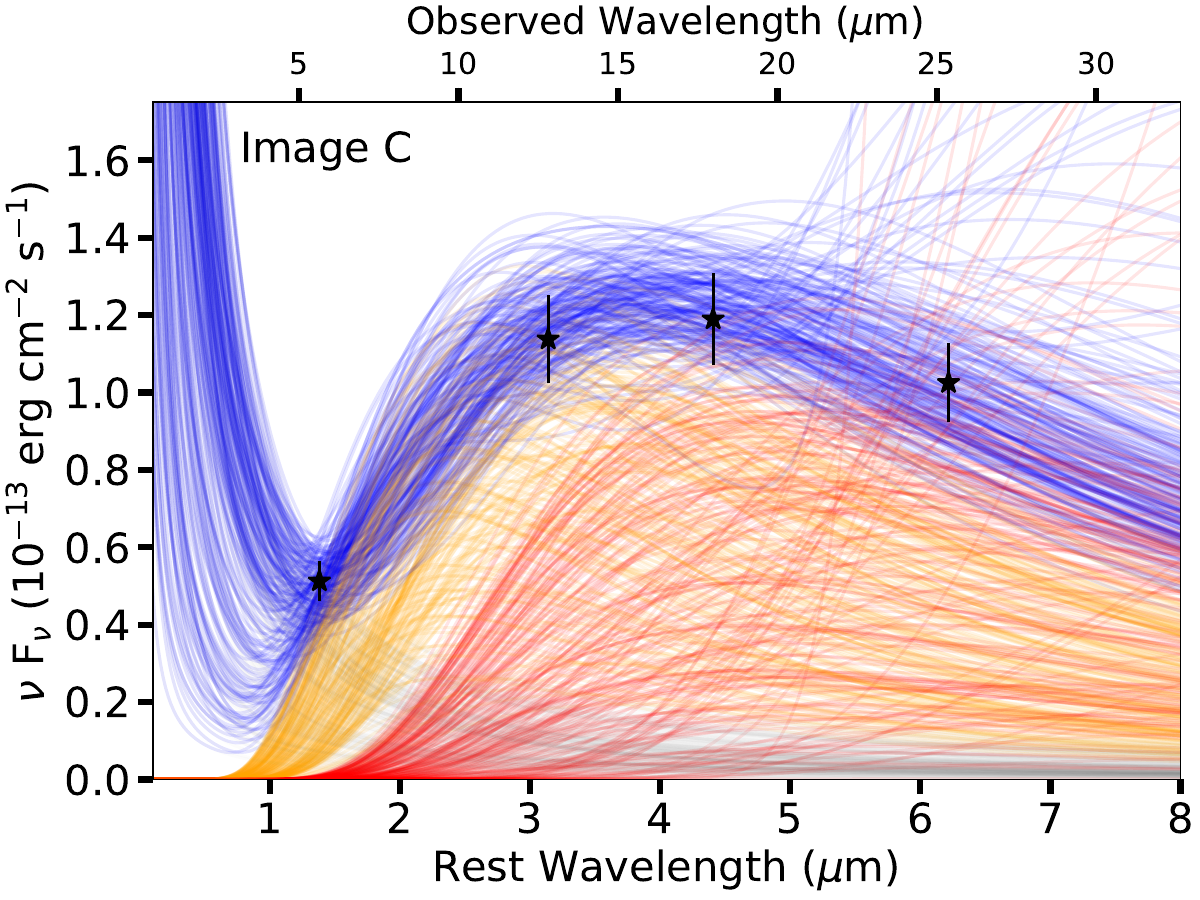}
    \includegraphics[width=0.49\textwidth]{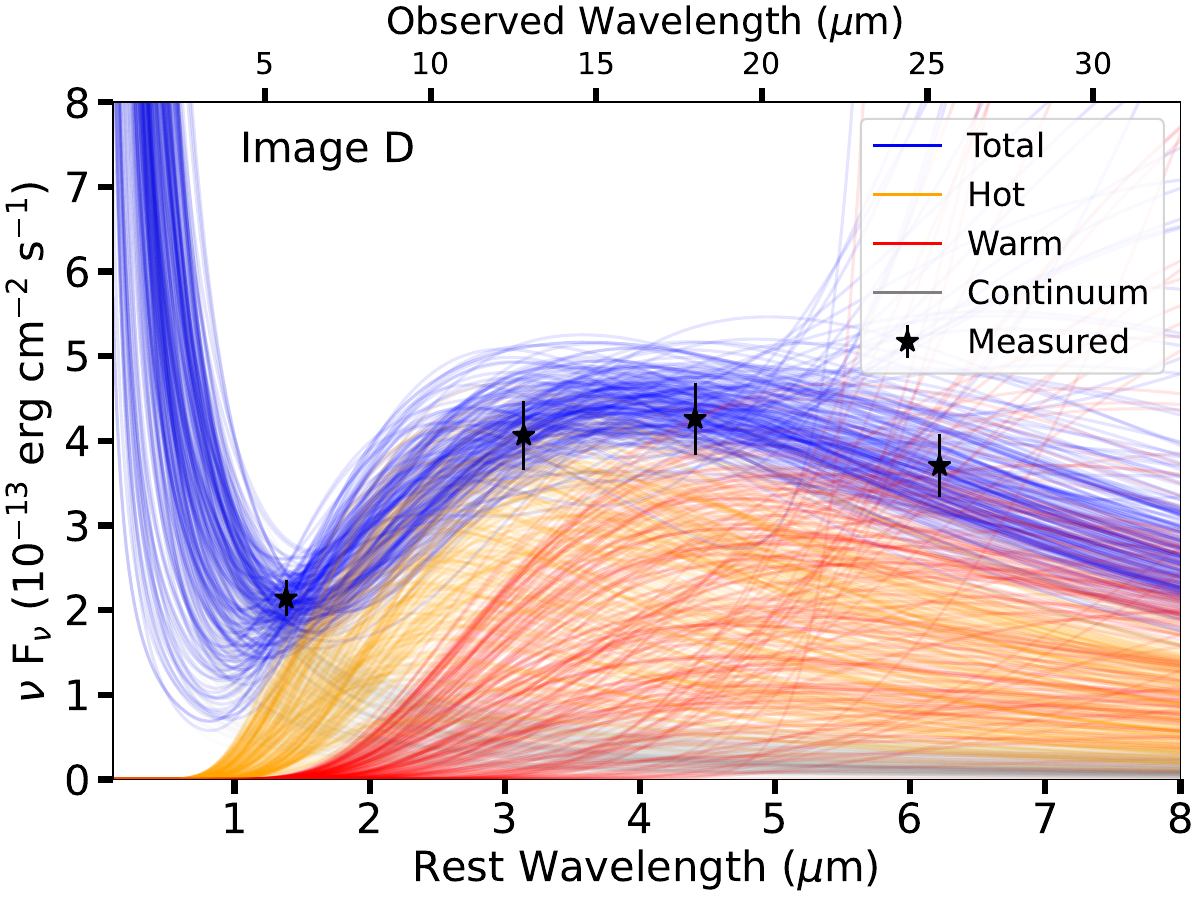}
    \caption{Posterior predictive distributions for the SED fitting for J0659.}
\end{figure*}

\begin{figure*}
    \includegraphics[width=\textwidth]{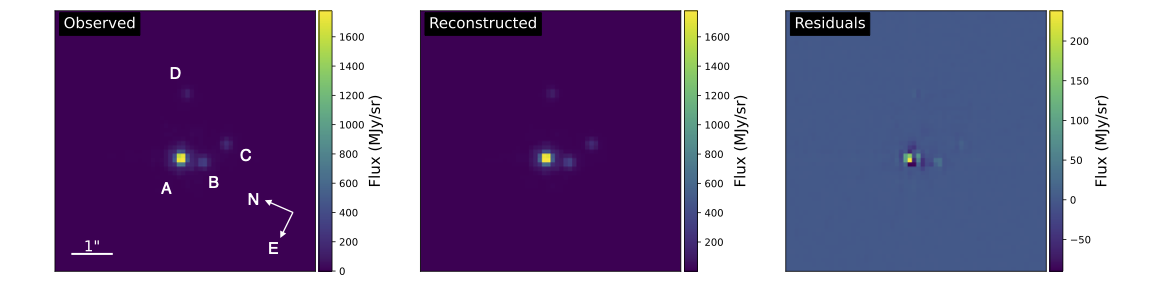}
     \caption{J1042 F560W.}
\end{figure*}
\begin{figure*}
    \includegraphics[width=\textwidth]{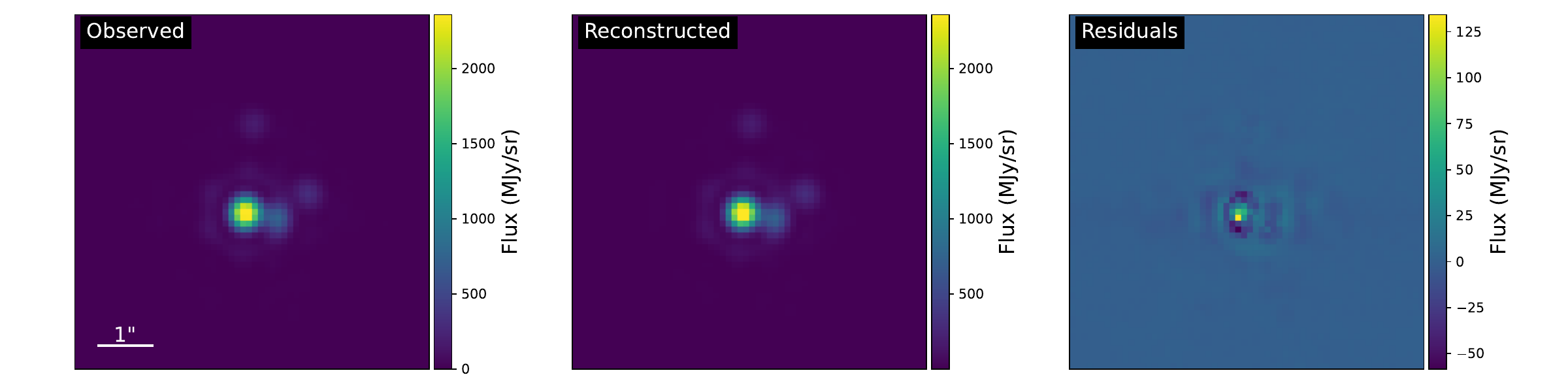}
    \caption{J1042 F1280W.}
\end{figure*}
\begin{figure*}
    \includegraphics[width=\textwidth]{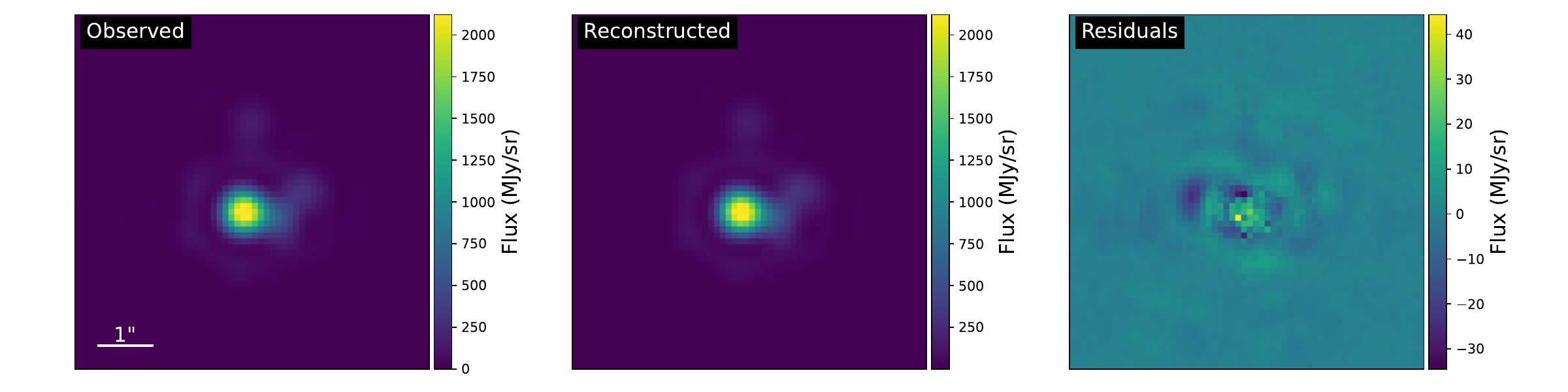}
    \caption{J1042 F1800W.}
\end{figure*}
\begin{figure*}
    \includegraphics[width=\textwidth]{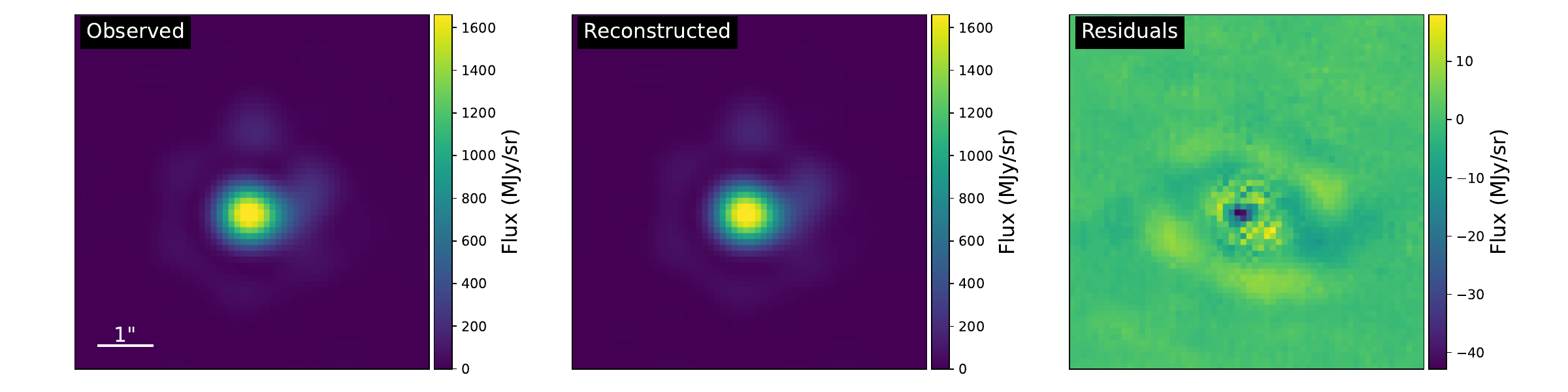}
    \caption{J1042 F2550W.}
\end{figure*}

\begin{figure*}
    \includegraphics[width=0.49\textwidth]{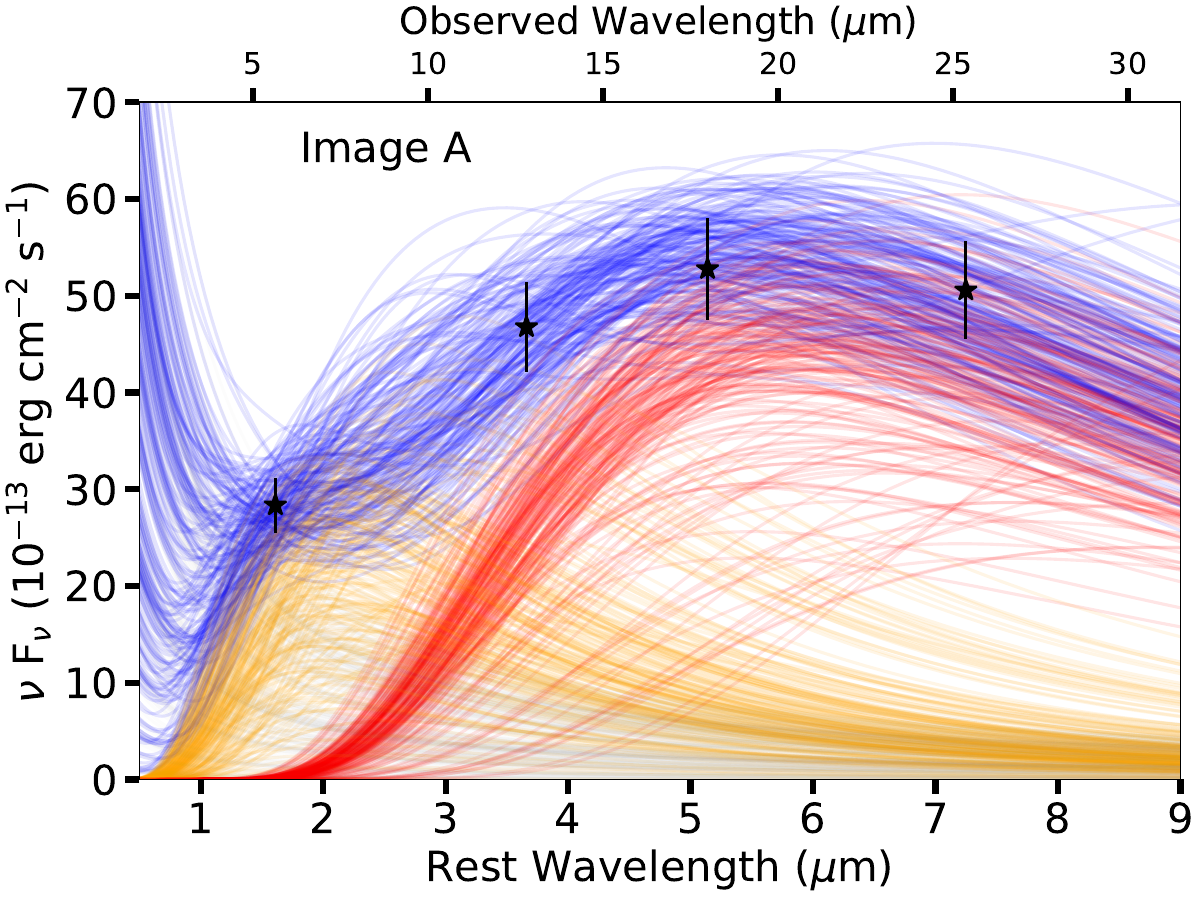}
    \includegraphics[width=0.49\textwidth]{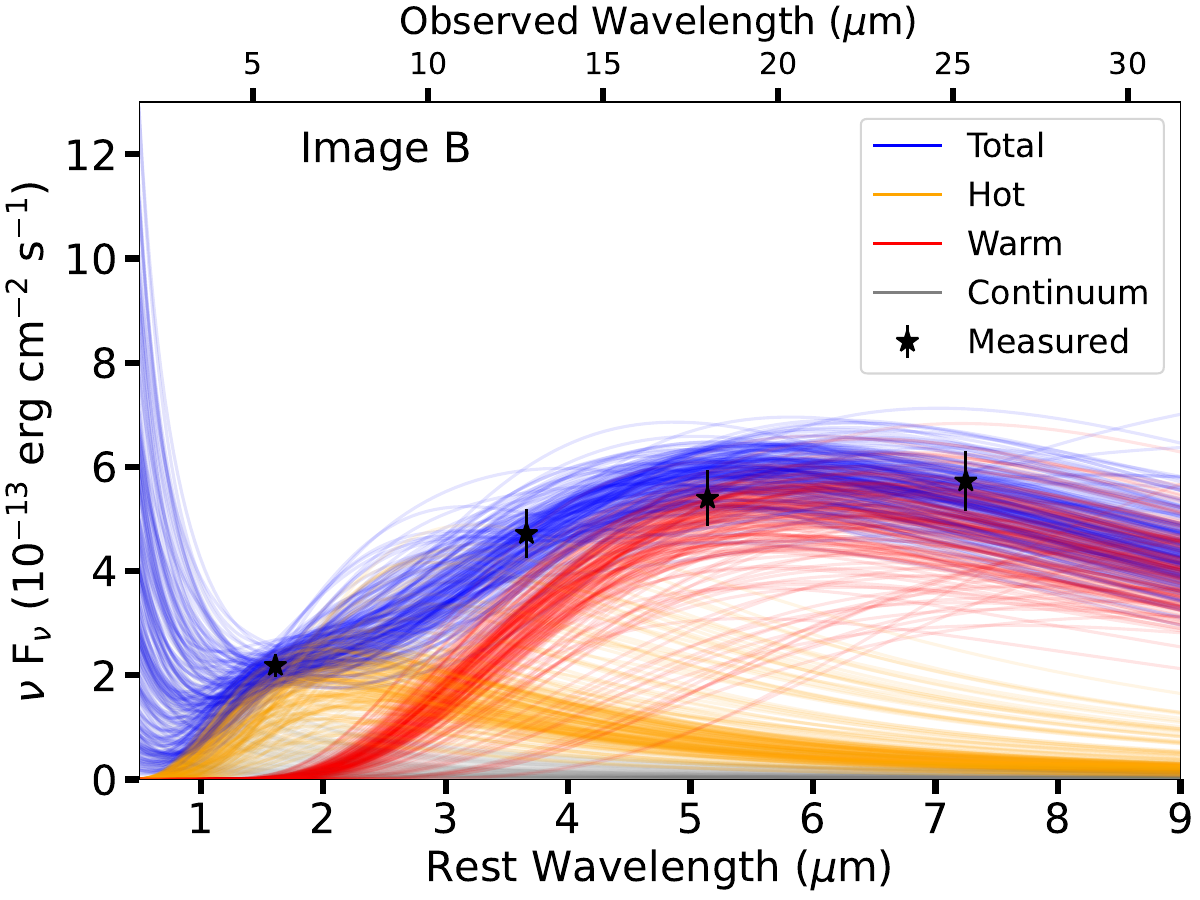}
    \includegraphics[width=0.49\textwidth]{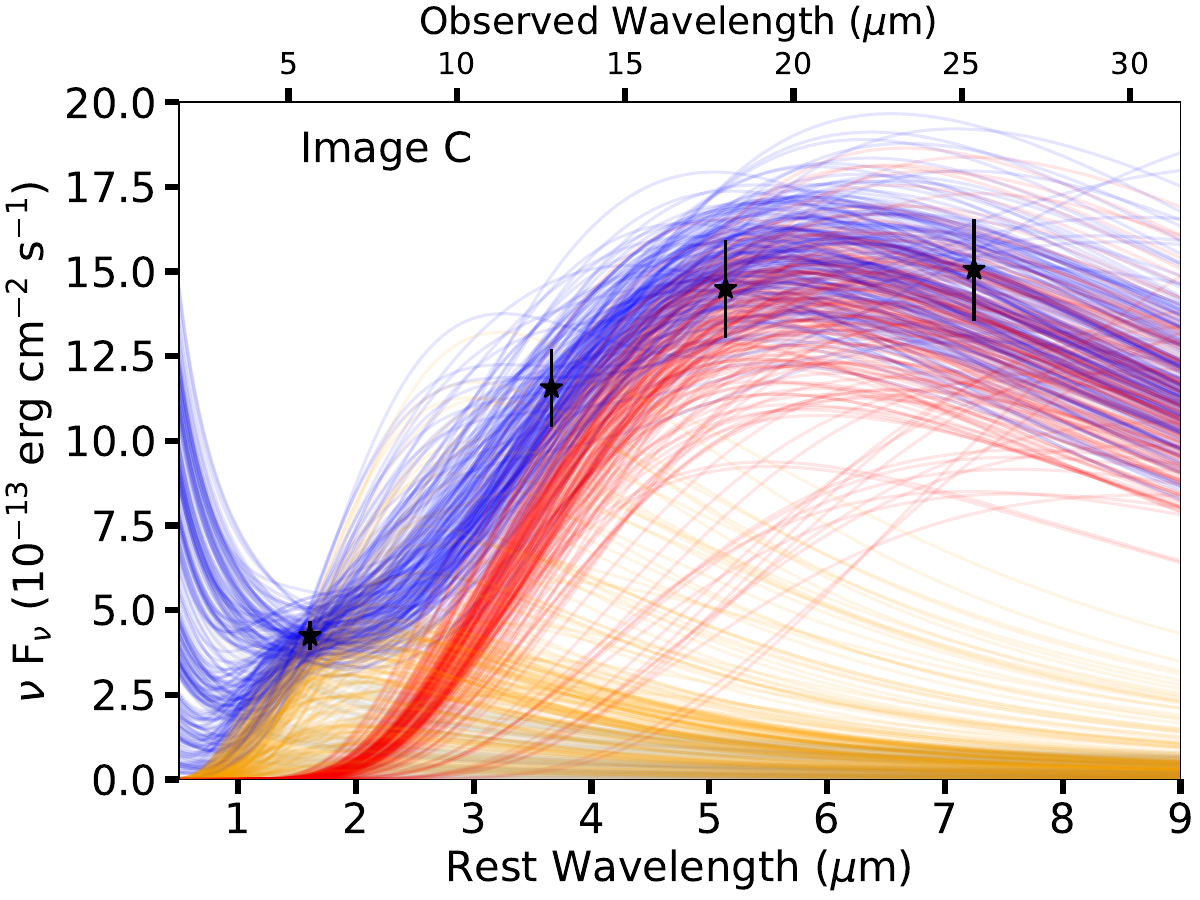}
    \includegraphics[width=0.49\textwidth]{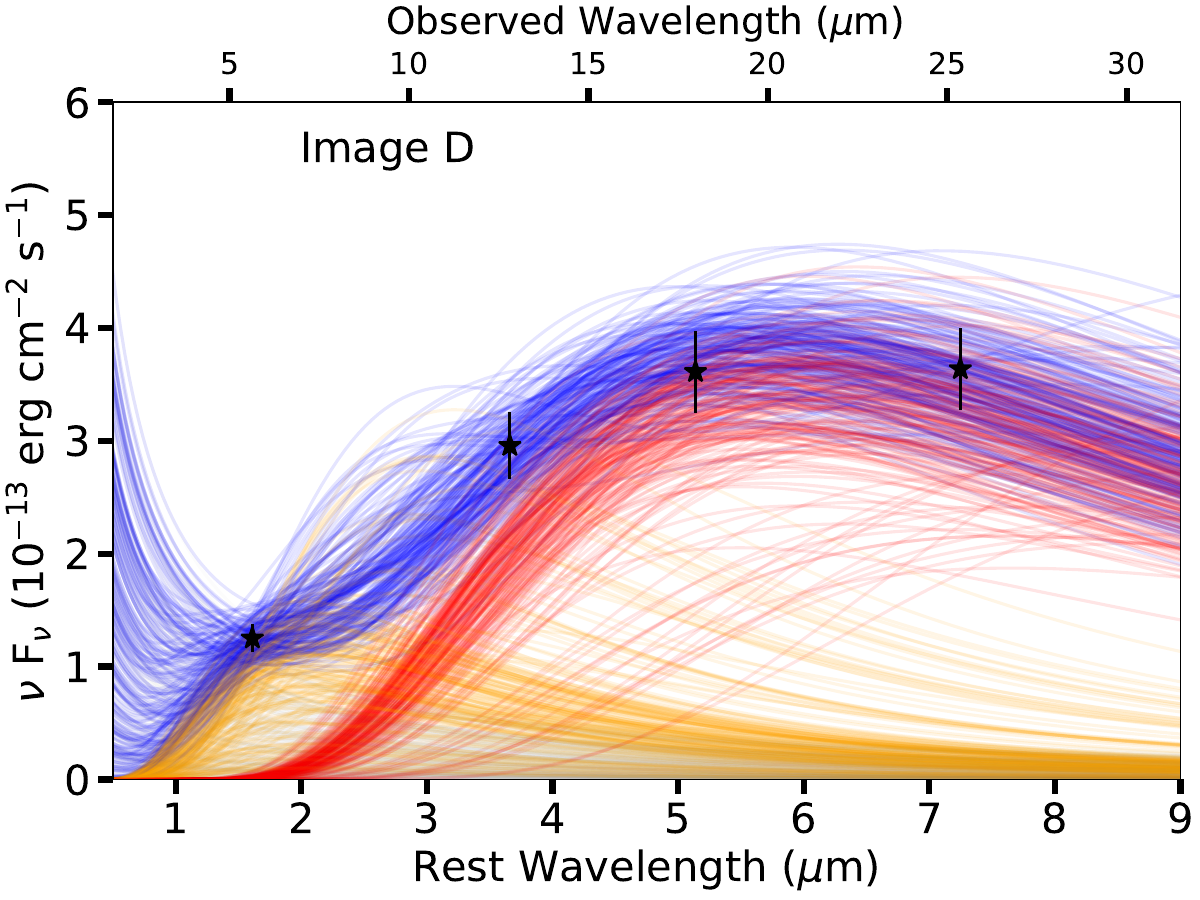}
    \caption{Posterior predictive distributions for the SED fitting for J1042.}
\end{figure*}

\begin{figure*}
    \includegraphics[width=\textwidth]{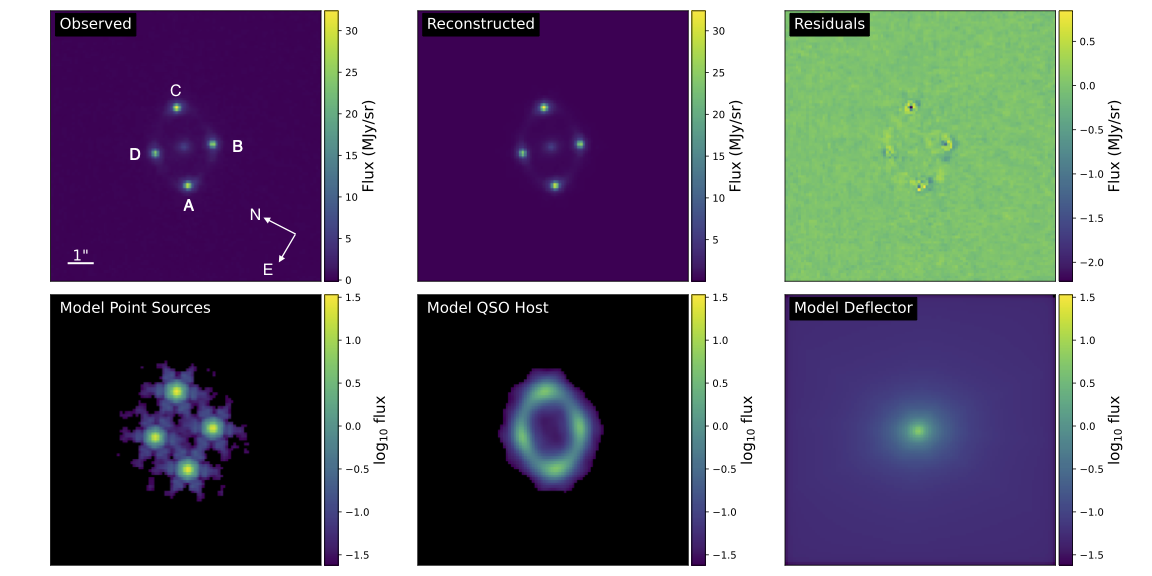}
    \caption{J1537 F560W.}
\end{figure*}
\begin{figure*}
    \includegraphics[width=\textwidth]{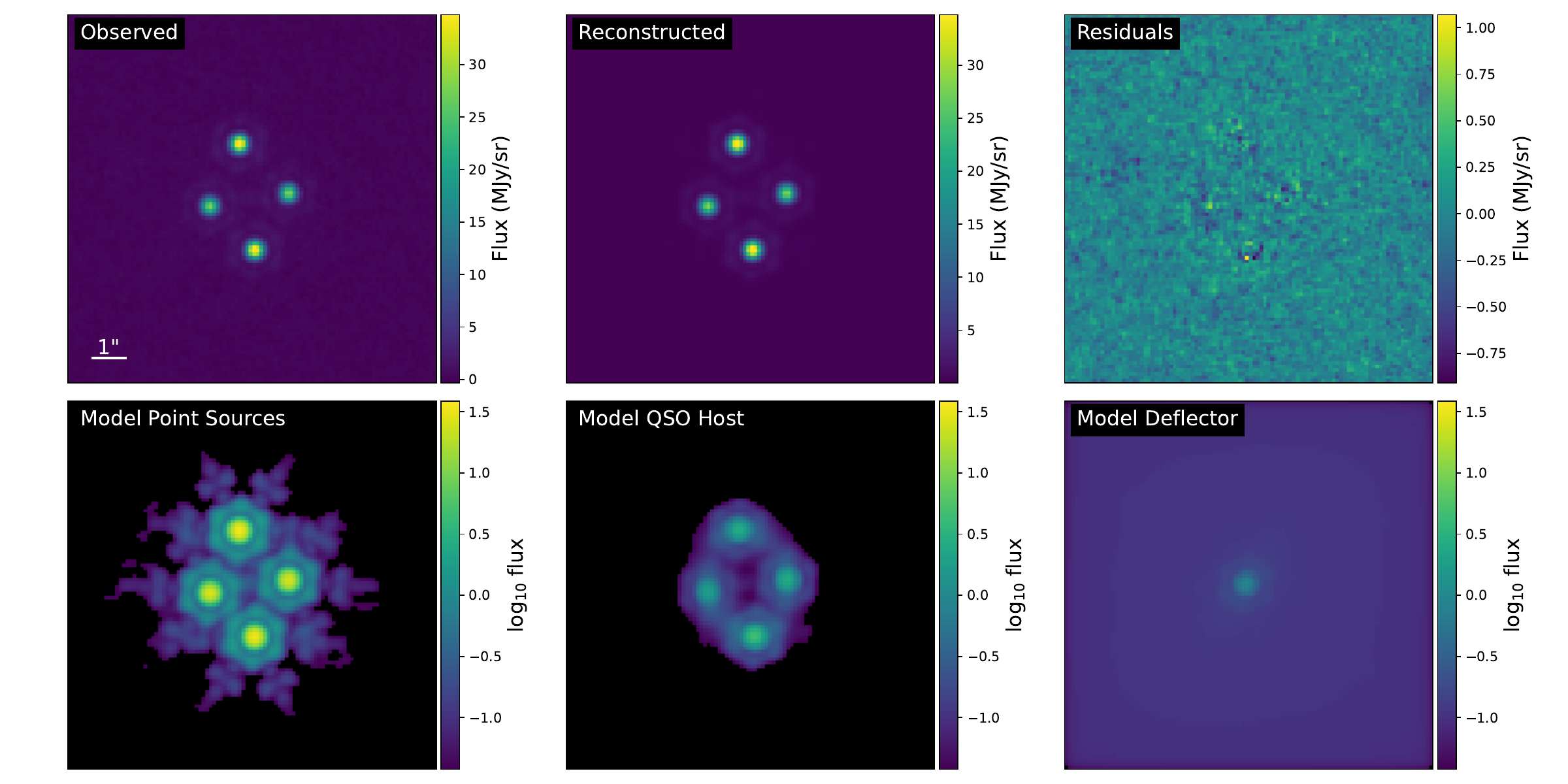}
    \caption{J1537 F1280W.}
\end{figure*}
\begin{figure*}
    \includegraphics[width=\textwidth]{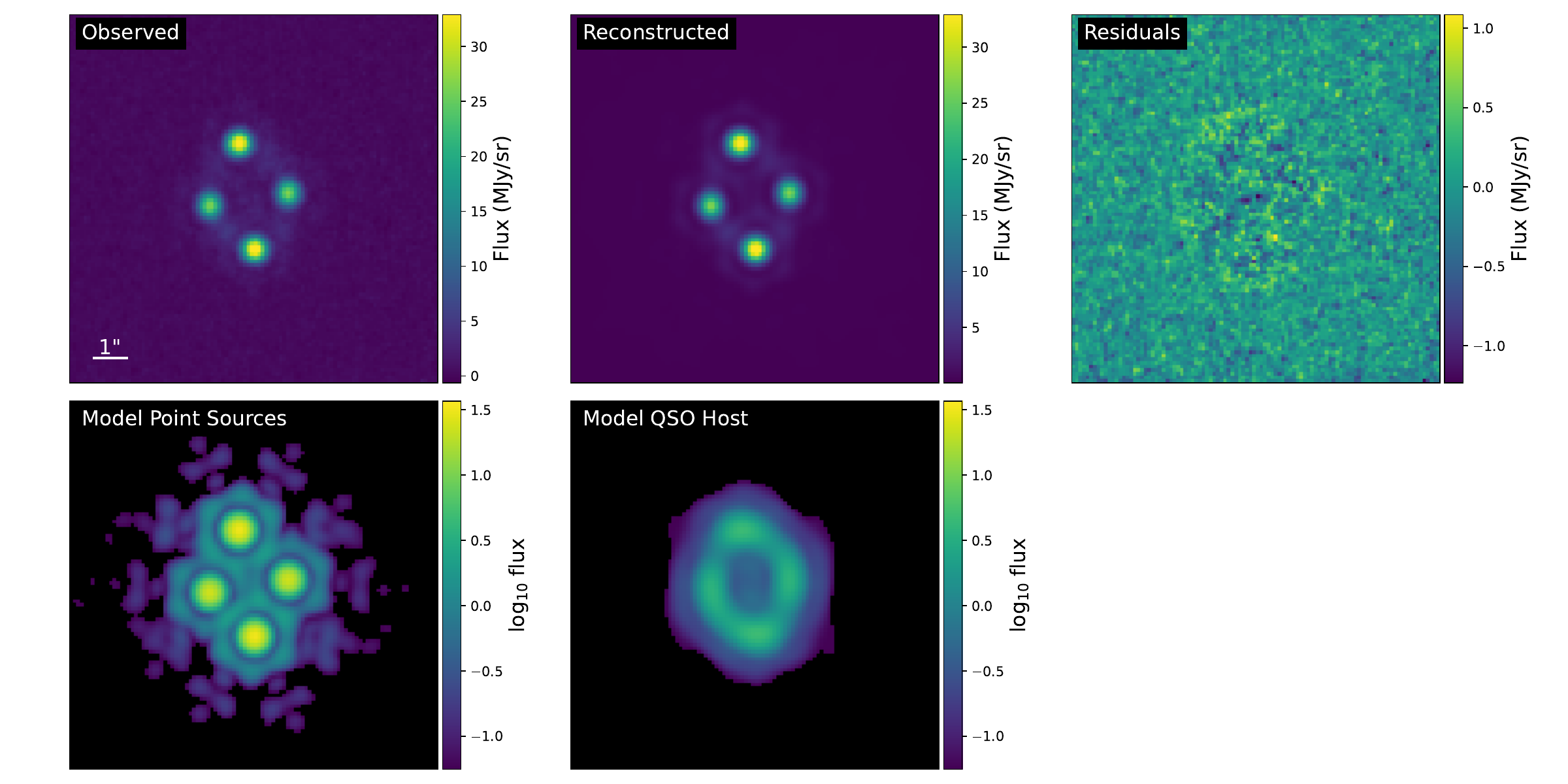}
    \caption{J1537 F1800W.}
\end{figure*}
\begin{figure*}
    \includegraphics[width=\textwidth]{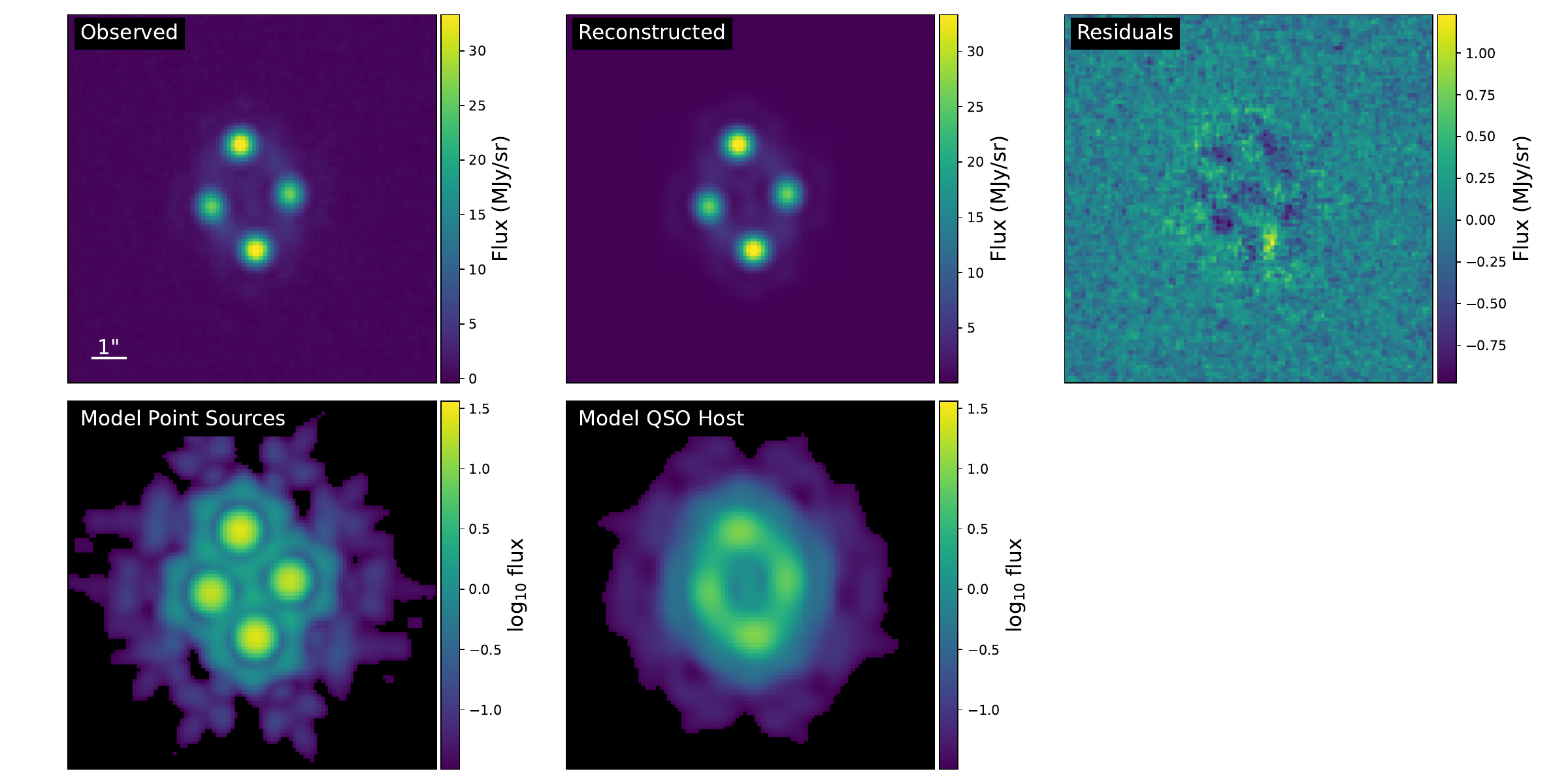}
    \caption{J1537 F2100W.}
\end{figure*}

\begin{figure*}
    \includegraphics[width=0.49\textwidth]{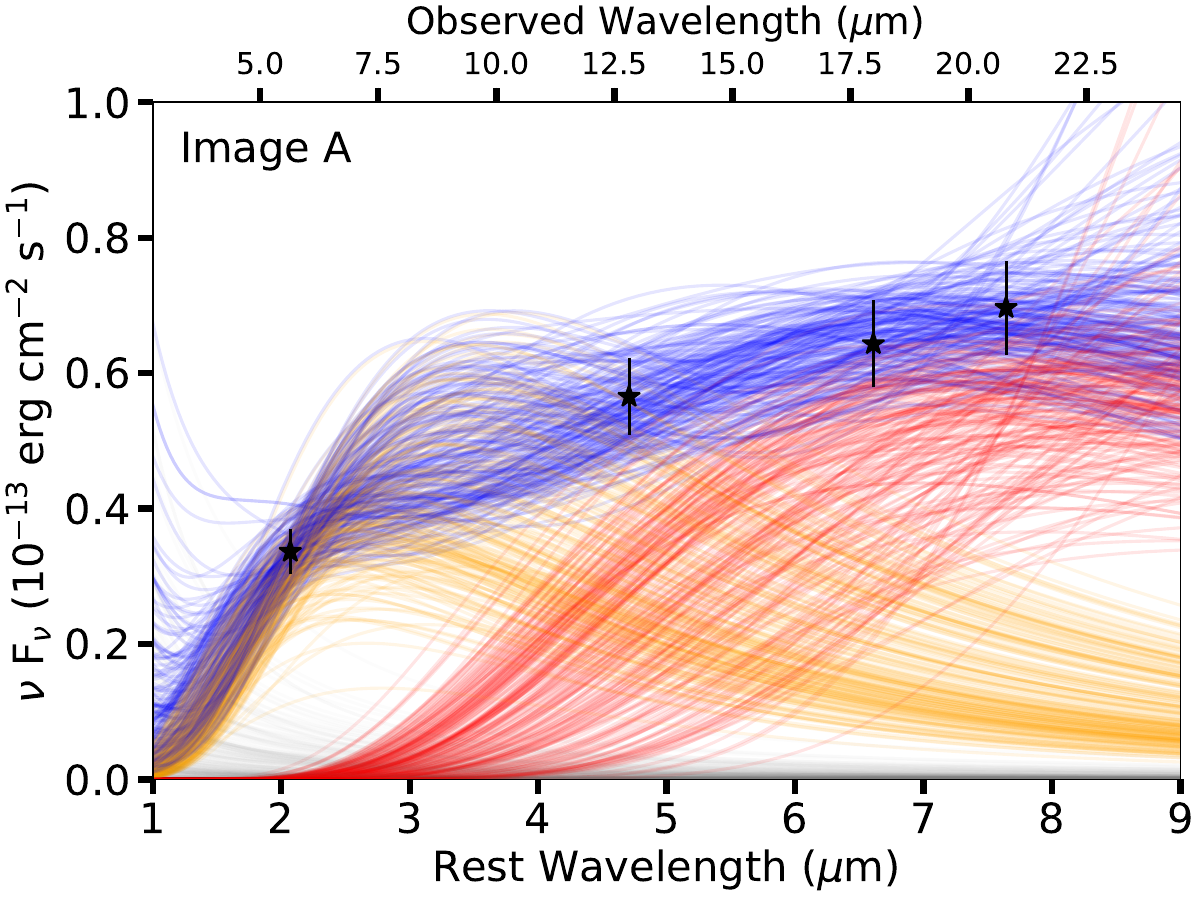}
    \includegraphics[width=0.49\textwidth]{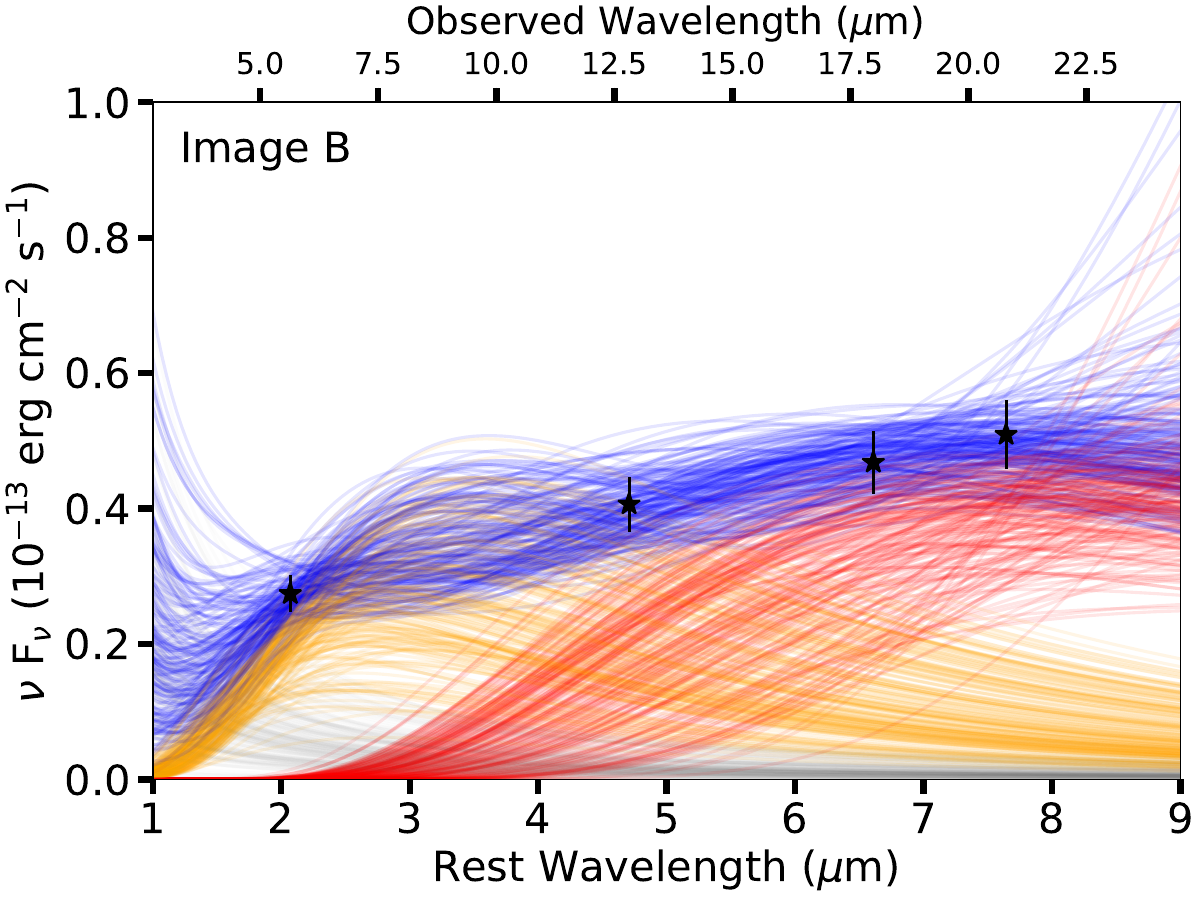}
    \includegraphics[width=0.49\textwidth]{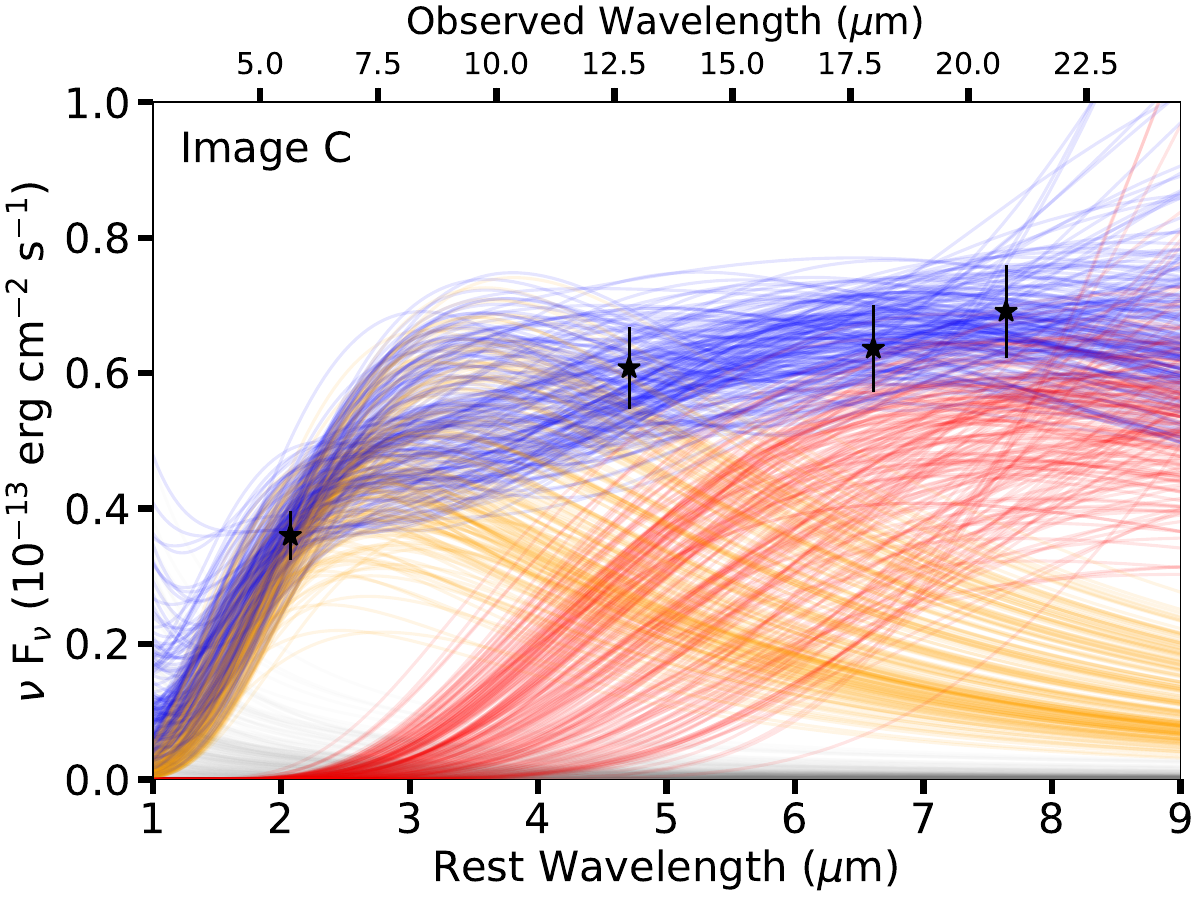}
    \includegraphics[width=0.49\textwidth]{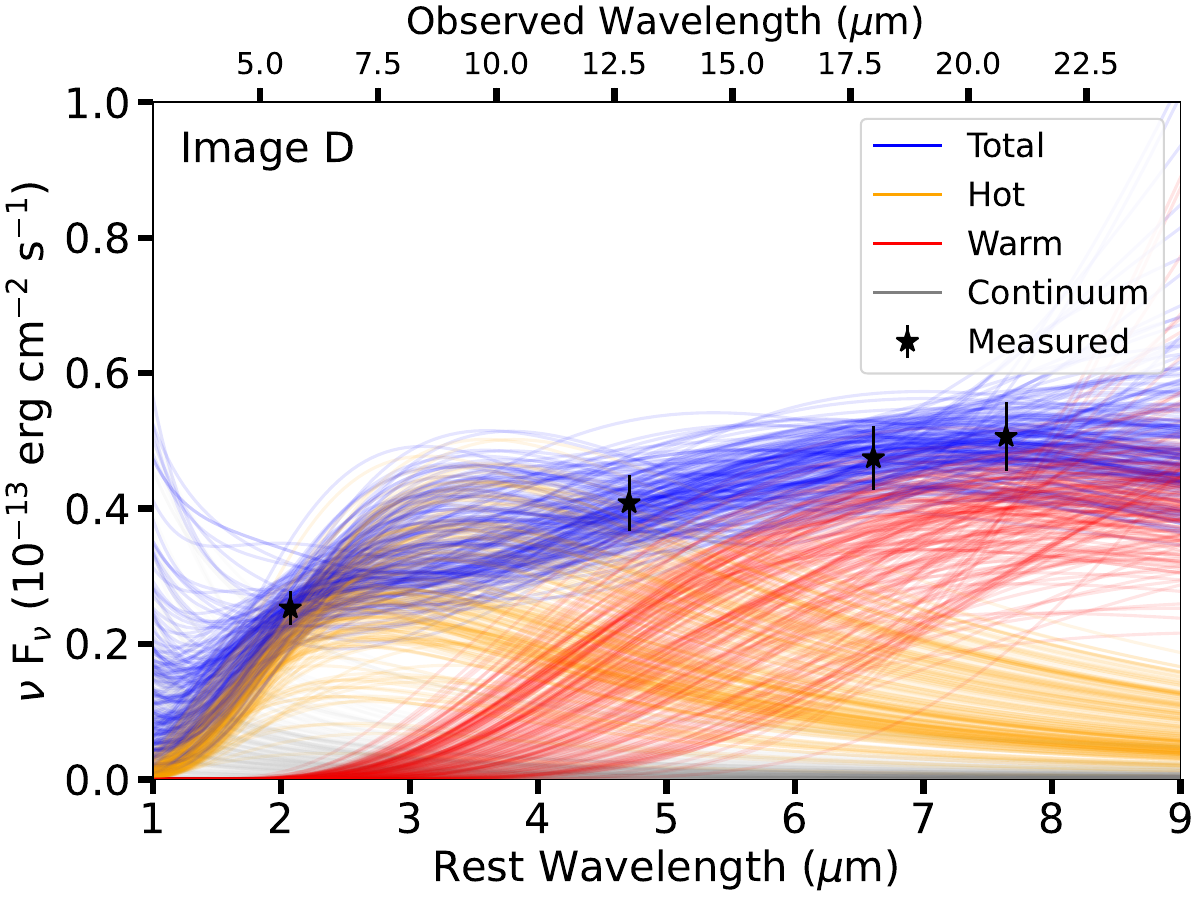}
    \caption{Posterior predictive distributions for the SED fitting for J1537.}
\end{figure*}

\begin{figure*}
    \includegraphics[width=\textwidth]{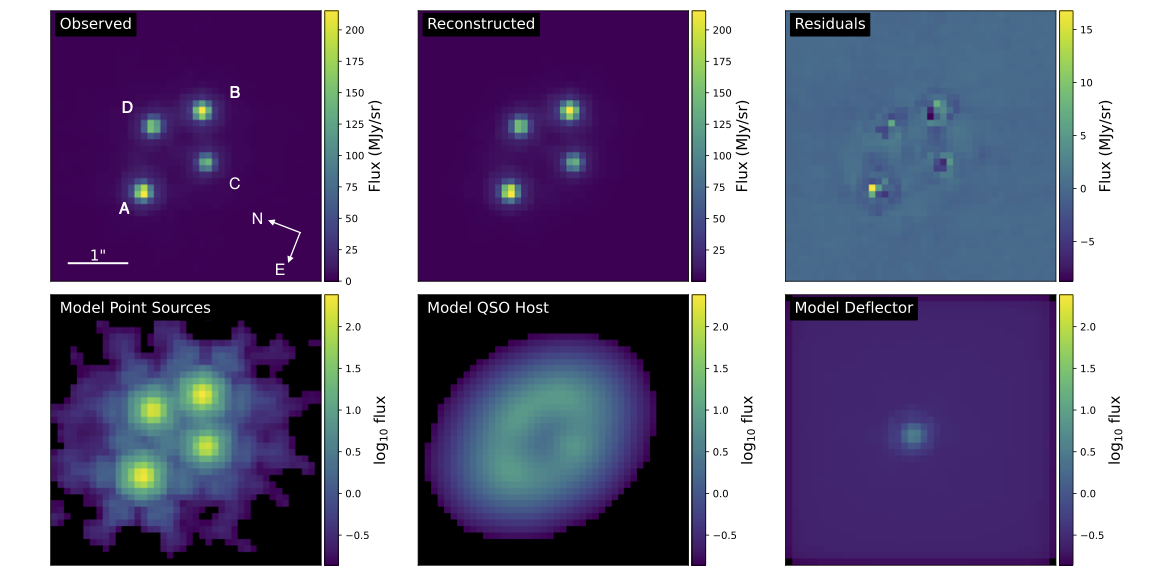}
    \caption{J1606 F560W.}
\end{figure*}
\begin{figure*}
    \includegraphics[width=\textwidth]{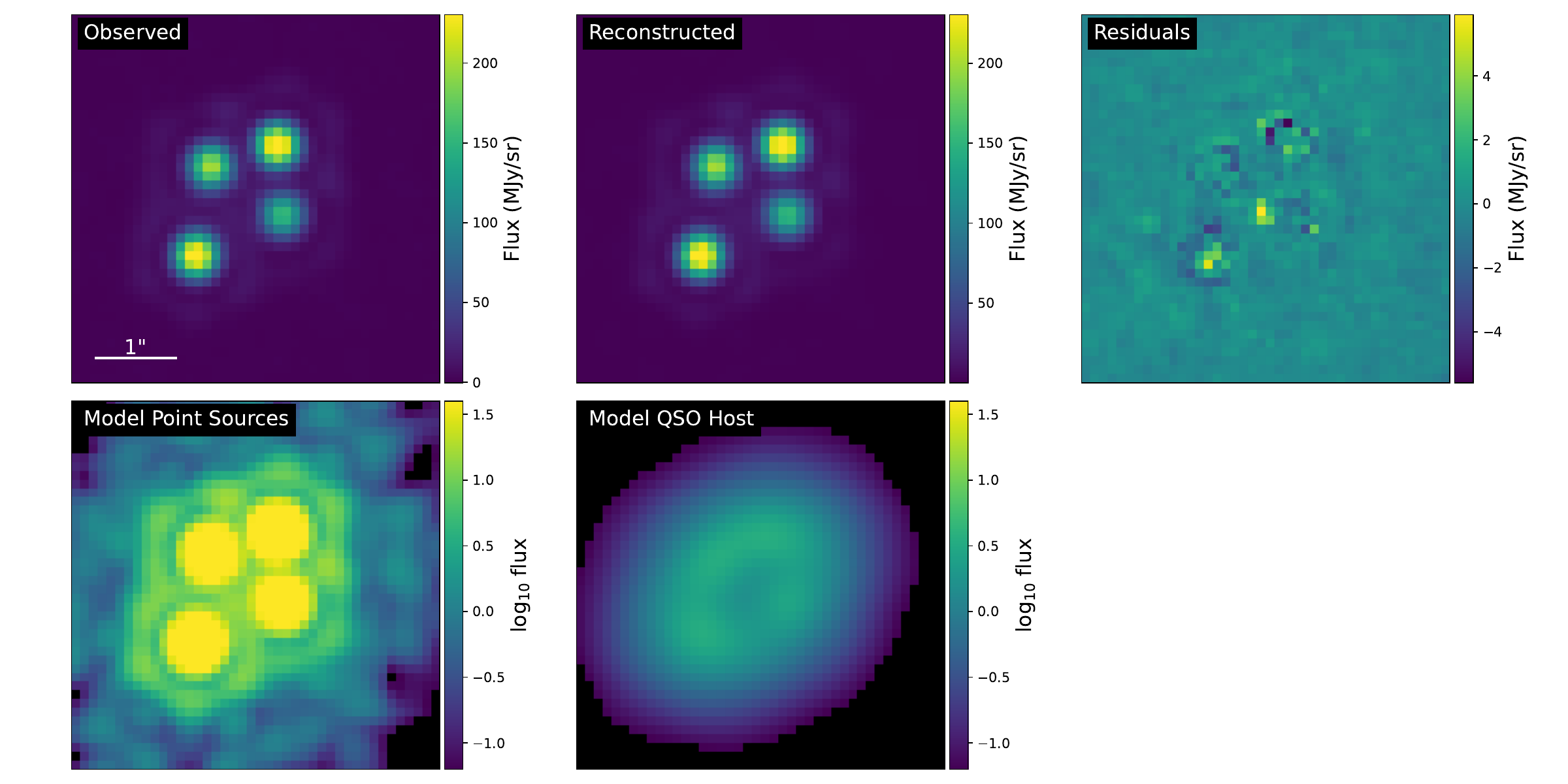}
    \caption{J1606 F1280W.}
\end{figure*}
\begin{figure*}
    \includegraphics[width=\textwidth]{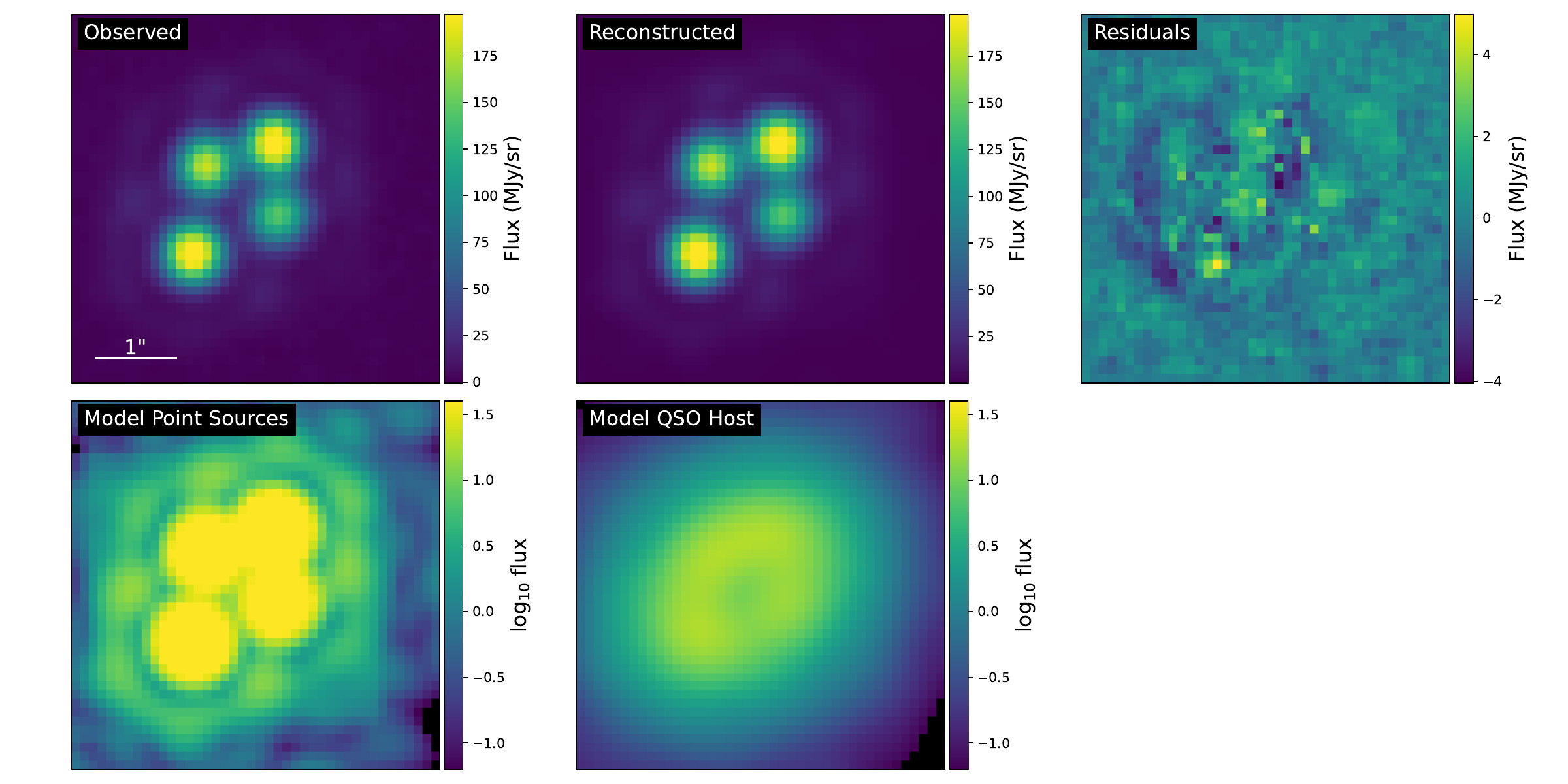}
    \caption{J1606 F1800W.}
\end{figure*}
\begin{figure*}
    \includegraphics[width=\textwidth]{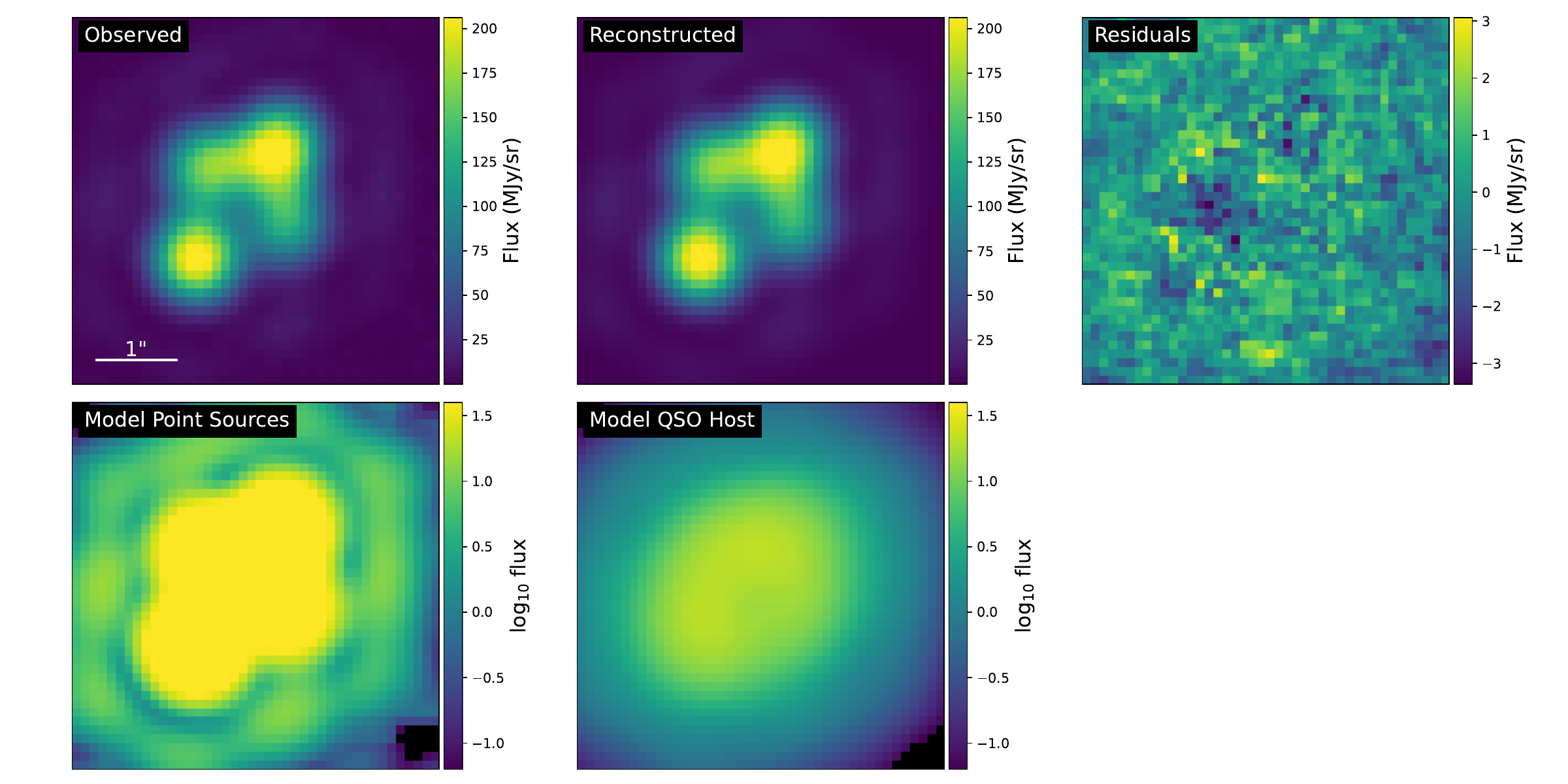}
    \caption{J1606 F2550W.}
\end{figure*}

\begin{figure*}
    \includegraphics[width=0.49\textwidth]{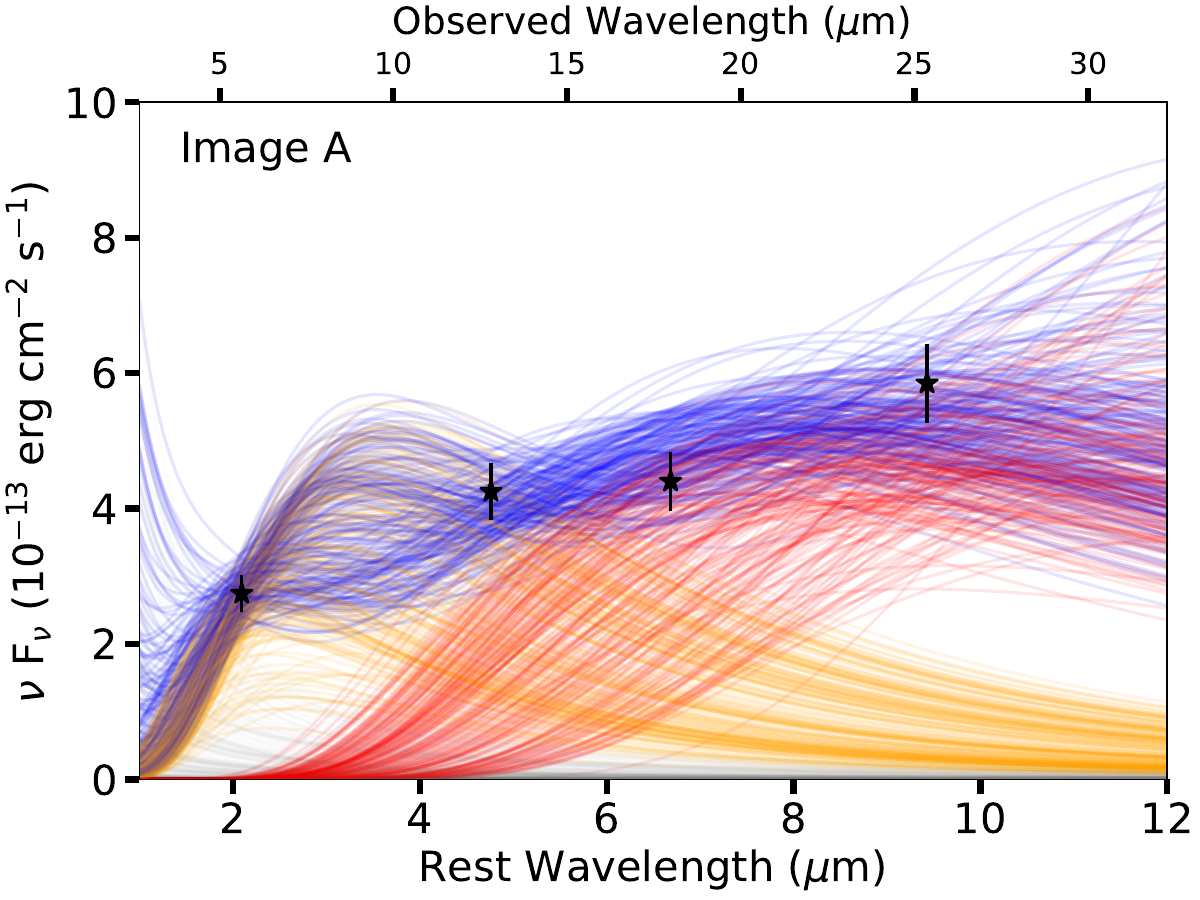}
    \includegraphics[width=0.49\textwidth]{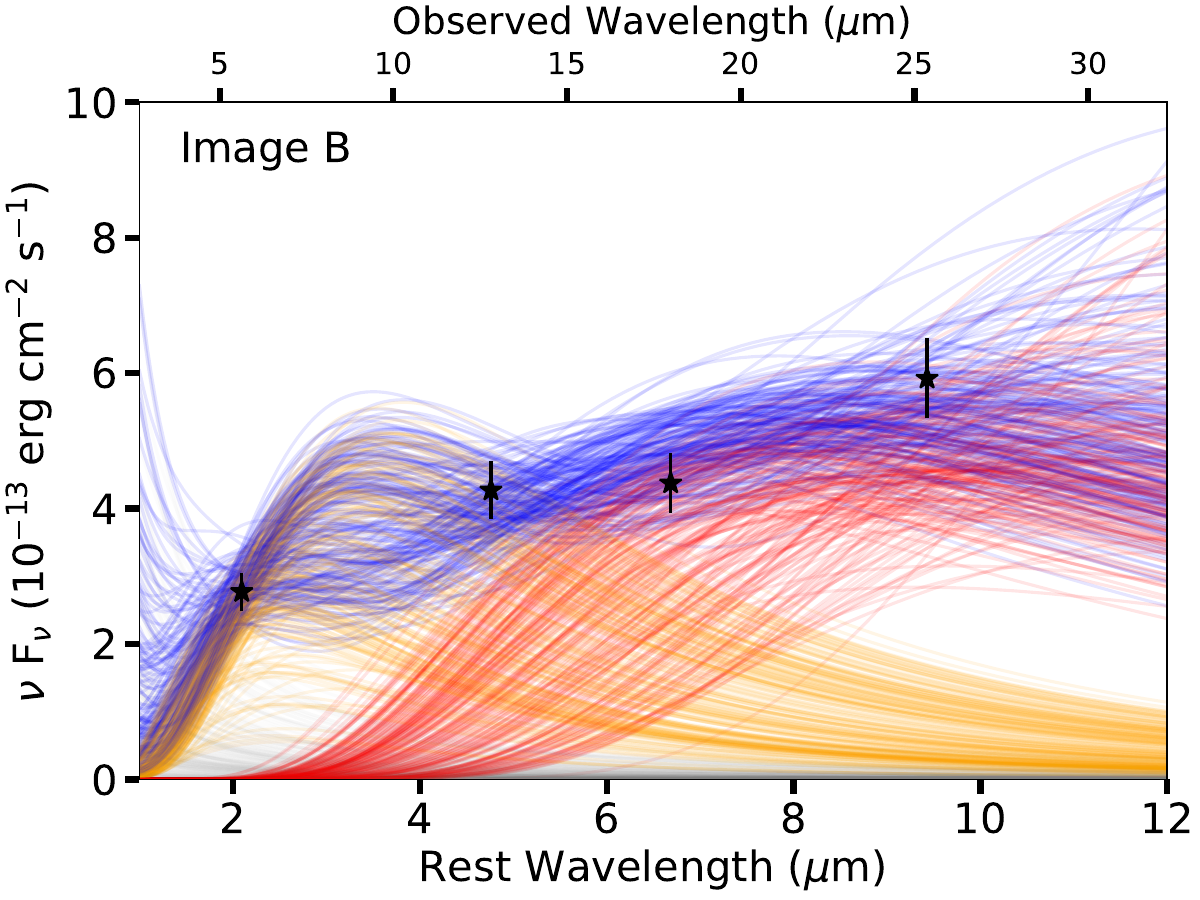}
    \includegraphics[width=0.49\textwidth]{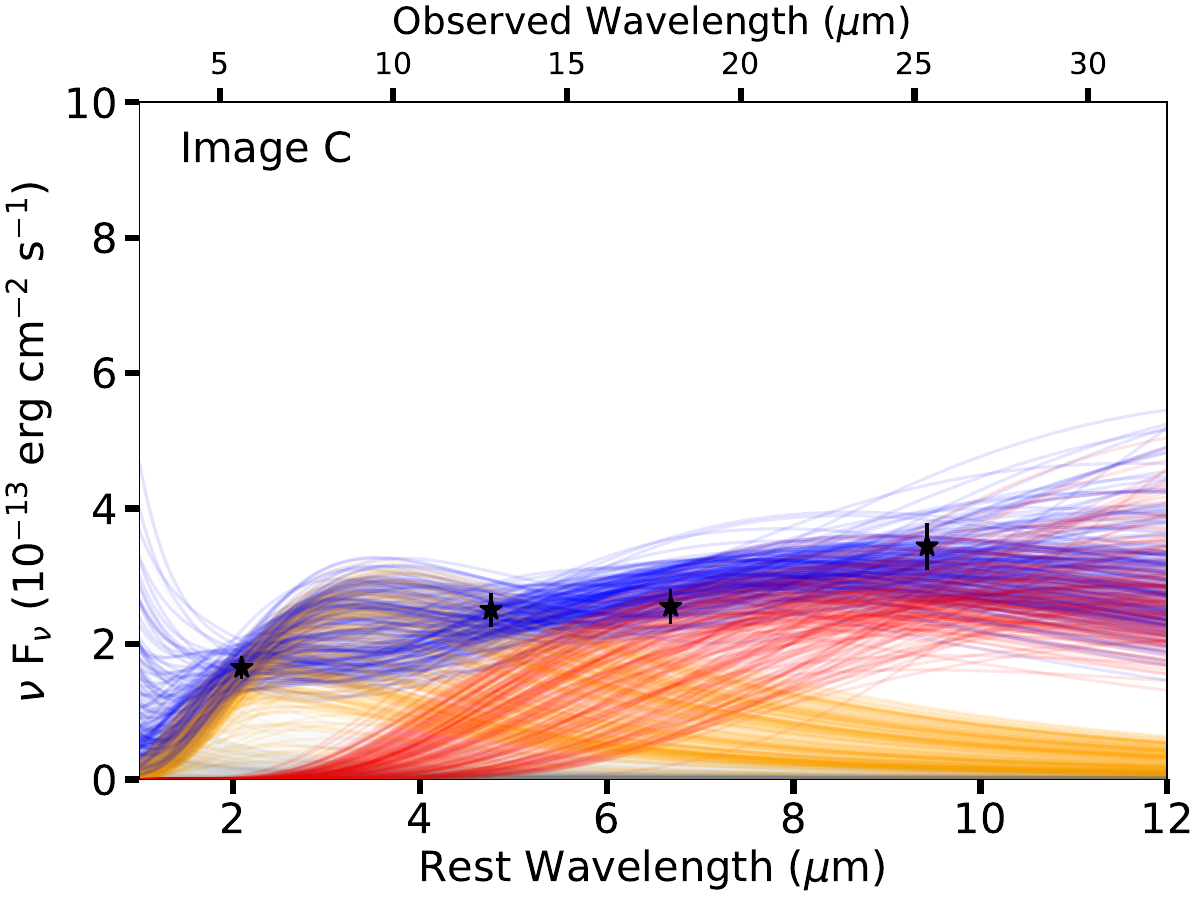}
    \includegraphics[width=0.49\textwidth]{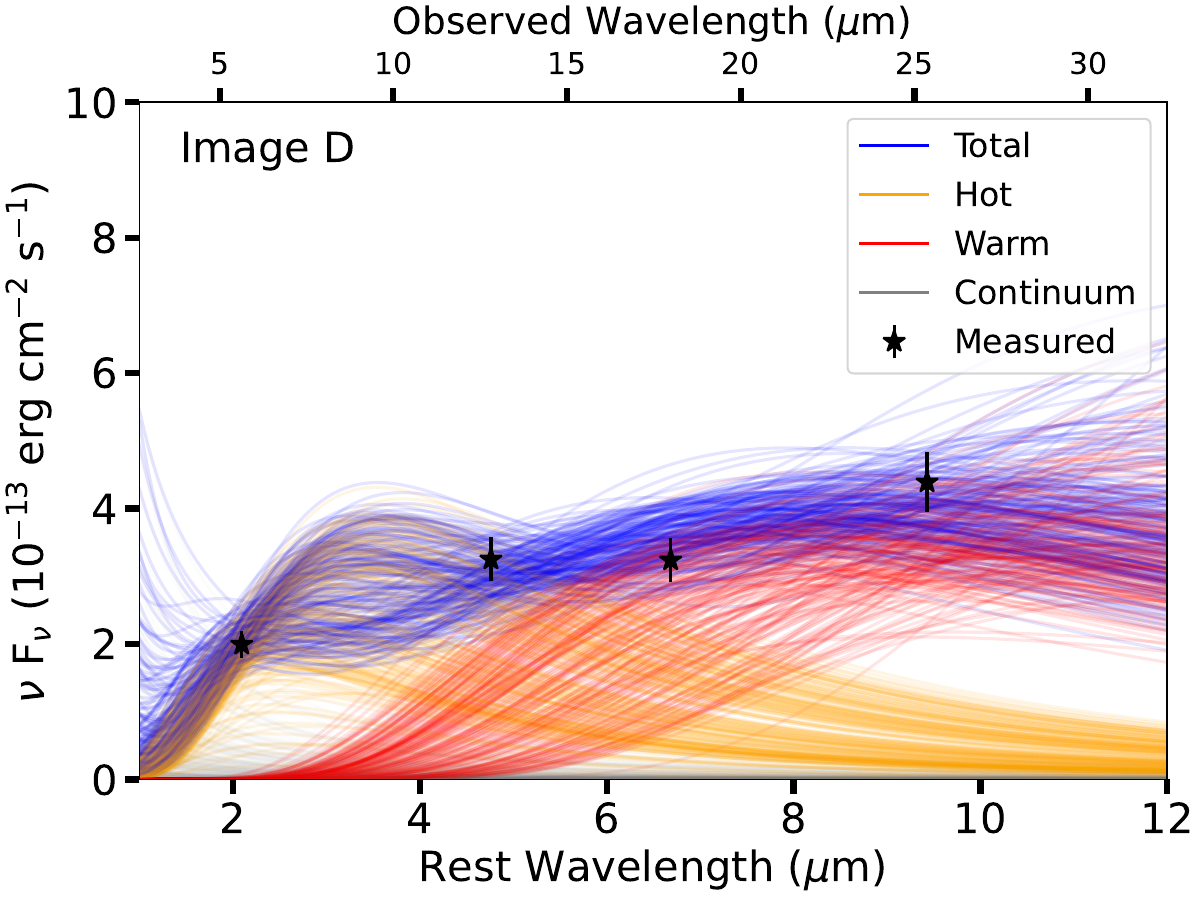}
    \caption{Posterior predictive distributions for the SED fitting for J1606.}
\end{figure*}

\begin{figure*}
    \includegraphics[width=\textwidth]{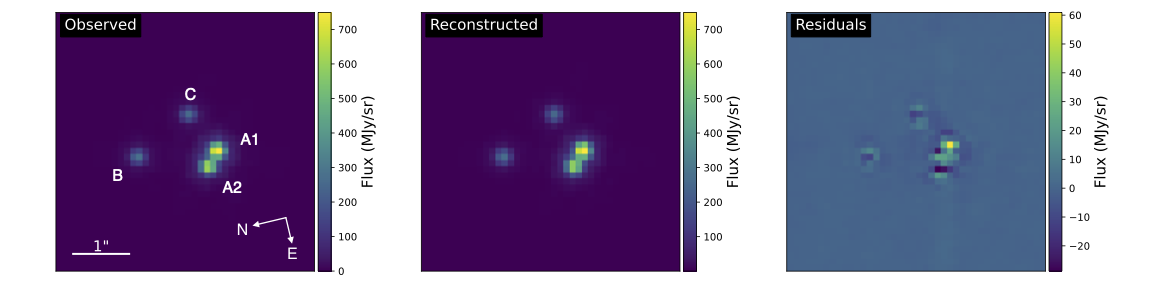}
    \caption{J2026 F560W.}
\end{figure*}
\begin{figure*}
    \includegraphics[width=\textwidth]{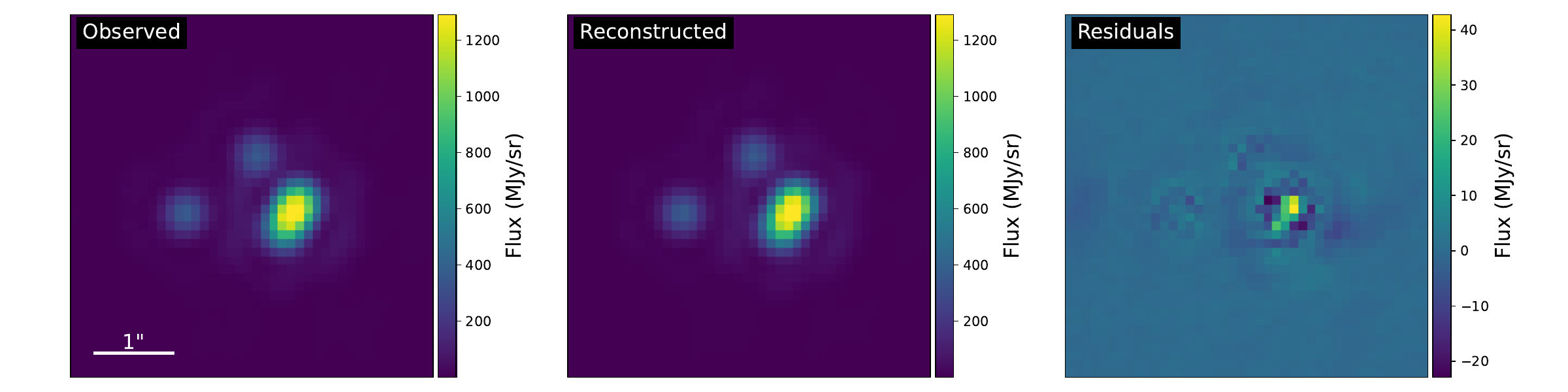}
    \caption{J2026 F1280W.}
\end{figure*}
\begin{figure*}
    \includegraphics[width=\textwidth]{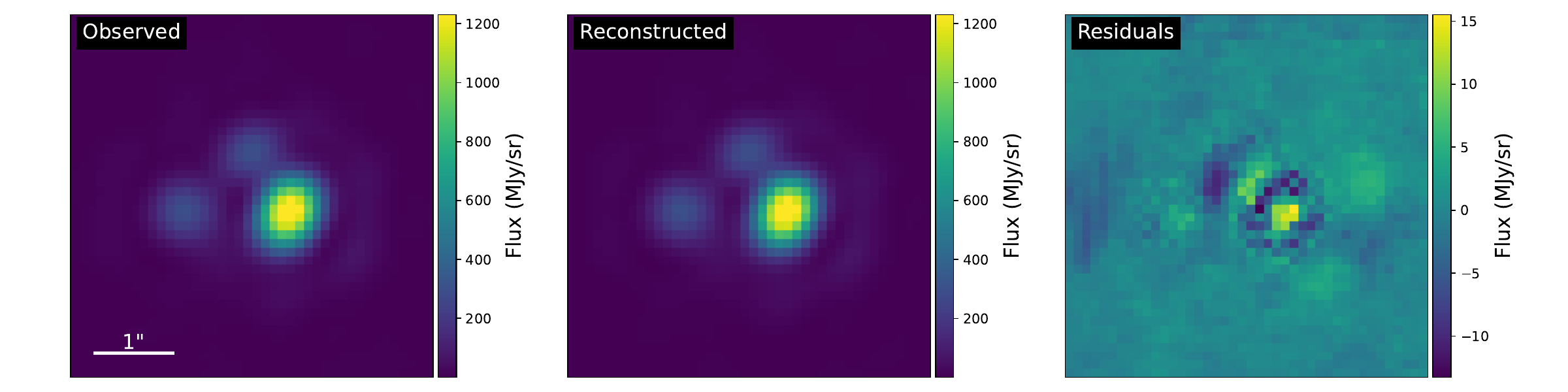}
    \caption{J2026 F1800W.}
\end{figure*}
\begin{figure*}
    \includegraphics[width=\textwidth]{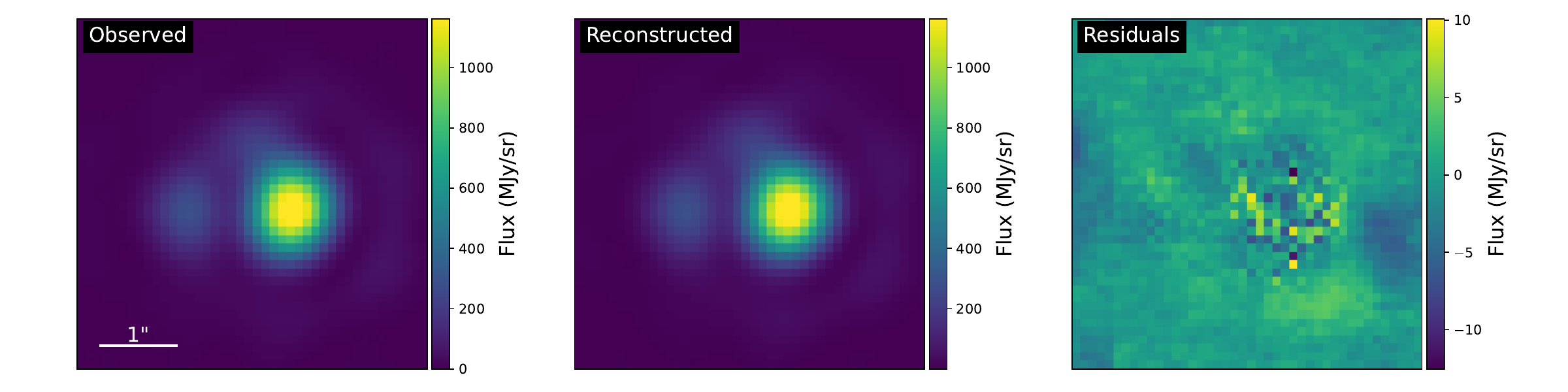}
    \caption{J2026 F2550W.}
\end{figure*}

\begin{figure*}
    \includegraphics[width=0.49\textwidth]{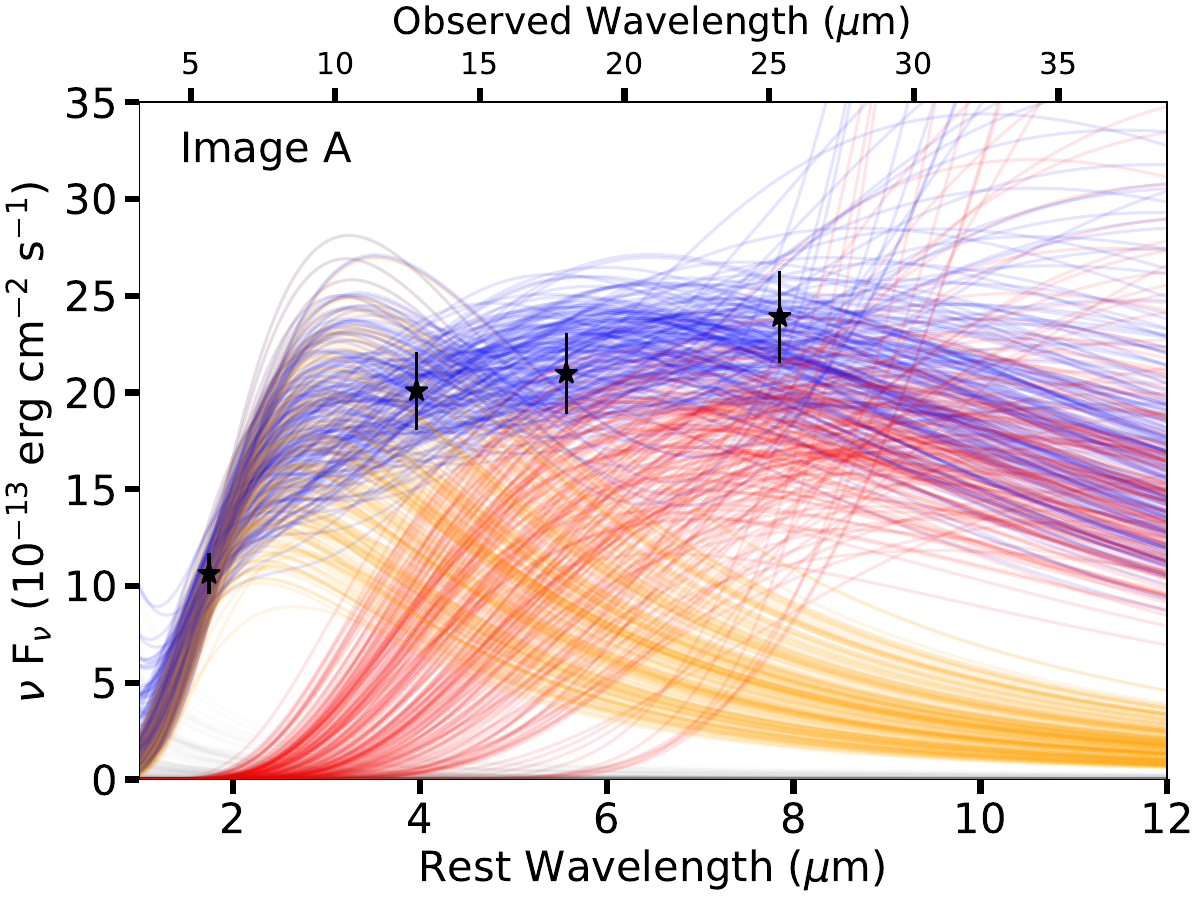}
    \includegraphics[width=0.49\textwidth]{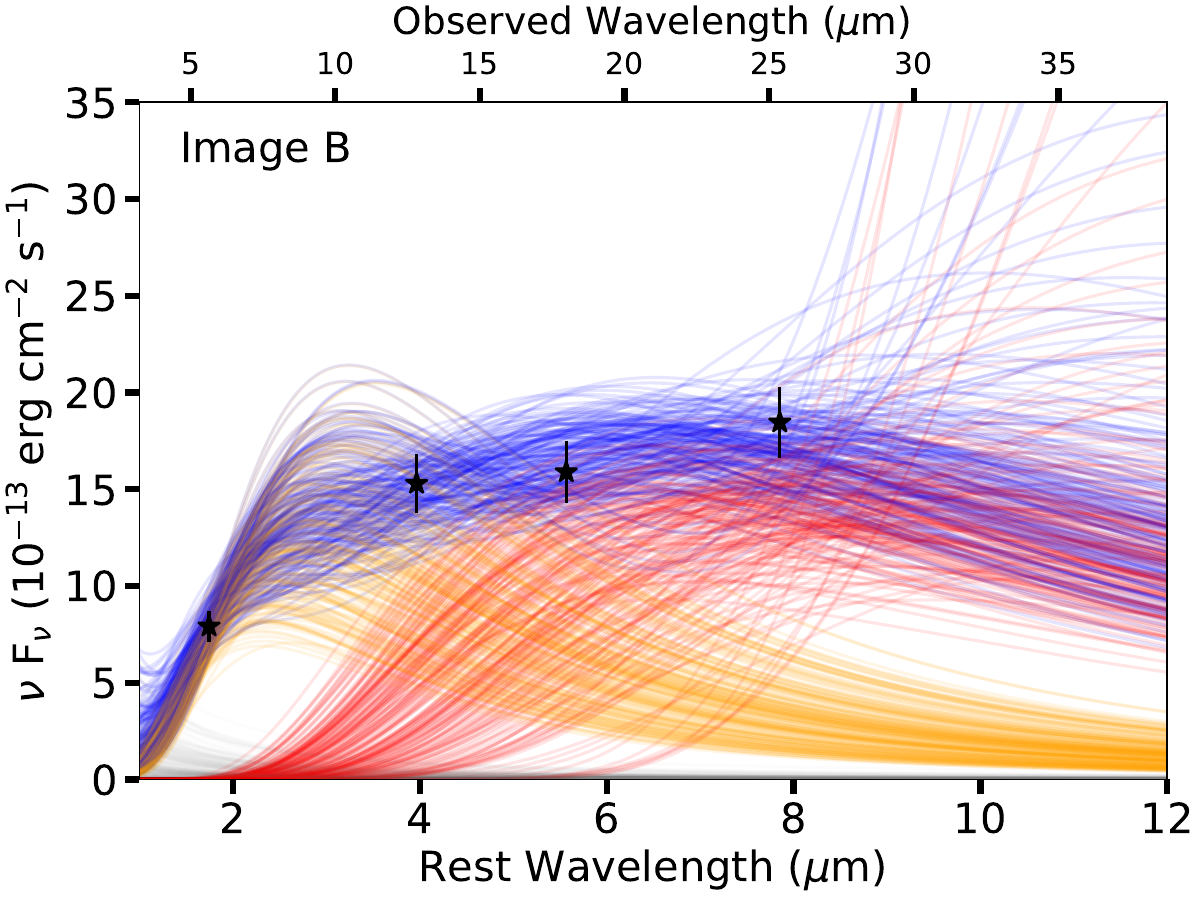}
    \includegraphics[width=0.49\textwidth]{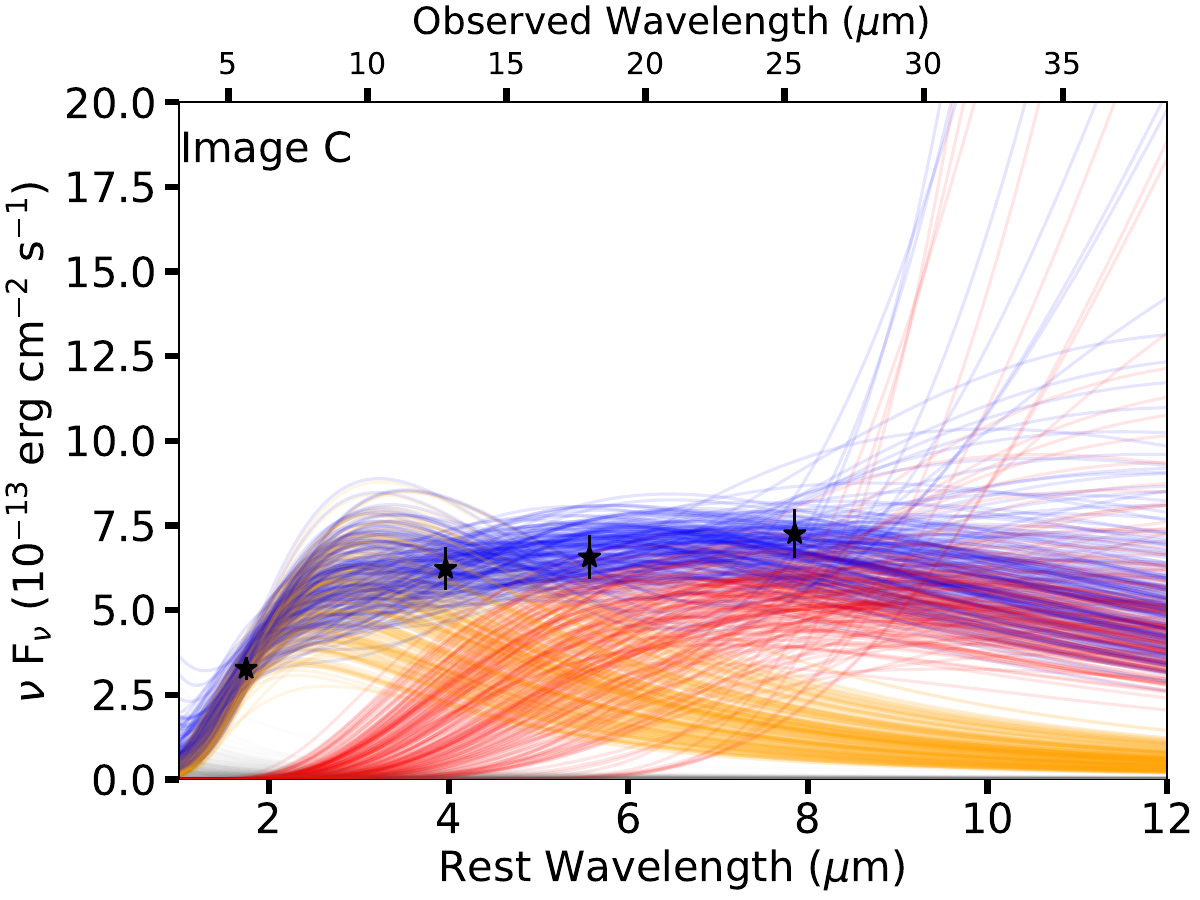}
    \includegraphics[width=0.49\textwidth]{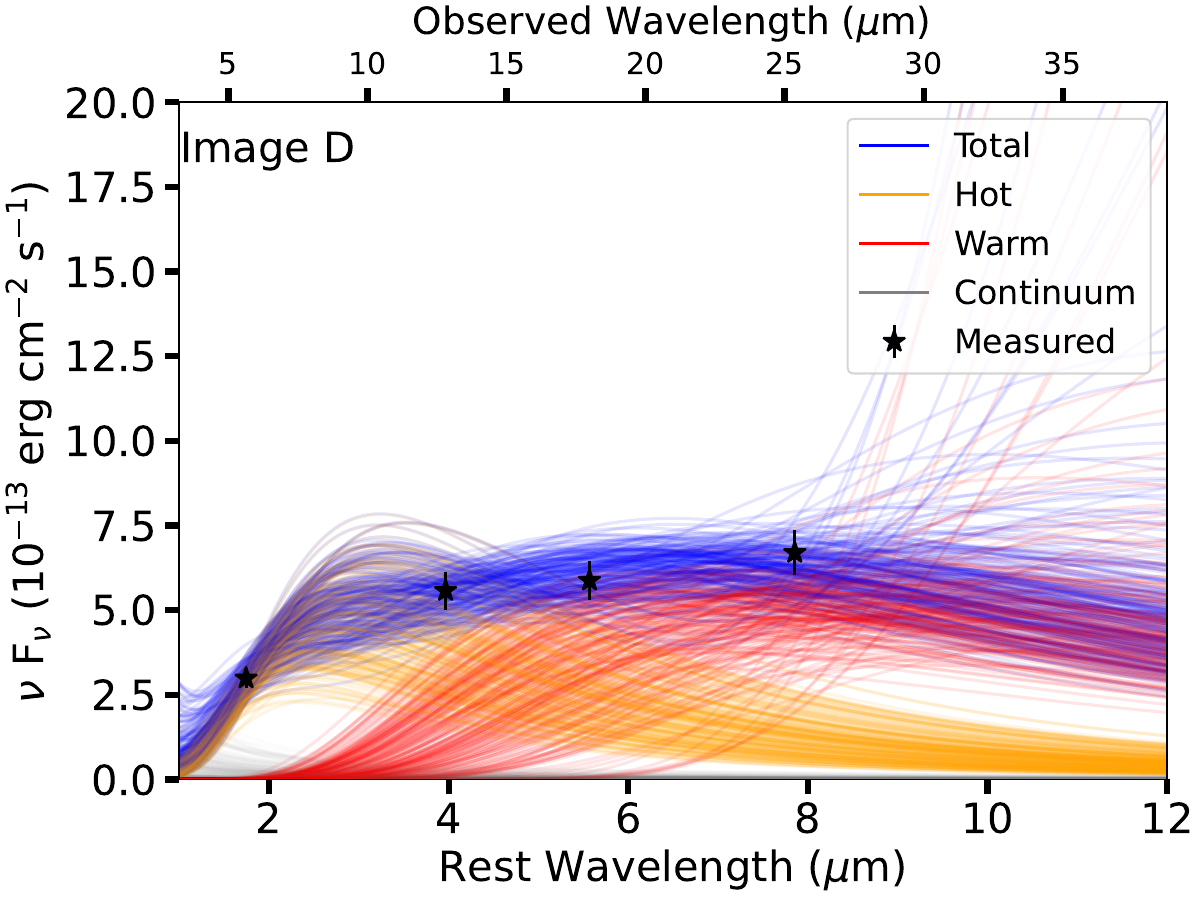}
    \caption{Posterior predictive distributions for the SED fitting for J2026.}
\end{figure*}

\begin{figure*}
    \includegraphics[width=\textwidth]{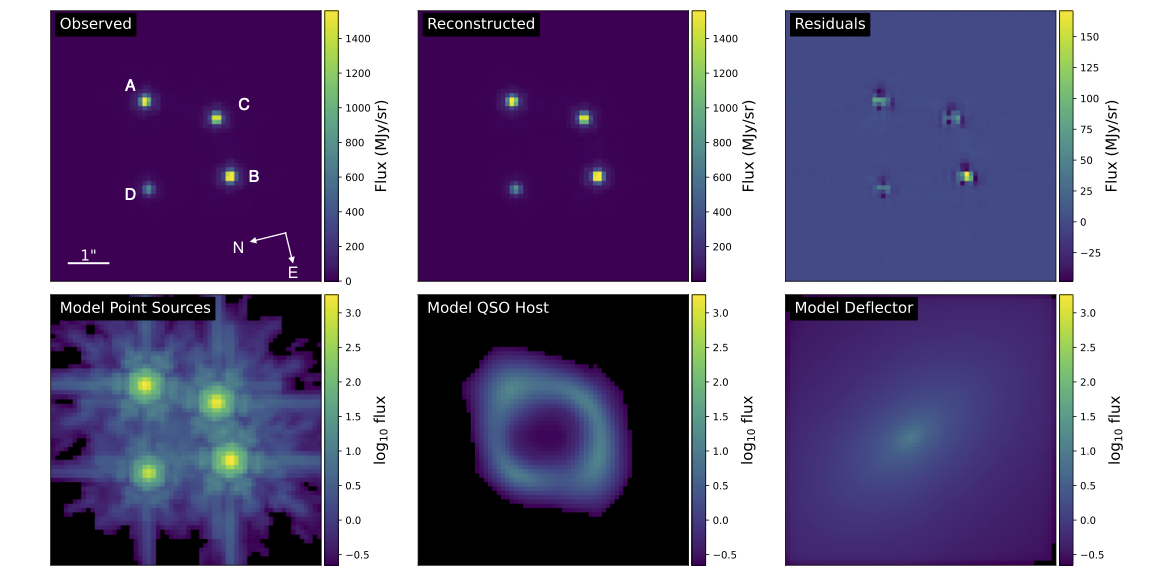}
    \caption{J2038 F560W}
\end{figure*}
\begin{figure*}
    \includegraphics[width=\textwidth]{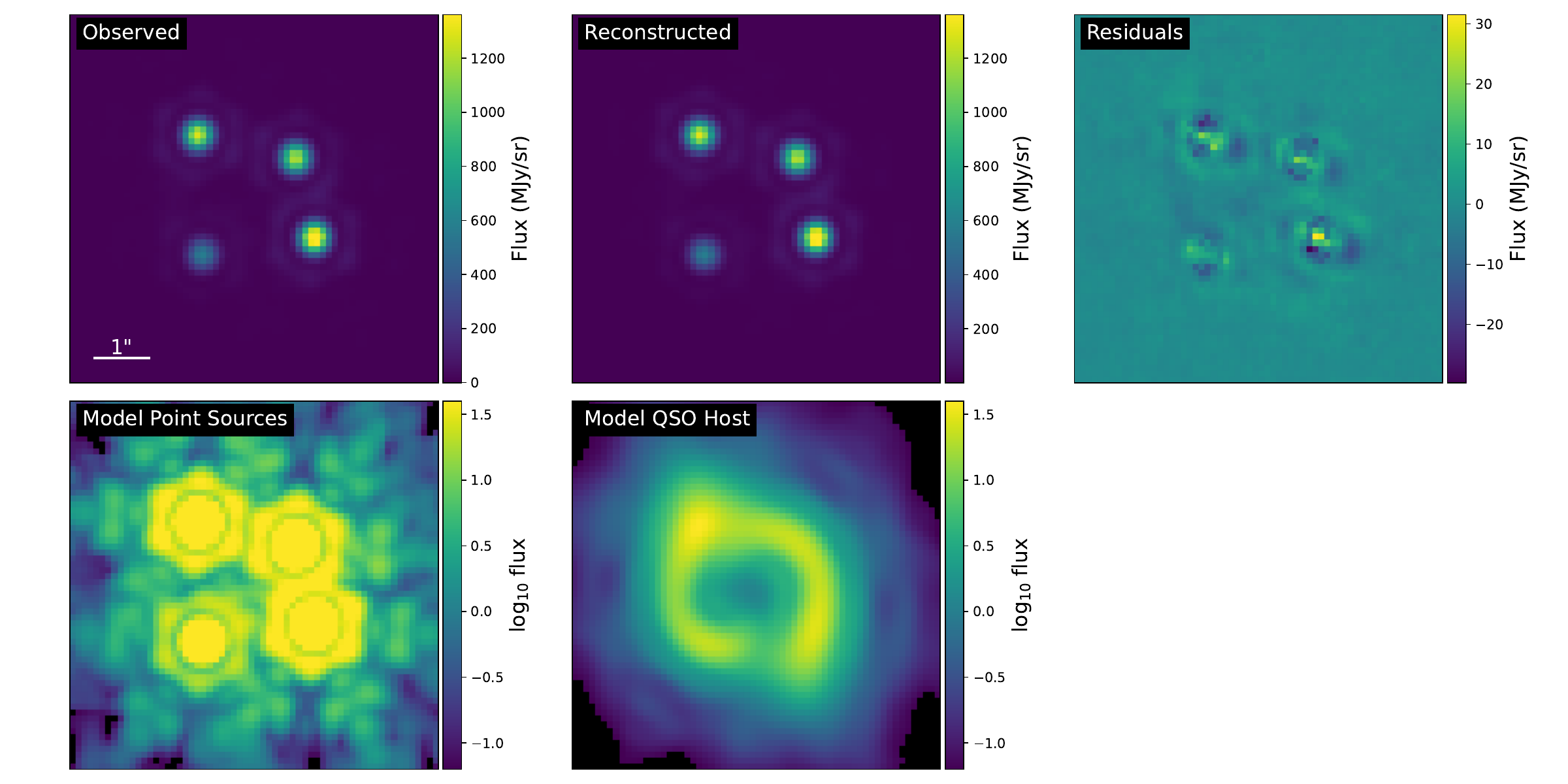}
    \caption{J2038 F1280W}
\end{figure*}
\begin{figure*}
    \includegraphics[width=\textwidth]{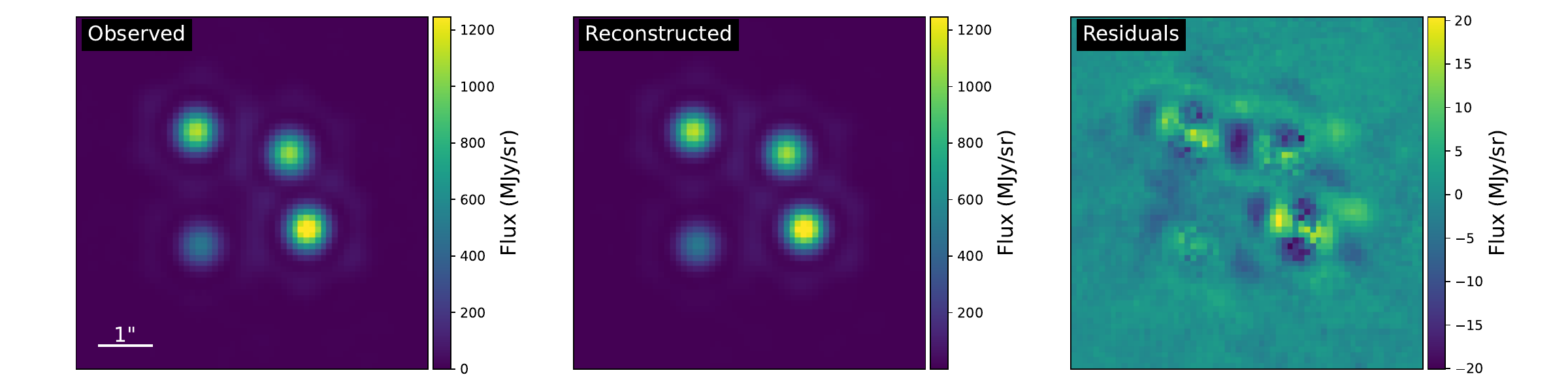}
    \caption{J2038 F1800W}
\end{figure*}
\begin{figure*}
    \includegraphics[width=\textwidth]{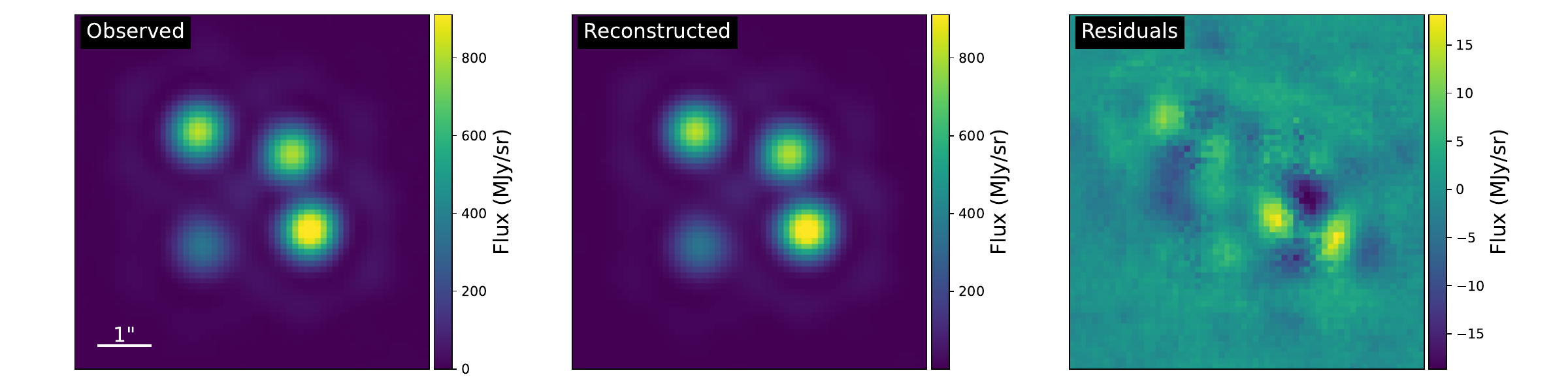}
    \caption{J2038 F2550W}
\end{figure*}

\begin{figure*}
    \includegraphics[width=0.49\textwidth]{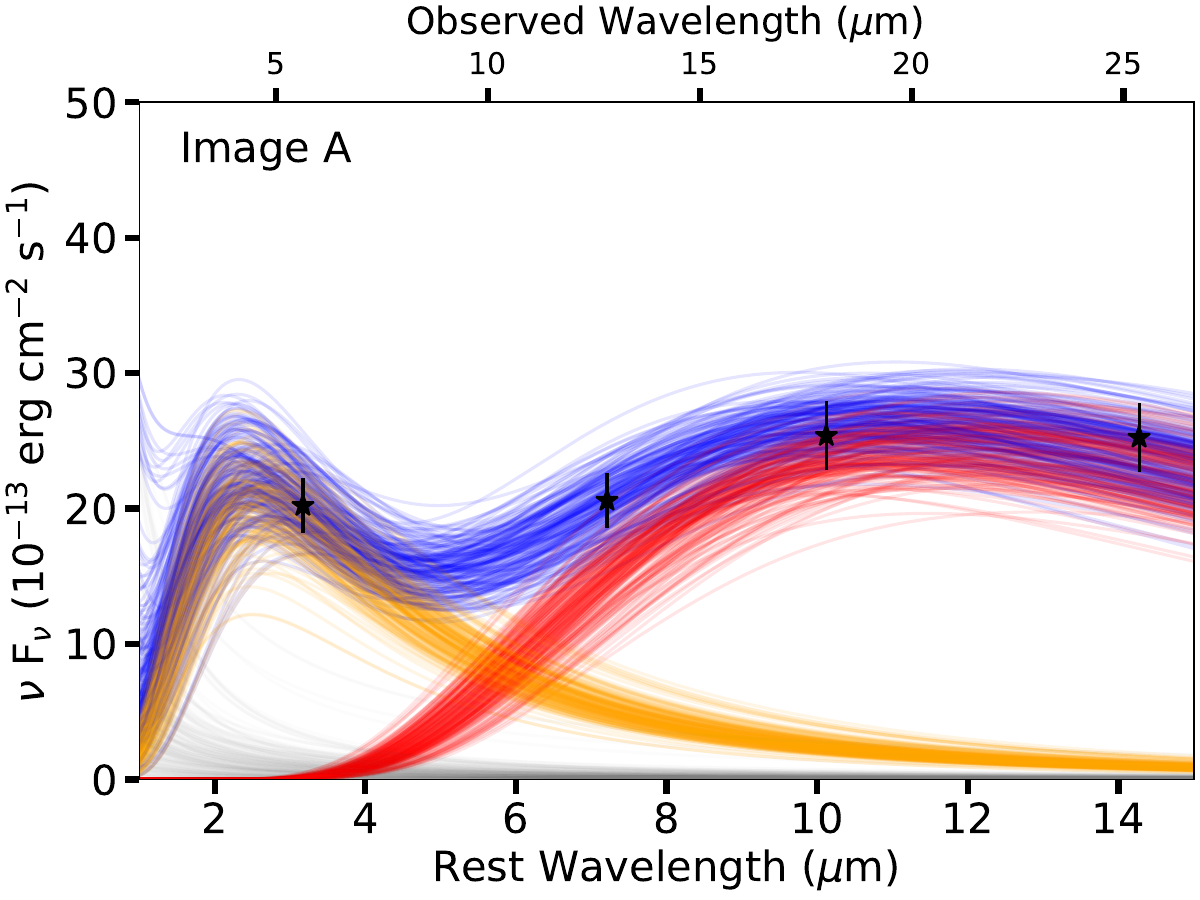}
    \includegraphics[width=0.49\textwidth]{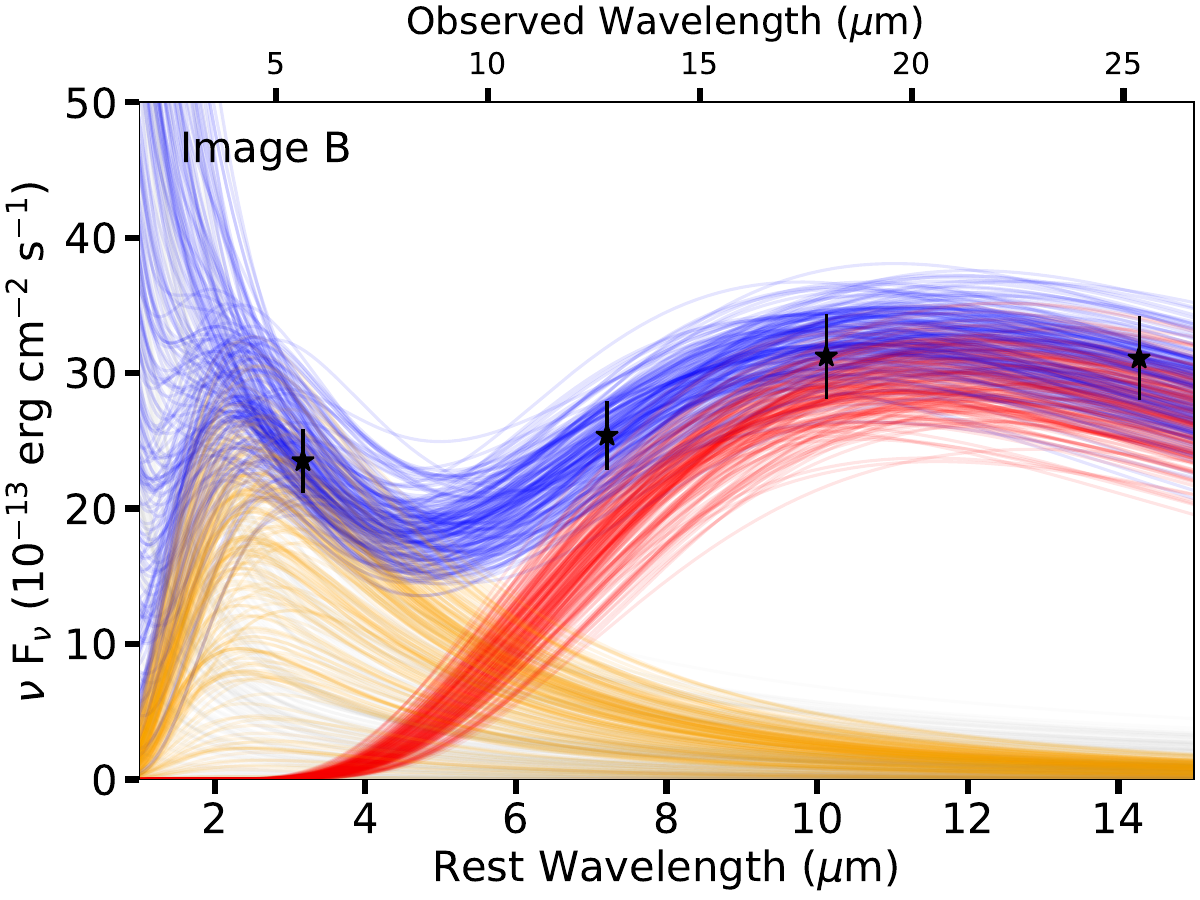}
    \includegraphics[width=0.49\textwidth]{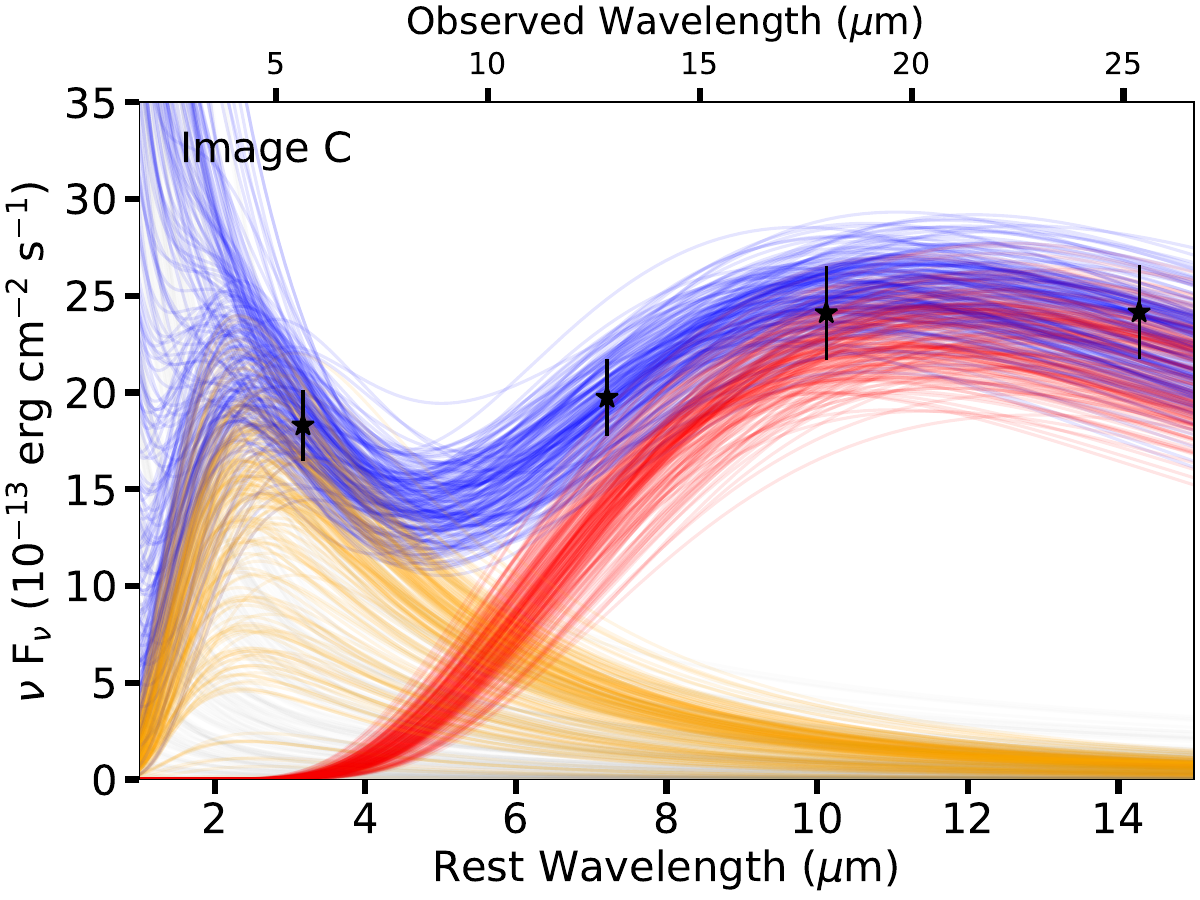}
    \includegraphics[width=0.49\textwidth]{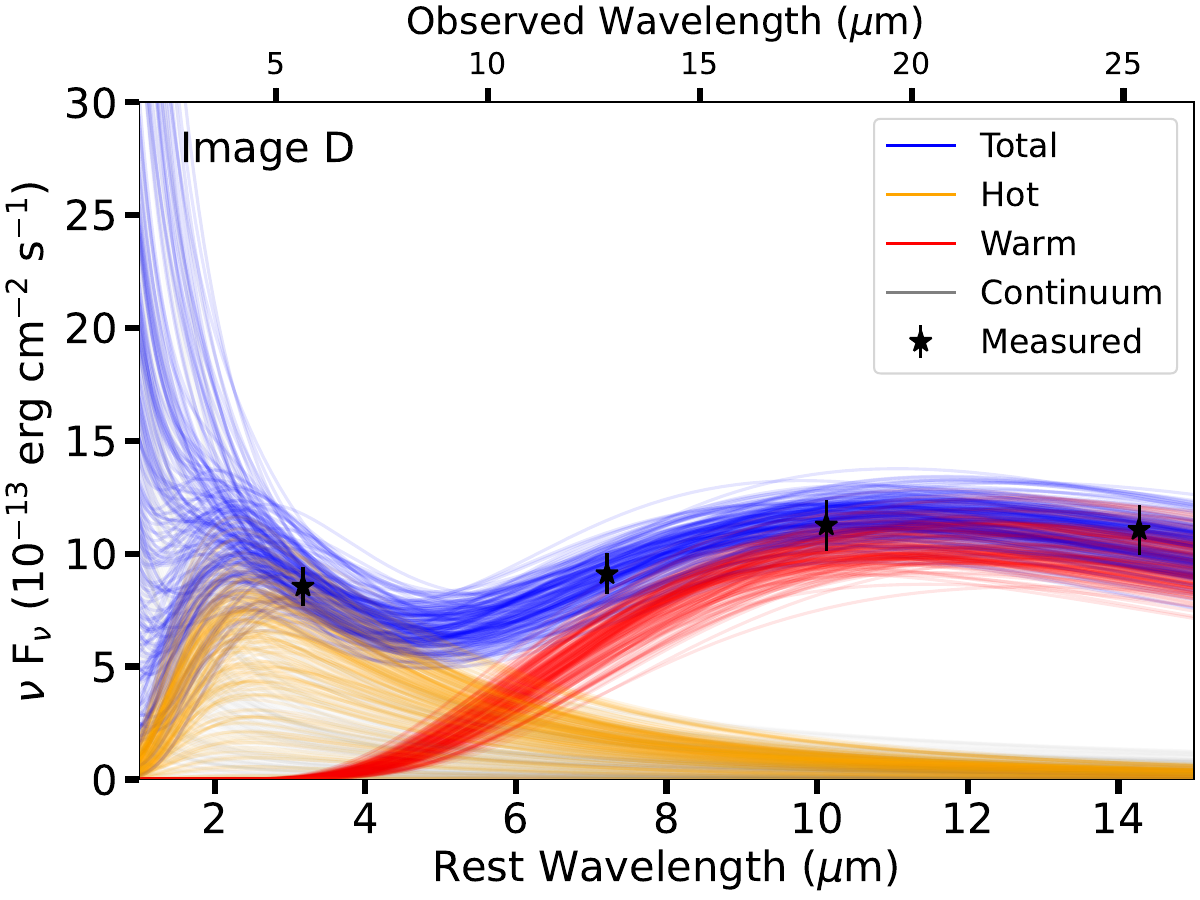}
    \caption{Posterior predictive distributions for the SED fitting for J2038}
\end{figure*}

% Don't change these lines
\bsp	% typesetting comment
\label{lastpage}
\end{document}